\documentclass[journal,twoside,web]{ieeecolor}
\usepackage{generic}
\usepackage{cite}
\usepackage{amsmath,amssymb,amsfonts,mathrsfs}
\usepackage{algorithm}
\usepackage{graphicx}
\let\labelindent\relax
\usepackage{textcomp,enumitem}
\usepackage{float}
\usepackage{subfig}
\usepackage{array,booktabs}
\usepackage[justification=centering]{caption}
\usepackage{bm}
\usepackage{algpseudocode}
\usepackage{color}
\usepackage{threeparttable,pifont}

\newtheorem{thm0}{Theorem}
\newtheorem{Theorem 1}[thm0]{Theorem}
\newtheorem{asm}{Assumption}
\newtheorem{Assumption 1}[asm]{Assumption}
\newtheorem{def0}{Definition}
\newtheorem{rmk}{Remark}
\newtheorem{Remark 1}[rmk]{Remark}
\newtheorem{lemma}{Lemma}
\newtheorem{Lemma 1}[lemma]{Lemma}
\newtheorem{Cor}{Corollary}
\newtheorem{Corollary 1}[Cor]{Corollary}

\newcolumntype{L}[1]{>{\raggedright\arraybackslash}m{#1}}
\newcolumntype{C}[1]{>{\centering\arraybackslash}m{#1}}
\newcommand{\diag}{\textrm{diag}}
\newcommand{\p}{\mathbb{P}}
\newcommand{\E}{\mathbb{E}}

\def\BibTeX{{\rm B\kern-.05em{\sc i\kern-.025em b}\kern-.08em
		T\kern-.1667em\lower.7ex\hbox{E}\kern-.125emX}}
\begin{document}
\title{\bf Differentially Private Gradient-Tracking-Based Distributed Stochastic Optimization over Directed Graphs \thanks{The work was supported by National Natural Science Foundation of China under Grants 62433020 and T2293770. The material in this paper was not presented at any conference.}}
\author{Jialong Chen, Jimin Wang, \IEEEmembership{Member, IEEE}, and Ji-Feng Zhang, \IEEEmembership{Fellow, IEEE}
\thanks{Jialong Chen is with the State Key Laboratory of Mathematical Sciences, Academy of Mathematics and Systems Science, Chinese Academy of Sciences, Beijing 100190, and also with the School of Mathematical Sciences, University of Chinese Academy of Sciences, Beijing 100049, China. (e-mail: chenjialong23@mails.ucas.ac.cn)}
\thanks{Jimin Wang is with the School of Automation and Electrical Engineering, University of Science and Technology Beijing, Beijing 100083, and also with the Key Laboratory of Knowledge Automation for Industrial Processes, Ministry of Education, Beijing 100083, China (e-mail: jimwang@ustb.edu.cn)}
\thanks{Ji-Feng Zhang is with the School of Automation and Electrical Engineering, Zhongyuan University of Technology, Zheng Zhou 450007; and also with the State Key Laboratory of Mathematical Sciences, Academy of Mathematics and Systems Science, Chinese Academy of Sciences, Beijing 100190, China. (e-mail: jif@iss.ac.cn)}}

\maketitle
\begin{abstract}
This paper proposes a differentially private gradient-tracking-based distributed stochastic optimization algorithm over directed graphs. In particular, privacy noises are incorporated into each agent’s state and tracking variable to mitigate information leakage, after which the perturbed states and tracking variables are transmitted to neighbors. We design two novel schemes for the step-sizes and the sampling number within the algorithm. The sampling parameter-controlled subsampling method employed by both schemes enhances the differential privacy level, and ensures a finite cumulative privacy budget even over infinite iterations. The algorithm achieves both almost sure and mean square convergence for nonconvex objectives. Furthermore, when nonconvex objectives satisfy the Polyak-{\L}ojasiewicz condition, \emph{Scheme (S1)} achieves a polynomial mean square convergence rate, and \emph{Scheme (S2)} achieves an exponential mean square convergence rate. The trade-off between privacy and convergence is presented. The effectiveness of the algorithm and its superior performance compared to existing works are illustrated through numerical examples of distributed training on the benchmark datasets ``MNIST'' and ``CIFAR-10''.
\end{abstract}\vspace{-0.3em}
\vspace{-0.8em}
\section{Introduction}
\IEEEPARstart{D}{istributed} optimization allows cooperative agents to compute and update their state variables through inter-agent communication to obtain an optimal solution of a common optimization problem (\!\!\!\cite{pu2020push}). Distributed stochastic optimization, a branch of distributed optimization, address scenarios where objectives are stochastic (\!\!\!\cite{bhavsar2023nonasymptotic}). This approach has found extensive applications across multiple domains, including distributed machine learning (\!\!\cite{shamir2014distributed}), cloud-based control systems~(\!\!\cite{chen2016dglb}), and the Internet of Things (\!\!\cite{wang2019design}). While it is frequently utilized in distributed stochastic optimization because of its adaptability in communication-efficient methods (\!\!\cite{doan2021}) and simplicity in algorithm structure (\!\!\cite{lu2023convergence}), stochastic gradient descent (SGD) does not guarantee the convergence over directed graphs (\!\!\cite[eq. (6)]{xie2018distributed}), and cannot achieve the exponential convergence rate (\!\!\cite[Th. 2]{qu2018harnessing}, \cite[eq. (2)]{xin2020variance}). To address these issues, the gradient-tracking method has been proposed over undirected graphs (\!\!~\cite{pu2021distributed,koloskova2021improved}). By developing tracking variables to track global stochastic gradients, \cite{xin2019distributed,pu2021distributed,koloskova2021improved} initially achieve the exponential convergence rate. The convergence analysis is further extended to directed graphs in \cite{zhao2024asymptotic,wang2023gradient,chen2024accelerated,lei2022distributed}. However, \cite{zhao2024asymptotic,chen2024accelerated,lei2022distributed} prove convergence under the assumption that weight matrices are row- and column-stochastic, which is often difficult to be satisfied in various practical scenarios (see e.g. \cite{chen2016dglb,wang2019design}). \cite{wang2023gradient} achieves the convergence by employing the two-time-scale step-sizes method, which removes the assumption that weight matrices are row- and column-stochastic, while requiring that the level sets of objectives are bounded.

\begin{table*}[tb]
	\centering
	\setlength{\tabcolsep}{0.5pt}
	\renewcommand{\arraystretch}{0.75}
	\caption{Comparison with existing works in distributed offline stochastic optimization}\label{table1}
	\begin{tabular}{C{2cm}C{2cm}C{2.5cm}C{2cm}C{3.4cm}C{5cm}}
		\toprule
		&Privacy budget&Convergence&Convergence rate&Graph topology& Key assumptions on objectives \par to achieve convergence\\
		\midrule
		\cite{xin2020variance} & N/A & Mean square & $O(\rho^K)$ & Row- and column-stochastic connected directed graphs& Strongly convex\\
		\midrule
		\cite{xin2019distributed} & N/A & \ding{55} & $O(\rho^K)$ & Strongly connected \par directed graphs& Strongly convex\\
		\midrule
		\cite{wang2023gradient} & N/A & Almost sure& - & Strongly connected \par directed graphs& Convex,\par bounded level sets\\
		\midrule
		\cite{ding2021differentially} & Per-iteration & Mean square& $O(\frac{1}{K^{\frac{1}{3}}})$ & Row- and column-stochastic undirected graphs& Strongly convex and nonconvex,\par bounded gradients\\
		\midrule
		\cite{wang2023decentralized} & Per-iteration & Almost sure& - & Row- and column-stochastic undirected graphs& Nonconvex,\par bounded gradients\\
		\midrule
		\cite{wang2023quantization} & Per-iteration & Mean square& $O(1)$ & Connected undirected graphs& Convex and nonconvex,\par bounded gradients\\
		\midrule
		\cite{chen2024differentially} & Finite cumulative & Mean square& $O(\frac{1}{K^{\frac{1}{3}}})$ & Row- and column-stochastic undirected graphs& Nonconvex,\par Polyak-{\L}ojasiewicz condition\\
		\midrule
		{\bf Scheme \emph{(S1)} (This work)} & Finite cumulative & Mean square \par\& Almost sure& $O(\frac{1}{K^{\frac{1}{3}}})$ & Directed graphs with spanning trees& Nonconvex,\par Polyak-{\L}ojasiewicz condition\\
		\midrule
		{\bf Scheme \emph{(S2)} (This work)} & Finite cumulative & Mean square \par\& Almost sure& $O(\rho^K)$ & Directed graphs with spanning trees& Nonconvex\\
		\bottomrule
	\end{tabular}
	\begin{tablenotes}
		\footnotesize
		\item ``N/A'' means privacy protection is not considered, ``\ding{55}'' means the convergence is not achieved, and ``-'' means the convergence rate is not given.
	\end{tablenotes}
	\vspace{-2.5em}
\end{table*}

When cooperative agents exchange information to address a distributed stochastic optimization problem, adversaries can infer stochastic gradients from the exchanged information, and further obtain agents' sensitive information through model inversion attacks (\!\!\!\cite{fredrikson2015model,zhu2019deep}). To address this issue, various privacy-preserving techniques have been developed (\!\!\cite{zhang2021}), such as homomorphic encryption (\!\!\cite{lu2018privacy,tan2023cooperative}), state decomposition (\!\!\cite{wang2019privacy}), random coupling weights (\!\!\cite{gao2023dynamics}), uncoordinated step-sizes (\!\!\cite{wang2024decentralized}), network augmentation (\!\!\cite{ramos2024privacy}), and adding noises (\!\!\cite{mo2017privacy,dwork2014algorithmic,le2014differentially,guo2025state}). Because of its simplicity of use and immunity to post-processing, differential privacy (\!\!\cite{dwork2014algorithmic,le2014differentially}) has attracted considerable interest and has been extensively applied in distributed optimization for both deterministic and stochastic objectives.  When objectives are deterministic,  based on the gradient-tracking method,  differentially private distributed optimization has been well developed in \cite{ding2022differentially,xuan2023gradient,xie2024differentially,wang2024tailoring,huang2024differential,huo2024differentially}.  Among others, \cite{ding2022differentially,xuan2023gradient,huang2024differential,wang2024tailoring,huo2024differentially} have successfully achieved the finite cumulative differential privacy budget over infinite iterations.  However, the difficulty caused by stochastic objectives makes the methods in \cite{ding2022differentially,xuan2023gradient,xie2024differentially,wang2024tailoring,huang2024differential,huo2024differentially} unsuitable to differentially private distributed stochastic optimization. In addition, to achieve convergence, (strongly) convex objectives (\!\!\cite{ding2022differentially,xuan2023gradient,wang2024tailoring,huang2024differential,huo2024differentially}) and nonconvex objectives with the Polyak-{\L}ojasiewicz condition (\!\!\cite{xie2024differentially}) are required. However, these requirements may be hard to be satisfied or verified in practice. 

When objectives are stochastic, a method based on SGD has been proposed for differentially private distributed stochastic optimization. Some interesting works can be found in \cite{ding2021differentially,kang2021weighted,xu2022dp,wang2023decentralized,wang2023quantization,liu2024distributed,yan2024killing}, while these works only give the per-iteration differential privacy budget, and thus, cannot protect the sensitive information over infinite iterations. Fortunately, by using sequentially acquired data samples inherent in online learning (\!\!\cite{chen2024locallya,chen2024locallyb}), the time-varying sampling number method (\!\!\cite{wang2024differentiallya}) and the sampling parameter-controlled subsampling method (\!\!\cite{chen2024differentially}), the finite cumulative differential privacy budget over infinite iterations is given. However, the differential privacy is tailored for distributed SGD in \cite{chen2024locallya,chen2024differentially,wang2024differentiallya}, respectively. Since the gradient-tracking method has shown advantages over the distributed-SGD method regarding the convergence over directed graphs, the differentially private gradient tracking-based distributed online stochastic optimization with streaming data over directed graphs is first studied in \cite{chen2024locallyb}. To the	best of our knowledge, differentially private gradient-tracking-based distributed offline stochastic optimization with pre-downloaded/local datasets over directed graphs has not been studied yet. As a result, the differentially private distributed {\it offline} stochastic optimization based on the gradient-tracking method is a challenging issue, especially on how to achieve the finite cumulative differential privacy budget even over infinite iterations, the almost sure and mean square convergence for nonconvex objectives without the Polyak-{\L}ojasiewicz condition, and the exponential mean square convergence rate. 

Summarizing the discussion above, we propose a new differentially private gradient-tracking-based distributed stochastic optimization algorithm with two schemes of step-sizes and the sampling number over directed graphs. \emph{Scheme (S1)} employs the polynomially decreasing step-sizes and the increasing sampling number with the maximum iteration number. \emph{Scheme~(S2)} employs constant step-sizes and the exponentially increasing sampling number with the maximum iteration number. Comparison with existing works in distributed offline stochastic optimization is presented in Table \ref{table1}, and the main contribution of this paper is as follows:
\begin{itemize}[leftmargin=*]
	\item The sampling parameter-controlled subsampling method is employed to enhance the differential privacy level of the algorithm. The algorithm with both schemes achieves the finite cumulative differential privacy budget even over infinite iterations. To the best of our knowledge, a finite cumulative differential privacy budget over infinite iterations is achieved in differentially private gradient-tracking-based distributed {\it offline} stochastic optimization for the first time.
	
	\item The almost sure and mean square convergence of the algorithm are given for nonconvex objectives without the Polyak-{\L}ojasiewicz condition. Furthermore, when nonconvex objectives satisfy the Polyak-{\L}ojasiewicz condition, the polynomial mean square convergence rate is achieved for \emph{Scheme (S1)}, and the exponential mean square convergence rate is achieved for \emph{Scheme (S2)}.
	
	\item Two schemes are shown to achieve the finite cumulative differential privacy budget over infinite iterations and mean square convergence simultaneously. For \emph{Scheme~(S1)}, the polynomial mean square convergence rate and the cumulative differential privacy budget are achieved simultaneously even over infinite iterations for general privacy noises, including decreasing, constant and increasing privacy noises. For \emph{Scheme~(S2)}, the exponential mean square convergence rate and the cumulative differential privacy budget are achieved simultaneously even over infinite iterations. 
\end{itemize}

The remainder of this paper is organized as follows: Section~\ref{section 2} presents preliminaries and the problem formulation. Section \ref{section 3} provides the algorithm with its convergence and privacy analysis. Section \ref{section 4} verifies the effectiveness of the algorithm through numerical examples of distributed training on the benchmark datasets ``MNIST'' and ``CIFAR-10''. Finally, Section \ref{section 5} concludes the paper.

\emph{Notation.} $\mathbb{R}$, $\mathbb{C}$, and $\mathbb{R}^{n}$ denote the set of real numbers, the set of complex numbers, and $n$-dimensional Euclidean space, respectively. $\mathbf{1}_n$ denotes a $n$-dimensional vector whose elements are all 1, and $\|x\|$ denotes the standard Euclidean norm of a vector $x$. $X\sim\text{Lap}(b)$ refers to a random variable that has a Laplacian distribution with the variance parameter $b>0$, and the probability density function of the random variable $X$ is given by $p(x;b)=\frac{1}{2b}\exp\left(-\frac{|x|}{b}\right)$. For a matrix $A\in\mathbb{R}^{n\times n}$, $A^\top$, $\rho(A)$ stand for its transpose and spectral radius, respectively. $\langle \cdot,\cdot \rangle$ denotes the inner product. $(\Omega,\mathcal{F},\p)$ denotes a probability space, $\p(B)$ and $\mathbb{E} X$ stand for the probability of an event $B\in\mathcal{F}$ and the expectation of the random variable $X$, respectively. $\otimes$ denotes the Kronecker product of matrices. $\lfloor a \rfloor$ denotes the largest integer which is not larger than $a$. For a differentiable function $f(x)$, $\nabla f(x)$ denotes its gradient at the point $x$. For a vector $x=[x_1,x_2,\dots,x_n]^\top\in\mathbb{R}^{n}$, the notation $\diag(x)$ denotes the diagonal matrix with diagonal elements being $x_1,x_2,\dots,x_n$. For a complex number $\varpi\in\mathbb{C}$, $\text{Re}(\varpi)$ stands for its real part. $\mathbb{I}_{\{\cdot\}}$ denotes the indicator function, whose value is 1 if its argument is true, and 0, otherwise.
\section{Preliminaries and problem formulation}\label{section 2}
\subsection{Graph theory}
In this paper, we consider a network of $n$ agents which exchange the information over two different directed graphs $\mathcal{G}_{\mathcal{R}}=(\mathcal{V},\mathcal{E}_{\mathcal{R}})$ and $\mathcal{G}_{\mathcal{C}}=(\mathcal{V},\mathcal{E}_{\mathcal{C}})$. $\mathcal{V}=\{1,2,\dots,n\}$ is the set of all agents, and $\mathcal{E}_{\mathcal{R}}$, $\mathcal{E}_{\mathcal{C}}$ are sets of directed edges in $\mathcal{G}_{\mathcal{R}}$, $\mathcal{G}_{\mathcal{C}}$, respectively. In our gradient-tracking algorithm, agents exchange state variables over $\mathcal{G}_{\mathcal{R}}$ and tracking variables over $\mathcal{G}_{\mathcal{C}}$. Directed graphs $\mathcal{G}_{\mathcal{R}}$ and $\mathcal{G}_{\mathcal{C}}$ are induced by the weight matrix $\mathcal{R}=(\mathcal{R}_{ij})_{i,j=1,\dots,n}$ and $\mathcal{C}=(\mathcal{C}_{ij})_{i,j=1,\dots,n}$, respectively. Any element $\mathcal{R}_{ij}$ of $\mathcal{R}$ is either strictly positive if Agent $i$ can receive Agent $j$'s state variable, or 0, otherwise. The same property holds for any element $\mathcal{C}_{ij}$ of $\mathcal{C}$. For any $i\in\mathcal{V}$, its in-neighbor and out-neighbor set of over $\mathcal{G}_{\mathcal{R}}$ are defined as $\mathcal{N}_{\mathcal{R},i}^{-}=\{\mbox{$j\in\mathcal{V}:$ }\mathcal{R}_{ij}>0\}$ and $\mathcal{N}_{\mathcal{R},i}^{+}=\{\mbox{$j\in\mathcal{V}:$ }\mathcal{R}_{ji}>0\}$, respectively. Similarly, Agent $i$'s in-neighbor and out-neighbor set over $\mathcal{G}_{\mathcal{C}}$ are defined as $\mathcal{N}_{\mathcal{C},i}^{+}$ and $\mathcal{N}_{\mathcal{C},i}^{+}$, respectively. Denote the in-Laplacian matrix of $\mathcal{R}$ and the out-Laplacian matrix of $\mathcal{C}$ as $\mathcal{L}_1=\diag(\mathcal{R}\cdot\mathbf{1}_n)-\mathcal{R}$ and $\mathcal{L}_2=\diag(\mathbf{1}_n^\top \cdot\mathcal{C})-\mathcal{C}$, respectively. Then, the assumption about directed graphs $\mathcal{G}_{\mathcal{R}}$, $\mathcal{G}_{\mathcal{C}}$ is given as follows:
\begin{asm}\label{asm1}
	Let $\mathcal{G}_{\mathcal{R}}$ and $\mathcal{G}_{\mathcal{C}^\top}$ be directed graphs induced by nonnegative matrices $\mathcal{R}$ and $\mathcal{C}^\top$, respectively. Then, both $\mathcal{G}_{\mathcal{R}}$ and $\mathcal{G}_{\mathcal{C}^\top}$ contain at least one spanning tree. Moreover, there exists at least one agent being a root of spanning trees in both $\mathcal{G}_{\mathcal{R}}$ and $\mathcal{G}_{\mathcal{C}^\top}$.
\end{asm}
\begin{rmk}
	Directed graphs in Assumption \ref{asm1} are allowed to have self-loops, which are commonly used in distributed stochastic optimization (see e.g. \cite{qu2018harnessing,xin2020variance,pu2021distributed,zhao2024asymptotic}). More importantly, directed graphs in Assumption \ref{asm1} are more general than connected undirected graphs in \cite{koloskova2021improved,ding2022differentially,xie2024differentially,ding2021differentially,kang2021weighted,xu2022dp,wang2023decentralized,wang2023quantization,wang2024differentiallya,chen2024differentially,chen2024locallya}, row- and column-stochastic connected directed graphs in \cite{doan2021,xin2020variance,pu2021distributed,zhao2024asymptotic,gao2023dynamics}, and strongly connected directed graphs in \cite{wang2023gradient,xin2019distributed,chen2024locallyb}. In addition, by \cite[Th. 3.8]{ren2005consensus}, Assumption~\ref{asm1} is a necessary condition for the consensus of Agents' state and tracking variables.
\end{rmk}
Based on Assumption \ref{asm1}, we have the following useful lemma for weight matrices $\mathcal{R}$ and $\mathcal{C}$:
\begin{lemma}\label{lemma1} 
	If Assumption~\ref{asm1} holds, then following statements hold:
	
	\indent(i) Let $\{\varpi_1^{(1)}\!\!\!,\dots,\varpi_n^{(1)}\!\}$ be the eigenvalues of the matrix $\mathcal{L}_1$ such that $|\varpi_1^{(1)}|\leq\dots\leq|\varpi_n^{(1)}|$, and $\{\varpi_1^{(2)}\!\!\!,\dots,\varpi_n^{(2)}\!\}$ be the eigenvalues of the matrix $\mathcal{L}_2$ such that $|\varpi_1^{(2)}|\leq\dots\leq|\varpi_n^{(2)}|$. Then, $\varpi_1^{(1)}=\varpi_1^{(2)}=0$ and $\text{Re}(\varpi_l^{(1)})>0,\text{Re}(\varpi_l^{(2)})>0$ for any $l=2,\dots,n$.
	
	\indent(ii) Let matrices $W_1=I_n-\frac{1}{n}\mathbf{1}_nv_1^\top$, $W_2=I_n-\frac{1}{n}v_2 \mathbf{1}_n^\top$, and step-sizes satisfy\vspace{-1em}
		\begin{align*}
			&0<\alpha_K<\min\{\min_{i\in\mathcal{V}}\{\frac{1}{\sum_{\hspace{-0.15em}j\in\mathcal{N}_{\mathcal{R},i}^{-}}\hspace{-0.65em}\mathcal{R}_{ij}}\},\min_{l=2,\dots,n}\{\frac{\text{Re}(\varpi_l^{(1)})}{1+|\varpi_l^{(1)}|^2}\}\},\cr &0<\beta_K<\min\{\min_{i\in\mathcal{V}}\{\!\frac{1}{\sum_{j\in\mathcal{N}_{\mathcal{C},i}^{+}}\hspace{-0.5em}\mathcal{C}_{ji}}\},\min_{l=2,\dots,n}\{\frac{\text{Re}(\varpi_l^{(2)})}{1+|\varpi_l^{(2)}|^2}\}\}.
		\end{align*}{\vskip -7pt}\noindent Then, there exist unique nonnegative vectors $v_1$, $v_2\in\mathbb{R}^n$ such that $v_1^\top(I_n-\alpha_K\mathcal{L}_1)=v_1^\top$, $(I_n-\beta_K\mathcal{L}_2)v_2=v_2$, $v_1^\top\mathbf{1}_n=n$, $v_2^\top\mathbf{1}_n=n$, $v_1^\top v_2>0$, and there exist $r_1, r_2>0$ such that $\rho(W_1-\alpha_K\mathcal{L}_1)\leq 1-r_1\alpha_K$, $\rho(W_2-\beta_K\mathcal{L}_2)\leq 1-r_2\beta_K$.
\end{lemma}
{\bf Proof.} See Appendix \ref{appendix b}. $\hfill\blacksquare$
\subsection{Problem formulation}
In this paper, the following distributed stochastic optimization problem is considered:
\begin{align}\label{problem}
	\smash{\min_{x\in\mathbb{R}^d}\!F(x)\!=\!\min_{x\in\mathbb{R}^d}\!\frac{1}{n}\!\sum_{i=1}^{n}\!f_{i}(x), f_{i}(x)\!=\!\mathbb{E}_{\xi_{i}\sim \mathscr{D}_i}[\ell_{i}(x,\xi_{i})],}
\end{align}
{\vskip 3pt}\noindent where $x$ is available to all agents, $\ell_{i}(x,\xi_{i})$ is a local objective which is private to Agent $i$, $\xi_i$ is a random variable drawn from an unknown probability distribution $\mathscr{D}_i$, and $\mathscr{D}_i$ is not required to be independent and identically distributed for any $i\in\mathcal{V}$. In practice, since the probability distribution $\mathscr{D}_i$ is difficult to obtain, it is usually replaced by the dataset $\mathcal{D}_i=\{\xi_{i,l}\in\mathbb{R}^r,l=1,\dots,D\}$. Then, \eqref{problem} can be rewritten as the following empirical risk minimization problem:
\begin{align}\label{problem2}
	\smash{\min_{x\in\mathbb{R}^d}\!F(x)\!=\!\min_{x\in\mathbb{R}^d}\!\frac{1}{n}\!\sum_{i=1}^{n}\!f_{i}(x), f_{i}(x)\!=\!\frac{1}{D}\sum_{l=1}^{D}\ell_{i}(x,\xi_{i,l}).}
\end{align}
{\vskip 3pt}\noindent When solving the problem \eqref{problem2}, a stochastic first-order oracle is often required (\!\!\!\cite{bubeck2015convex}), which returns a sampled gradient $g_i(x,\lambda_i)$ of the local objective $\ell(x,\lambda_i)$ for any $i\in\mathcal{V}$, $x\in\mathbb{R}^d$ and $\lambda_i$ generated by uniformly sampling $\xi_i$ from $\mathcal{D}_i$, i.e., $g_i(x,\lambda_i)=\nabla \ell_i(x,\lambda_i)$. Then, the following standard assumption is given:
\begin{asm}\label{asm2}
	(i) There exist $L_1,L_2>0,\tau\geq0$ satisfying $\|g_i(x,\!\lambda_i)-g_i(y,\!\lambda_i)\|\leq L_1\|x-y\|$, $\|g_i(x,\!\lambda_i)-g_i(x,\!\lambda_i^\prime)\|\leq L_2\|\lambda_i-\lambda_i^\prime\|^\tau$, $\forall i$$\in$$\mathcal{V}$, $\forall x,\!y$$\in$$\mathbb{R}^d$, $\forall \lambda_i,\!\lambda_i^\prime$$\in$$\mathbb{R}^r$.
	
	\noindent(ii) There exists $\sigma_g>0$ satisfying $\mathbb{E}[g_i(x,\lambda_i)]=\nabla f_i(x)$, $\mathbb{E}[\|g_i(x,\lambda_i)-\nabla f_i(x)\|^2]\leq \sigma_g^2$.
\end{asm}
\begin{rmk}\label{rmk2}
	Assumption \ref{asm2}(i) requires the sampled gradient $g_i(x,\lambda_i)$ is $L_1$-Lipschitz continuous with respect to $x$ and $(\tau,L_2)$-H\"{o}lder continuous with respect to $\lambda_i$, which is more general than \cite{wang2023decentralized,chen2024differentially} with $\tau=1$ and \cite{chen2024locallya} with $\tau=0$. Assumption~\ref{asm2}(ii) requires that each sampled gradient $g_i(x,\lambda_i)$ is unbiased with a bounded variance $\sigma_g^2$, which is standard for distributed stochastic optimization (see e.g. \cite{xin2019distributed,xin2020variance,pu2021distributed,koloskova2021improved,lei2022distributed,chen2024accelerated,ding2021differentially,xu2022dp,wang2023gradient,wang2023quantization,yan2024killing,wang2024differentiallya,chen2024locallya,chen2024differentially}).
\end{rmk}

Next, assumptions for the nonconvex and strongly convex global objective are respectively given as follows:
\begin{asm}\label{asm3}
	There exists $x^*\in\mathbb{R}^d$ such that $F(x^*)$ is the global minimum of the nonconvex global objective $F(x)$. Moreover, the Polyak-{\L}ojasiewicz condition holds, i.e., there exists $\mu>0$ such that \mbox{$2\mu(F(x)-F(x^*))\leq\|\nabla F(x)\|^2$, $\forall x\in \mathbb{R}^d$.}
\end{asm}
\begin{rmk}\label{rmk3}
	Assumption \ref{asm3} requires the gradient $\nabla F(x)$ to grow faster than a quadratic function as we move away from the global minimum, which is commonly used (see e.g. \cite{kang2021weighted,lu2023convergence,xie2024differentially,chen2024accelerated,chen2024differentially}).
\end{rmk}
\begin{rmk}
	There exists functions that satisfy Assumptions \ref{asm2}, \ref{asm3} simultaneously. We give two examples. One example is $l_i(x,\xi_i)=\frac{1}{2n}\|\mathbf{A}x-\mathbf{d}\|^2+\frac{\|x\|\xi_i}{1+\|x\|}$, where $x\in\mathbb{R}^d$, the matrix $\mathbf{A}\in\mathbb{R}^{m\times d}$ has full column rank, $\mathbf{d}\in\mathbb{R}^d$ is a constant vector, and $\xi_i\sim N(0,4)$ is a Gaussian noise. Denote $\rho(\mathbf{A}),\Theta_{\mathbf{A}^\top\mathbf{A}}$$>$$0$ as the spectral radius of $\mathbf{A}$ and the minimum eigenvalue of $\mathbf{A}^\top\mathbf{A}$, respectively. Then, by \cite[Th. 2]{karimi2016linear}, $l_i(x,\xi_i)$ satisfies Assumption \ref{asm2} with $L_1$$=$$\frac{\rho(\mathbf{A})^2}{2n}$, $L_2$$=$$1$, $\tau$$=$$1$, $\sigma_g$$=$$2$, and $F(x)$ satisfies Assumption \ref{asm3} with $\mu$$=$$2\Theta_{\mathbf{A}^\top\mathbf{A}}^2$. Another example is $l_i(x,\xi_i)=x^2+(3+\xi_i)(\sin x)^2+2\xi_i\cos x$, where $x\in\mathbb{R}$, and $\xi_i\sim\text{Lap}(\frac{1}{2})$ is a Laplacian noise. Then, by \cite[Subsec. 2.2]{karimi2016linear}, $l_i(x,\xi_i)$ satisfies Assumption \ref{asm2} with $L_1$$=$$8$, $L_2$$=$$2$, $\tau$$=$$1$, $\sigma_g$$=$$\frac{5}{2}$, and $F(x)$ satisfies Assumption \ref{asm3} with $\mu$$=$$\frac{n}{32}$.
\end{rmk}

In practice, since finding the exact optimal solution is computationally expensive and time-consuming, suboptimal solutions within a given error $\varphi>0$ are often preferred. Inspired by~\cite{bhavsar2023nonasymptotic}, the $\varphi$-suboptimal solution and the oracle complexity are defined as follows:
\begin{def0}\label{def1}
	Let $\varphi>0$, $K=0,1,\dots$, $x_K$$=$$[x_{1,K}^\top,\dots$, $x_{n,K}^\top]^\top$ be the output of an algorithm. Then, $x_K$ is a $\varphi$-suboptimal solution if $\E\|\nabla F(x_{i,K+1})\|^2<\varphi$, $\forall i\in\mathcal{V}$.
\end{def0}

\begin{def0}\label{def2}
	Let $\varphi>0$, $N(\varphi)=\min\{K\!:x_K\text{ is a $\varphi$-}$ $\text{suboptimal solution}\}$, and $m_k$ be the sampling number at the $k$-th iteration. Then, the oracle complexity of the algorithm is $\sum_{k=0}^{N(\varphi)}m_k$.
\end{def0}
\subsection{Local differential privacy}
As shown in \cite{wang2023decentralized,wang2023quantization,chen2024locallya}, there are two kinds of adversary models widely used in the privacy-preserving issue for distributed stochastic optimization:
\begin{itemize}[leftmargin=*]
	\item A \emph{semi-honest} adversary. This kind of adversary is defined as an agent within the network which has access to certain internal information (such as state variable $x_{i,k}$ and tracking variable $y_{i,k}$ of Agent $i$), follows the prescribed protocols and accurately computes iterative state and tracking correctly. However, it aims to infer the sensitive information of other agents.
	\item An \emph{eavesdropper}. This kind of adversary refers to an external adversary who has capability to  wiretap and monitor all communication channels, allowing them to capture distributed messages from any agent. This enables the eavesdropper to infer the sensitive information of agents.
\end{itemize}

When cooperative agents exchange information to solve the empirical risk minimization problem~\eqref{problem2}, these two kinds of adversaries can use the model inversion attack (\!\!\cite{fredrikson2015model}) to infer sampled gradients, and further obtain the sensitive information in agents' data samples from sampled gradients (\!\!\cite{zhu2019deep}). Inspired by \cite{chen2024locallya,chen2024locallyb}, a symmetric binary relation called \emph{adjacency relation} is defined as follows:
\begin{def0}\label{def3}
	For any $i\in\mathcal{V}$, let $\mathcal{D}_i$$=$$\{\xi_{i,l},l=1,\dots, D\}$, $\mathcal{D}_i^\prime$$=$$\{\xi_{i,l}^{\prime},l=1,\dots,D\}$ be two sets of Agent $i$'s data samples. If there exist $C>0$ and exactly one pair of data samples $\xi_{i,l_0},\xi_{i,l_0}^\prime$ in $\mathcal{D}_i,\mathcal{D}_i^\prime$ such that for any $x\in\mathbb{R}^d$, $l=1,\dots,D$,\vspace{-0.5em}
	\begin{align}\label{eq3}
		\begin{cases}
			0\!<\!\|g_i(x,\xi_{i,l})\!-\!g_i(x,\xi_{i,l}^\prime)\|_1\!\leq\!C,&\hspace{-0.8em}\text{ if }l=l_0;\cr
			\!\|g_i(x,\xi_{i,l})\!-\!g_i (x,\xi_{i,l}^\prime)\|_1\!=\!0,&\hspace{-0.8em}\text{ if }l\neq l_0,
		\end{cases}\!\!\!\!
	\end{align} then $\mathcal{D}_i$ and $\mathcal{D}_i^\prime$ are said to be adjacent, denoted by $\text{Adj}(\mathcal{D}_i$,$\mathcal{D}_i^\prime)$.
\end{def0}
\begin{rmk}
	The constant $C$ is an upper bound of the magnitude of sampled gradients when changing one data sample in $\text{Adj}(\mathcal{D}_i,\mathcal{D}_i^\prime)$. The larger the constant $C$ is, the larger the allowed magnitude of sampled gradients between adjacent datasets is. As long as there exist $\mathcal{D}_i,\mathcal{D}_i^\prime$ satisfying the adjacency relation defined by a constant $C$, then the privacy analysis in Subsection \ref{III-C} holds for $\text{Adj}(\mathcal{D}_i,\mathcal{D}_i^\prime)$. For more details, please refer to \cite[Subsec. II-D]{chen2024differentially}. 
\end{rmk}
\begin{rmk}\label{rmk6}
	Definition \ref{def3} allows us to avoid the assumption of bounded gradients required in \cite{ding2021differentially,kang2021weighted,wang2023decentralized,wang2023quantization,liu2024distributed,chen2024locallyb}. Since $\mathcal{D},\mathcal{D}^\prime$ have finite data samples, it follows that $\max_{\xi\in\mathcal{D}\cup\mathcal{D}^\prime}\|\xi\|^\tau$$<$$\infty$. Then, for any $C$$\geq$$(2^\tau$$+$$1)$$\sqrt{d}$$L_2$$\max_{\xi\in\mathcal{D}\cup\mathcal{D}^\prime}\|\xi\|^\tau$ and $x\in\mathbb{R}^d$, by \cite[Ths. 2.8, 2.13]{adams2003sobolev} and Assumption 2(i), we have\vspace{-0.4em}
	\begin{align*}
		\begin{cases}
			\begin{matrix}
				\|g_i(x,\xi_{i,l})-g_i(x,\xi_{i,l}^\prime)\|_1\leq \sqrt{d}L_2\|\xi_{i,l}-\xi_{i,l}^\prime\|^\tau\\
				\leq(2^\tau\! +\!1)\sqrt{d}L_2\max_{\xi\in\mathcal{D}\cup\mathcal{D}^\prime}\|\xi\|^\tau\leq C,
			\end{matrix}&\hspace{-0.8em}\text{ if }l=l_0;\cr
			\|g_i(x,\xi_{i,l})-g_i (x,\xi_{i,l}^\prime)\|_1=0,&\hspace{-0.8em}\text{ if }l\ne l_0.
		\end{cases}
	\end{align*}{\vskip -5pt}\noindent Thus, there exists a constant $C$ such that \eqref{eq3} holds for any $x\in\mathbb{R}^d$ no matter whether gradients are bounded or not.
\end{rmk}
\begin{rmk}
	Different from the adjacency relation defined in differentially private distributed optimization (\!\!\cite{ding2022differentially,xuan2023gradient,xie2024differentially,wang2024tailoring,huang2024differential,huo2024differentially}), Definition~\ref{def3} is given with respect to data samples. Moreover, by allowing one data sample of each agent to be different, Definition \ref{def3} corresponds to local differential privacy in the offline setting, inspired by the online setting in \cite[Def. 2]{chen2024locallya}, \cite[Def. 2]{chen2024locallyb}, and then is more stringent than the one of \cite{ding2021differentially,kang2021weighted,xu2022dp,wang2023decentralized,wang2023quantization,liu2024distributed,yan2024killing,wang2024differentiallya,chen2024differentially}, which only allow one data sample of one agent to be different.
\end{rmk}

Next, the definition of differential privacy is given to show the privacy-preserving level of the algorithm:
\begin{def0}\label{def4}
	(\!\!\!\cite{dwork2014algorithmic}) Let $\varepsilon_i\geq0$ be the differential privacy budget of Agent $i$. Then, the mechanism $\mathcal{M}$ achieves $\varepsilon_i$-local differential privacy for $\text{Adj}(\mathcal{D}_i,\mathcal{D}_i^\prime)$ if $\p(\mathcal{M}(\mathcal{D}_i)\!\in\!\mathcal{O})$$\leq$$ e^{\varepsilon_i}\p(\mathcal{M}(\mathcal{D}_i^\prime)\!\in\!\mathcal{O})$ holds for any Borel-measurable observation set $\mathcal{O}\!\subseteq\!\text{Range}(\mathcal{M})$.
\end{def0}
\begin{rmk}
	As shown in \cite{ding2021differentially,kang2021weighted,xu2022dp,wang2023decentralized,wang2023quantization,liu2024distributed,yan2024killing,wang2024differentiallya,chen2024locallya,chen2024differentially}, the differential privacy budget $\varepsilon_i$ measures the similarity of the mechanism $\mathcal{M}$'s output distributions under under $\text{Adj}(\mathcal{D}_i,\mathcal{D}_i^\prime)$. The smaller the differential privacy budget $\varepsilon_i$ is, the higher the differential privacy level is.
\end{rmk}
\begin{rmk}
	Both $\varepsilon$-local differential privacy and $(\varepsilon, \delta)$-local differential privacy has been used in differentially private distributed stochastic optimization. $\varepsilon$-local differential privacy is achieved by Laplacian noises, while $(\varepsilon, \delta)$-local differential privacy is achieved by Gaussian noises. To simplify the analysis, $\varepsilon$-local differential privacy is used in this paper. If $(\varepsilon, \delta)$-local differential privacy is used, then the framework of the convergence and privacy analysis still holds.
\end{rmk}

{\bf Problem of interest:} In this paper, we first aim to propose a new differentially private gradient-tracking-based algorithm for the problem \eqref{problem2} over directed graphs; then design schemes of step-sizes and the sampling number to enhance the differential privacy level, achieve the almost sure and mean square convergence for nonconvex objectives without the Polyak-{\L}ojasiewicz condition, and further accelerate the convergence rate.\vspace{-1em}
\section{Main results}\label{section 3}
\subsection{The proposed algorithm}
In this subsection, we propose a differentially private gradient-tracking-based distributed stochastic optimization algorithm over directed graphs. Detailed steps are given in Algorithm~\ref{algorithm1}.
\begin{algorithm}[!htbp]
	\caption{Differentially\hspace{-0.05em} private\hspace{-0.05em} gradient-tracking-based\hspace{-0.05em} distributed\hspace{-0.05em} stochastic\hspace{-0.05em} optimization\hspace{-0.05em} algorithm\hspace{-0.05em} over\hspace{-0.05em} directed\hspace{-0.05em} graphs}
	\label{algorithm1}
	\renewcommand{\algorithmicensure}{\textbf{Initialization:}}
	\begin{algorithmic}[1]
		\Ensure $x_{i,0}\in\mathbb{R}^d$ for any $i\in\mathcal{V}$, $m_K$ different data samples $\lambda_{i,0,1},\dots$, $\lambda_{i,0,m_K}$ in $\mathcal{D}_i$, $y_{i,0}=g_{i,0}=\frac{1}{m_K}\sum_{l=1}^{m_K}$ $g_i(x_{i,0},\lambda_{i,0,l})$ for any $i\in\mathcal{V}$, weight matrices $\mathcal{R}=(\mathcal{R}_{ij})_{i,j=1,\dots,n}$, $\mathcal{C}=(\mathcal{C}_{ij})_{i,j=1,\dots,n}$, the maximum iteration number $K$, step-sizes $\alpha_K,\beta_K,\gamma_K$ and the sampling number $m_K$.
		
		\hspace{-3.7em}{\bf for} $k=0,1,\dots,K$, {\bf do}
		\State Agent $i$ adds independent $d$-dimensional Laplacian noises $\zeta_{i,k}$, $\eta_{i,k}$ to its state variable $x_{i,k}$ and tracking variable $y_{i,k}$, respectively: $\breve{x}_{i,k}=x_{i,k}+\zeta_{i,k}$, $\breve{y}_{i,k}=\;y_{i,k}+\eta_{i,k}$, where each coordinate of $\zeta_{i,k}$, $\eta_{i,k}$ has the distribution $\text{Lap}(\sigma_k^{(\zeta_i)})$ and $\text{Lap}(\sigma_k^{(\eta_i)})$, respectively.
		\State Agent $i$ broadcasts its perturbed state variable $\breve{x}_{i,k}$ to all its out-neighbors in $\mathcal{N}_{\mathcal{R},i}^{+}$, and broadcasts its perturbed tracking variable $\breve{y}_{i,k}$ to all its out-neighbors in $\mathcal{N}_{\mathcal{C},i}^{+}$.
		\State Agent $i$ receives $\breve{x}_{j,k}$ from all its in-neighbors in $\mathcal{N}_{\mathcal{R},i}^{-}$ and $\breve{y}_{j,k}$ from all its in-neighbors in $\mathcal{N}_{\mathcal{C},i}^{-}$.
		\State Agent $i$ updates its state variable by\vspace{-5pt}
		\begin{align}\label{alg,1}
			\hspace{-1.75em}x_{i,k\!+\!1}\!=\!(1\!-\!\alpha_K\hspace{-0.75em}\sum_{j\in\mathcal{N}_{\mathcal{R},i}^{-}}\hspace{-0.75em}\mathcal{R}_{ij})x_{i,k}\!+\!\alpha_K\hspace{-0.75em}\sum_{j\in\mathcal{N}_{\mathcal{R},i}^{-}}\hspace{-0.75em}\mathcal{R}_{ij}\breve{x}_{j,k}\!-\!\gamma_K y_{i,k}.
		\end{align}\vspace{-12pt}
		\State Agent $i$ takes $m_K$ different samples $\lambda_{i,k\!+\!1,1}$, $\dots$, $\lambda_{i,k\!+\!1,m_K}$ uniformly from $\mathcal{D}_i$ to generate sampled gradients $g_i(x_{i,k\!+\!1},\lambda_{i,k\!+\!1,1})$, $\dots$, $g_i(x_{i,k\!+\!1},\lambda_{i,k\!+\!1,m_K})$. Then, Agent $i$ puts these data samples back into $\mathcal{D}_i$.
		\State Agent $i$ computes the averaged sampled gradient by\vspace{-6pt}
		\begin{align}\label{avg_sampled_grad}
			g_{i,k+1}=\frac{1}{m_K}\sum_{l=1}^{m_K} g_i(x_{i,k\!+\!1},\lambda_{i,k\!+\!1,l}).
		\end{align}\vspace{-10pt}
		\State Agent $i$ updates its tracking variable by\vspace{-5pt}
		\begin{align}\label{alg,2}
			\hspace{-1.75em}y_{i,k\!+\!1}\!\!=\!\!(1\!\!-\!\!\beta_K\hspace{-0.75em}\sum_{j\in\mathcal{N}_{\mathcal{C},i}^{+}}\hspace{-0.75em}\mathcal{C}_{ji})y_{i,k}\!+\!\beta_K\hspace{-0.75em}\sum_{j\in\mathcal{N}_{\mathcal{C},i}^{-}}\hspace{-0.75em}\mathcal{C}_{ji}\breve{y}_{j,k}\!+\!g_{i,k\!+\!1}\!-\!g_{i,k}.
		\end{align}{\vskip -3pt}

		\vspace{-0.9em}
		\hspace{-3.7em}{\bf end for}
		\item[]\hspace{-1.7em}{\bf Return} $x_{1,K+1},\dots,x_{n,K+1}$
	\end{algorithmic}
\end{algorithm}

For the convenience of the analysis, let $x_k=[x_{1,k}^\top,\dots,$ $x_{n,k}^\top]^\top$ $y_k=[y_{1,k}^\top,\dots,y_{n,k}^\top]^\top$, $\zeta_k=[\zeta_{1,k}^\top,\dots,\zeta_{n,k}^\top]^\top$, $\eta_k=[\eta_{1,k}^\top,\dots,$ $\eta_{n,k}^\top]^\top$, $g_k=[g_{1,k}^\top,\dots,g_{n,k}^\top]^\top$. Then, \eqref{alg,1} and \eqref{alg,2} can be written in the following compact form:\vspace{-0.5em}
\begin{align}
	x_{k\!+\!1}\!=&(\!(I_n\!\!-\!\alpha_K\mathcal{L}_1)\!\otimes\! I_d)x_k\!+\!\alpha_K\!(\mathcal{R}\!\otimes\! I_d)\zeta_k\!-\!\gamma_K y_k,\label{eq7}\\
	y_{k\!+\!1}\!=&(\!(I_n\!\!-\!\beta_K\mathcal{L}_2)\!\otimes\! I_d)y_k\!+\!\beta_K\!(\mathcal{C}\!\otimes\! I_d)\eta_k\!+\!g_{k\!+\!1}\!-\!g_k.\label{eq8}
\end{align}
\vspace{-2.5em}
\subsection{Convergence analysis}
In this subsection, we will give the convergence analysis of Algorithm \ref{algorithm1}. First, we give the following key lemma:
\begin{lemma}\label{lemma a10}
	For any $K=0,1,\dots$, $k=0,\dots,K$, let $V_k=$ $[\E\|(W_1\!\otimes\! I_d)x_k\|^2,\E\|(W_2\!\otimes\! I_d)y_k\|^2,\E(F(x_k)\!-\!F(x^*))]^\top$. Under Assumptions \ref{asm1}-\ref{asm3}, if step-sizes $\alpha_K,\beta_K,\gamma_K$ satisfy the following conditions:\vspace{-1em}
	\begin{align*}
		&0<\alpha_K<\min\{\min_{i\in\mathcal{V}}\{\frac{1}{\sum_{\hspace{-0.15em}j\in\mathcal{N}_{\mathcal{R},i}^{-}}\hspace{-0.65em}\mathcal{R}_{ij}}\},\min_{l=2,\dots,n}\{\frac{\text{Re}(\varpi_l^{(1)})}{1+|\varpi_l^{(1)}|^2}\}\},\cr
		\noalign{\vskip -3pt} &0<\beta_K<\min\{\min_{i\in\mathcal{V}}\{\!\frac{1}{\sum_{j\in\mathcal{N}_{\mathcal{C},i}^{+}}\hspace{-0.5em}\mathcal{C}_{ji}}\},\min_{l=2,\dots,n}\{\frac{\text{Re}(\varpi_l^{(2)})}{1+|\varpi_l^{(2)}|^2}\}\}\cr
		\noalign{\vskip -5pt}
		&0<\gamma_K<\frac{n}{4(v_1^\top v_2)L_1},
	\end{align*}{\vskip -3pt}\noindent then the following inequality holds:\vspace{-0.4em}
	\begin{align}\label{41b}
		\E V_{k+1}\leq A_K\E V_k + u_k,
	\end{align}
	{\vskip -8pt}\noindent where \vspace{-0.8em}
		\begin{align*}
			u_k=\left[\begin{array}{c}
				u_k^{(1)}\\
				u_k^{(2)}\\
				u_k^{(3)}
			\end{array}\right],A_K=\left[\begin{array}{ccc}
				\!\!A_K^{(11)}&\!\!\!\!A_K^{(12)}\!\!&\!\!\!\!A_K^{(13)}\!\!\\
				\!\!A_K^{(21)}&\!\!\!\!A_K^{(22)}\!\!&\!\!\!\!A_K^{(23)}\!\!\\
				\!\!A_K^{(31)}&\!\!\!\!A_K^{(32)}\!\!&\!\!\!\!A_K^{(33)}\!\!
			\end{array}\right],
		\end{align*}
		{\vskip -6pt}\noindent $u_k^{(1)}$$=$$2nd\rho(\mathcal{R})^2\alpha_K^2\max_{i\in\mathcal{V}}\{(\sigma_k^{(\zeta_i)})^2\}$$+$$\frac{2(1+r_1\alpha_K)\|v_2\|^2\gamma_K^2\sigma_g^2}{n^2r_1\alpha_Km_K}$ $+$$\frac{4d(1+r_1\alpha_K)\|v_2\|^2\rho(\mathcal{C})^2\beta_K^2\gamma_K^2}{n^3r_1\alpha_K}\sum_{l=0}^{k-1}\max_{i\in\!\mathcal{V}}\{(\sigma_k^{(\eta_i)})^2\}$, $u_k^{(2)}$$=$$\frac{12d(1+r_2\beta_K)\|v_2\|^2\rho(\mathcal{C})^2\beta_K\gamma_K^2L_1^2}{nr_2}\sum_{l=0}^{k-1}\max_{i\in\!\mathcal{V}}\{(\sigma_k^{(\eta_i)})^2\}$ $+$$\frac{(2n\!+\!3nr_2\beta_K\!+\!(6\!+\!6r_2\beta_K)\!\|v_2\|^2\!\gamma_K^2L_1^2)\sigma_g^2}{r_2m_K\beta_K}$$+$$2nd\rho(\mathcal{C})^2\!\beta_K^2\!\max_{i\in\mathcal{V}}$$\{\!(\sigma_k^{(\eta_i)})^2\}$ $+$$\frac{4(1\!+\!r_2\beta_K)nd\rho(\mathcal{R})^2\!\alpha_K^2L_1^2\!\max_{i\in\mathcal{V}}\{\!(\sigma_k^{(\zeta_i)})^2\}}{r_2\beta_K}$, $u_k^{(3)}$$=$$\frac{(v_1^\top\! v_2)(3n+2(v_1^\top \!v_2)\gamma_KL_1)\gamma_K\sigma_g^2}{2n^2m_K}$$+$$\frac{2d\|v_1\|^2\rho(\mathcal{R})^2\alpha_K\max_{i\in\mathcal{V}}\{(\sigma_k^{(\zeta_i)})^2\}}{n}$ $+$$\frac{(v_1^\top\! v_2)d\rho(\mathcal{C})^2(3n+2(v_1^\top \!v_2)\gamma_KL_1)\beta_K^2\gamma_K}{n^3}\sum_{l=0}^{k-1}\max_{i\in\!\mathcal{V}}\{(\sigma_k^{(\eta_i)})^2\}$,  $A_K^{(11)}$$=$$1$$-$$r_1\alpha_K$$+$$\frac{4(1+r_1\alpha_K)\|v_2\!\|^2\gamma_K^2\!L_1^2}{n^3r_1\alpha_K}$,$A_K^{(12)}$$=$$\frac{2(1+r_1\alpha_K)\gamma_K^2}{r_1\alpha_K}$, $A_K^{(13)}$$=$$\frac{8(1\!+\!r_1\alpha_K)\|v_2\|^2\!\gamma_K^2\!L_1}{n^2r_1\alpha_K}$, $A_K^{(31)}$$=$$\frac{(v_1^{\!\top}\! v_2)(3n+4(v_1^{\!\top}\! v_2)\gamma_K\!L_1)\gamma_KL_1^2}{2n^3}$, $A_K^{(22)}$$=$$1$$-$$r_2\beta_K$$+$$\frac{6(1+r_2\beta_K)\gamma_K^2L_1^2}{r_2\beta_K}$,$A_K^{(23)}$$=$$\frac{24(1\!+\!r_2\beta_K)\|v_2\|^2\gamma_K^2\!L_1^3}{r_2\beta_K}$, $A_K^{(21)}$$=$$\frac{6(1+r_2\beta_K)(n\rho(\mathcal{L}_1)^2\alpha_K^2+2\|v_2\|^2\gamma_K^2L_1^2)L_1^2}{nr_2\beta_K}$,$A_K^{(32)}$$=$$\frac{3\|v_1\|^2\gamma_K}{2n(v_1^\top\! v_2)}$, $A_K^{(33)}$$=$$1$$-$$\frac{(v_1^\top v_2)\mu\gamma_K}{n}$$+$$\frac{4(v_1^\top v_2)^2\gamma_K^2L_1}{n^2}$.
\end{lemma}
{\bf Proof.} See Appendix \ref{appendix c0}. $\hfill\blacksquare$

\indent Next, we give two different schemes of step-sizes and the sampling number for Algorithm \ref{algorithm1}:

\noindent\emph{Scheme (S1)}: For any $K=0,1,\dots$,
\begin{enumerate}[topsep=0em,label=(\Roman*)]
	\item step-sizes: $\alpha_K$$=$$\frac{a_1}{(K+1)^{p_\alpha}}$, $\beta_K$$=$$\frac{a_2}{(K+1)^{p_\beta}}$, $\gamma_K$$=$$\frac{a_3}{(K+1)^{p_\gamma}}$,
	\item the sampling number: $m_K=\lfloor a_4 K^{p_m}\rfloor+1$,
\end{enumerate}where $a_1,a_2,a_3,a_4>0,p_\alpha,p_\beta,p_\gamma>0,p_m\geq0$.

\noindent \emph{Scheme (S2)}: For any $K=0,1,\dots$,
\begin{enumerate}[topsep=0em,label=(\Roman*)]
	\item step-sizes: $\alpha_K=\alpha,\beta_K=\beta,\gamma_K=\gamma$ are constants,
	\item the sampling number: $m_K=\lfloor p_m^K\rfloor+1$,
\end{enumerate}where $\alpha,\beta,\gamma>0$, $p_m\geq0$.

To get the almost sure and mean square convergence of Algorithm \ref{algorithm1}, we need the following assumptions:
\begin{asm}\label{asm4}
	Under \emph{Scheme (S1)}, step-sizes $\alpha_K$, $\beta_K$, $\gamma_K$, the sampling number $m_K$, and privacy noise parameters $\sigma_k^{(\zeta_i)}$$=$$(k+1)^{p_{\zeta_i}}$, $\sigma_k^{(\eta_i)}$$=$$(k+1)^{p_{\eta_i}}$ satisfy the following conditions:
	\begin{align*}
		&a_1<\min\{\min_{i\in\mathcal{V}}\{\frac{1}{\sum_{\hspace{-0.15em}j\in\mathcal{N}_{\mathcal{R},i}^{-}}\hspace{-0.65em}\mathcal{R}_{ij}}\},\min_{l=2,\dots,n}\{\frac{\text{Re}(\varpi_l^{(1)})}{1+|\varpi_l^{(1)}|^2}\}\},\cr
		\noalign{\vskip -2pt} &a_2<\min\{\min_{i\in\mathcal{V}}\{\!\frac{1}{\sum_{j\in\mathcal{N}_{\mathcal{C},i}^{+}}\hspace{-0.5em}\mathcal{C}_{ji}}\},\min_{l=2,\dots,n}\{\frac{\text{Re}(\varpi_l^{(2)})}{1+|\varpi_l^{(2)}|^2}\}\},\cr
		\noalign{\vskip -2pt}
		&a_3\!<\!\frac{n}{4(v_1^\top v_2)L},\frac{1}{2}\!<\!p_\beta\!<\!p_\alpha\!<\!p_\gamma\!<\!1,p_m-p_\beta\!\geq\!1,\cr
		\noalign{\vskip -2pt}
		& 2p_\gamma\!-\!p_\alpha\!\geq\!1,2p_\alpha\!\!-\!p_\beta\!\!-\!2\max\{\max_{i\in\mathcal{V}}\{p_{\zeta_i}\},\!0\}\!\geq\!1,\cr
		&p_\gamma\!\!+\!2p_\beta\!\!-\!2\max\{\max_{i\in\mathcal{V}}\{p_{\eta_i}\},\!0\}\!\geq\!2.
	\end{align*}
\end{asm}
\begin{asm}\label{asm5}
	Under \emph{Scheme (S2)}, step-sizes $\alpha,\beta,\gamma$, the sampling number $m_K$, and privacy noise parameters $\sigma_k^{(\zeta_i)}=p_{\zeta_i}^K$, $\sigma_k^{(\eta_i)}=p_{\eta_i}^K$ satisfy the following conditions:
	\begin{align*}
		&\beta<\min\{\min_{i\in\mathcal{V}}\{\!\frac{1}{\sum_{j\in\mathcal{N}_{\mathcal{C},i}^{+}}\hspace{-0.5em}\mathcal{C}_{ji}}\},\min_{l=2,\dots,n}\{\frac{\text{Re}(\varpi_l^{(2)})}{1+|\varpi_l^{(2)}|^2}\}\},\cr
		\noalign{\vskip -3pt}
		&\alpha<\min\{\min_{i\in\mathcal{V}}\{\frac{1}{\sum_{\hspace{-0.15em}j\in\mathcal{N}_{\mathcal{R},i}^{-}}\hspace{-0.65em}\mathcal{R}_{ij}}\},\min_{l=2,\dots,n}\{\frac{\text{Re}(\varpi_l^{(1)})}{1+|\varpi_l^{(1)}|^2}\},\cr
		&~~~~~~~~~~\frac{\sqrt{2}(v_1^\top v_2)r_2\beta}{12\rho(\mathcal{L}_1)\|v_1\|L_1}\},0<p_{\zeta_i},p_{\eta_i}<1,p_m>1,\cr
		\noalign{\vskip -2pt}
		&\gamma\!\!<\!\!\min\{1,\frac{n}{20(v_1^\top v_2)L_1},Q_1\alpha,Q_2\beta\},
	\end{align*} where
	\begin{align*}
		Q_1=&\min\{\frac{n\sqrt{3n}r_1}{24\|v_2\|L_1},\frac{r_1}{2\|v_2\|L_1}\sqrt{\frac{\mu}{12L_1+2\mu}+\frac{\mathbb{I}_{\{\mu=0\}}}{2}}\},\cr
		Q_2=&\min\{\frac{\sqrt{3}r_2}{6nL_1},\frac{\sqrt{3}(v_1^\top v_2)r_2}{36\|v_1\|\|v_2\|L_1},\frac{\sqrt{6}(v_1^\top v_2)r_1r_2}{144\rho(\mathcal{L}_1)\|v_1\|\|v_2\|L_1},\cr
		&~~~~\frac{\sqrt{6}(v_1^\top v_2)r_2}{12\|v_1\|\|v_2\|L_1}\sqrt{\frac{\mu}{36L_1+7\mu}+\frac{\mathbb{I}_{\{\mu=0\}}}{7}}\}.
		\end{align*}
\end{asm}
\begin{thm0}\label{thm1}
	If Assumptions \ref{asm1}, \ref{asm2}, \ref{asm4} hold under \emph{Scheme~(S1)}, then $\liminf_{K\to\infty}\|\nabla F(x_{i,K\!+\!1})\|^2=0$ a.s., $\liminf_{K\to\infty}\E\|\nabla F(x_{i,K\!+\!1})\|^2=0$, $\forall i\in\mathcal{V}$. If Assumptions \ref{asm1}, \ref{asm2}, \ref{asm5} hold under \emph{Scheme (S2)}, then $\lim_{K\to\infty}\|\nabla F(x_{i,K\!+\!1})\|^2=0$ a.s., $\lim_{K\to\infty}\E\|\nabla F(x_{i,K\!+\!1})\|^2=0$, $\forall i\in\mathcal{V}$. 
\end{thm0}
{\bf Proof.} See Appendix \ref{appendix c1}. \hfill$\blacksquare$
\begin{rmk}
	Algorithm \ref{algorithm1} achieves the almost sure and mean square convergence for nonconvex objectives without the Polyak-{\L}ojasiewicz condition. The condition imposed on objectives is weaker than (strongly) convex objectives (\!\!\cite{xin2020variance,pu2021distributed,koloskova2021improved,xin2019distributed,lei2022distributed,zhao2024asymptotic,wang2023gradient,ding2022differentially,xuan2023gradient,wang2024tailoring,huang2024differential,huo2024differentially}) or the Polyak-{\L}ojasiewicz condition (\!\!\cite{chen2024accelerated,xie2024differentially}). Thus, Algorithm \ref{algorithm1} has wider applicability than \cite{xin2020variance,pu2021distributed,koloskova2021improved,xin2019distributed,lei2022distributed,zhao2024asymptotic,chen2024accelerated,wang2023gradient,ding2022differentially,xuan2023gradient,wang2024tailoring,huang2024differential,xie2024differentially,huo2024differentially}. 
\end{rmk}
The polynomial mean square convergence rate and the oracle complexity of Algorithm \ref{algorithm1} with \emph{Scheme (S1)} are given as follows:
\begin{thm0}\label{thm2}
	Under Assumptions \ref{asm1}-\ref{asm3} and \ref{asm4}, Algorithm~\ref{algorithm1} with \emph{Scheme (S1)} achieves the following polynomial mean square convergence rate for any $i\in\mathcal{V}$:
	\begin{align}\label{eq9}
		\smash{\E\|\nabla F(x_{i,K\!+\!1})\|^2\!=\!O\!\left(\!\frac{1}{(K\!+\!1)^{\theta-p_\gamma}}\!\right),}
	\end{align}where $\theta$$=$$\min\{p_m-p_\beta,2p_\alpha-p_\beta-2\max\{\max_{i\in\mathcal{V}}\{p_{\zeta_i}\},0\}$, $2p_\beta-2\max\{\max_{i\in\mathcal{V}}\{p_{\eta_i}\},0\}\}$. Furthermore, for any $\varphi>0$, if $p_\alpha\!=\!\max\{1\!-\!\frac{\varphi}{5},\frac{9}{10}\}$, $p_\beta\!=\!\max\{\frac{2}{3}(1\!-\!\frac{\varphi}{5}),\frac{3}{5}\},p_\gamma\!=\!\max\{1\!-\!\frac{\varphi}{10},\frac{9}{10}\},p_m\!=\!\max\{2-\frac{\varphi}{10},\frac{39}{20}\},p_{\zeta_i}\!=\!p_{\eta_i}\!=\!\max\{\frac{\varphi}{10},\frac{1}{20}\}$, then the oracle complexity of Algorithm \ref{algorithm1} with \emph{Scheme (S1)} is $O(\varphi^{-\frac{177+3\max\{1-2\varphi,0\}}{9-11\max\{1-2\varphi,0\}}})$.
\end{thm0}
{\bf Proof.} See Appendix \ref{appendix c}. \hfill$\blacksquare$
\begin{rmk}
	In Theorem \ref{thm2}, the polynomial mean square convergence rate is given for privacy noises with decreasing, constant (see e.g. \cite{ding2021differentially,wang2023decentralized,kang2021weighted,liu2024distributed}), and increasing variances (see e.g. \cite{chen2024locallya,wang2024differentiallya,chen2024differentially}). This is non-trivial even without considering the privacy protection. For example, let step-sizes $\alpha_K=\frac{1}{(K+1)^{0.96}}$, $\beta_K=\frac{1}{(K+1)^{0.7}}$, $\gamma_K=\frac{1}{(K+1)^{0.98}}$. Then, Theorem \ref{thm2} holds as long as privacy noise parameters $\sigma_k^{(\zeta_i)}$, $\sigma_k^{(\eta_i)}$ have the increasing rate no more than $O(k^{0.19})$.
\end{rmk}
\begin{rmk}
	The key to achieving the polynomial mean square convergence rate without the assumption of bounded gradients is to use polynomially decreasing step-sizes and the increasing sampling number, which reduces the effect of stochastic gradient noises and privacy noises. This is different from \cite{doan2021,ding2021differentially,kang2021weighted,xu2022dp,lu2023convergence,wang2023decentralized,wang2023quantization,liu2024distributed,chen2024locallya}, where the assumption of bounded gradients is  required.
\end{rmk}

Next, the exponential mean square convergence rate and the oracle complexity of Algorithm \ref{algorithm1} with \emph{Scheme (S2)} are given:
\begin{thm0}\label{thm3}
	Under Assumptions \ref{asm1}-\ref{asm3} and \ref{asm5}, Algorithm~\ref{algorithm1} with \emph{Scheme (S2)} achieves the following exponential mean square convergence rate for any $i\in\mathcal{V}$:\vspace{-0.4em}
	\begin{align*}
		\E\|\nabla F(x_{i,K\!+\!1})\|^2\!=\!O(\max\{\rho(A_K),\frac{1}{p_m},\max_{i\in\mathcal{V}}\{p_{\zeta_i}^2,p_{\eta_i}^2\}\}^K).
	\end{align*}
	{\vskip -4pt}\noindent Furthermore, for any $\varphi>0$, if $\beta=\min\{\frac{1}{2},\frac{n}{40(v_1^\top v_2)L},\min_{i\in\mathcal{V}}$ $\{\!\frac{1}{2\sum_{j\in\mathcal{N}_{\mathcal{C},i}^{+}}\!\!\!\!\mathcal{C}_{ji}}\!\},\min_{l=2,\dots,n}\{\frac{\text{Re}(\varpi_l^{(2)})}{2+2|\varpi_l^{(2)}|^2}\}\}$, $\alpha=\min\{\beta,\min_{i\in\mathcal{V}}$ \mbox{$\{\!\frac{1}{2\sum_{j\in\mathcal{N}_{\mathcal{R},i}^{-}}\!\!\!\!\mathcal{R}_{i j}}\!\},\min_{l=2,\dots,n}\{\frac{\text{Re}(\varpi_l^{(1)})}{2+2|\varpi_l^{(1)}|^2}\},\frac{\sqrt{2}(v_1^\top v_2)r_2\beta}{12\rho(\mathcal{L}_1)\|v_1\|L_1}$$\}$, $\gamma=$} \mbox{$\min\{\frac{1}{2},\frac{n}{40(v_1^{\!\top}v_2)L},\frac{Q_1\alpha}{2},\frac{Q_2\beta}{2}\}$, $p_m=\max\{\frac{1}{\varphi},\frac{1}{\rho(A_K)}\}$, $p_{\zeta_i}=$} $p_{\eta_i}$$=$$\min\{\varphi,\rho(A_K)\}$, then the oracle complexity of Algorithm~\ref{algorithm1} with \emph{Scheme (S2)} is $O(\frac{|\ln \varphi|}{\varphi})$.
\end{thm0}
{\bf Proof.} See Appendix \ref{appendix e}. \hfill$\blacksquare$
\begin{rmk}
	By Theorems \ref{thm2}, \ref{thm3}, \emph{Scheme (S2)} achieves the exponential mean square convergence rate, while \emph{Scheme (S1)} and methods in  \cite{doan2021,ding2021differentially,kang2021weighted,xu2022dp,lu2023convergence,wang2023gradient,wang2023decentralized,wang2023quantization,zhao2024asymptotic,yan2024killing,liu2024distributed,chen2024locallya,wang2024differentiallya,chen2024differentially} achieve the polynomial mean square convergence rate. For example, when the index of convergence rate is $\frac{1}{K+1}\sum_{k=0}^{K}\E(F(\bar{x}_{k})-F(x^*))$, methods in \cite{wang2023decentralized,wang2023quantization} achieve convergence rates of $O(\frac{1}{\sqrt{K}})$ and $O(1)$, respectively. Since the method in \cite{wang2023decentralized} is the same as the one in \cite{koloskova2020unified}, by \cite[Th. 2]{koloskova2020unified},the method in \cite{wang2023decentralized} achieves the convergence rate of $O(\frac{1}{\sqrt{K}})$. By \cite[Th. 2]{wang2023quantization}, the method in \cite{wang2023quantization} achieves the convergence rate of $O(1)$. Thus, \emph{Scheme (S2)} is suitable for the scenario where the convergence rate is prioritized. However, by Theorem \ref{thm1}, \emph{Scheme (S1)} achieves the almost sure and mean square convergence under decreasing, constant, and increasing privacy noises, while \emph{Scheme (S2)} achieves the almost sure and mean square convergence only under decreasing privacy noises. This shows the trade-off of Algorithm \ref{algorithm1} between the convergence rate and the added privacy noises.
\end{rmk}
\begin{rmk}
	When the global objective $F(x)$ is strongly convex (i.e., there exists $s>0$ such that  $F(y)\geq F(x)+\langle \nabla F(x),y-x \rangle+\frac{s}{2}\|y-x\|^2$, $\forall x,y\in\mathbb{R}^d$), by \cite[Lemma 6.9]{bubeck2015convex}, we have $2s(F(x)-F(x^*))\leq\|\nabla F(x)\|^2$. Then Assumption \ref{asm3} is satisfied with $\mu=s$,  and thus, Theorems \ref{thm2}, \ref{thm3} also hold for strongly convex objectives. Hence, we provides a general frame for Algorithm \ref{algorithm1}'s convergence rate analysis under both nonconvex objectives with Polyak-{\L}ojasiewicz conditions and strongly convex objectives.
\end{rmk}
\begin{rmk}
	The oracle complexity of \emph{Scheme (S2)} shows that the sampling number required to achieve the desired accuracy is lower than existing works (see e.g. \cite{lei2022distributed}). By Theorem \ref{thm3}, the oracle complexity of \emph{Scheme (S2)} is $O(\frac{|\ln\varphi|}{\varphi})$, which is smaller than the oracle complexity $O(\frac{1}{\varphi^2})$ of the gradient-tracking-based algorithm in \cite{lei2022distributed}. For example, when the error $\varphi=0.02$, $O(10^2)$ data samples are required in \emph{Scheme (S2)}, while $O(10^3)$ data samples are required in the gradient-tracking-based algorithm in \cite{lei2022distributed}. Moreover, the increasing sampling number in both \emph{Schemes (S1)} and \emph{(S2)} is feasible in machine learning scenarios, such as the speech recognition problem (\!\!\cite{byrd2012sample}), the simulated annealing problem (\!\!\cite{smith2017don}), and the noun-phrase chunking problem (\!\!\cite{friedlander2012hybrid}).
\end{rmk}
\subsection{Privacy analysis}\label{III-C}
\vspace{-0.3em}
In the following, the definition of the sensitivity is provided to compute the cumulative differential privacy budget $\varepsilon_i$ for any $i\in\mathcal{V}$.
\begin{def0}\label{def5}
	Given $\text{Adj}(\mathcal{D}_i$,$\mathcal{D}_i^\prime)$ for any $i$$\in$$\mathcal{V}$ and query $q$. For any $k$$=$$0,\dots,K$, let $\mathcal{D}_{i,k}$$=$$\{\lambda_{i,k,l},l$$=$$1,\dots,m_K\}$, $\mathcal{D}_{i,k}^\prime$$=$ $\{\lambda_{i,k,l}^{\prime},l$$=$$1,\dots,m_K\}$ be the data samples taken from $\mathcal{D}_i,\mathcal{D}_i^\prime$ at the $k$-th iteration, respectively. Then, Agent $i$'s sensitivity in Algorithm \ref{algorithm1} at the $k$-th iteration is defined follows:\vspace{-0.6em}
	\begin{align}\label{eq:definition of sensitivity,1}
		\hspace{-1.2em}\Delta_{i,k}^{q}\!\!\triangleq\!\!\begin{cases}
			\sup\limits_{\substack{\mathcal{O}\subseteq\mathbb{R}^{2nd},\\(\breve{x}_0,\breve{y}_0)\in\mathcal{O},\\\text{Adj}(\mathcal{D}_i,\mathcal{D}_i^\prime)}}\hspace{-0.3em}\|q(\mathcal{D}_{i,0})-q(\mathcal{D}_{i,0}^\prime)\|_1,&\hspace{-0.6em}\text{if }k=0;\\
			\sup\limits_{\substack{\mathcal{O}\subseteq\mathbb{R}^{2nd},\\(\breve{x}_{k\!-\!1},\breve{y}_{k\!-\!1})\in\mathcal{O},\\\text{Adj}(\mathcal{D}_i,\mathcal{D}_i^\prime)}}\hspace{-1.5em}\|q(\mathcal{D}_{i,k})-q(\mathcal{D}_{i,k}^\prime)\|_1,&\hspace{-0.6em}\text{if }k=1,\dots,K.
		\end{cases}
	\end{align}
\end{def0}\vspace{0.3em}
\begin{rmk}\label{rmk4}
	Definition \ref{def5} captures the magnitude by which Agent $i$'s data sample can change the query $q$ in the worst case. It is the key quantity showing how many noises should be added to achieve $\varepsilon_{i,k}$-local differential privacy for Agent $i$ at the $k$-th iteration. In Algorithm~\ref{algorithm1}, the query $q(\mathcal{D}_{i,k})$$=$$[x_{i,k}^\top,y_{i,k}^\top]^\top$, and the mechanism $\mathcal{M}(\mathcal{D}_{i,k})$$=$$[\breve{x}_{i,k}^\top,\breve{y}_{i,k}^\top]^\top$.
\end{rmk}

The following lemma gives the sensitivity $\Delta_{i,k}^{q}$ of Algorithm~\ref{algorithm1} for any $k=0,\dots,K$.

\begin{lemma}\label{lemma4}
	If Assumption \ref{asm2}(i) holds, then the sensitivity of Algorithm \ref{algorithm1} at the $k$-th iteration satisfies $\Delta_{i,k}^{q}=\|\Delta x_{i,k}\|_1+\|\Delta y_{i,k}\|_1$, where $\|\Delta x_{i,k}\|_1$ and $\|\Delta y_{i,k}\|_1$ are given as follows:\vspace{-0.5em}
{\fontsize{9.5}{1}\selectfont\begin{align*}
		\hspace{-1em}\|\Delta x_{i,k}\|_1\!\!\leq\!&
			\begin{cases}
				0,&\hspace{-0.6em}\text{if }k\!=\!0;\cr
				\gamma_K\hspace{-0.5em}\sum\limits_{l=0}^{k-1}\!\!|1\!\!-\!\!\alpha_K\hspace{-0.35em}\sum_{j\in\mathcal{N}_{\mathcal{R},i}^{-}}\hspace{-0.6em}\mathcal{R}_{ij}|^{k\!-\!l\!-\!1}\|\Delta y_{i,l}\|_1,&\hspace{-0.6em}\text{if }k\!=\!1,\dots,K,\cr
			\end{cases}\cr
			\hspace{-1em}\|\Delta y_{i,k}\|_1\!\!\leq\!&
			\begin{cases}
				\frac{C}{m_K},&\hspace{3.2em}\text{if }k\!=\!0;\cr
				\begin{matrix}
					&\hspace{-1.5em}\sum\limits_{l=0}^{k-1}|1\!\!-\!\!\beta_K\hspace{-0.25em}\sum_{j\in\mathcal{N}_{\mathcal{C},i}^{+}}\hspace{-0.6em}\mathcal{C}_{ji}|^l\frac{2C}{m_K}\\
					&+|1\!\!-\!\!\beta_K\hspace{-0.25em}\sum_{j\in\mathcal{N}_{\mathcal{C},i}^{+}}\hspace{-0.6em}\mathcal{C}_{ji}|^k\frac{C}{m_K},
				\end{matrix}&\hspace{3.2em}\text{if }k\!=\!1,\dots,K.
		\end{cases}
	\end{align*}}
\end{lemma}\noindent{\bf Proof:} See Appendix \ref{appendix f}. \hfill$\blacksquare$
\begin{lemma}\label{lemma5}
If Assumption \ref{asm2}(i) holds, then for any $K=0,1,\dots$, Algorithm \ref{algorithm1} achieves $\varepsilon_i$-local differential privacy for Agent $i$ over $K$ iterations, where $\varepsilon_i$$=$$\sum_{k=0}^{K}(\frac{\|\Delta x_{i,k}\|_1}{\sigma_k^{(\zeta_i)}}$$+$$\frac{\|\Delta y_{i,k}\|_1}{\sigma_k^{(\eta_i)}})$.
\end{lemma}\vspace{0.3em}
\noindent{\bf Proof.} See Appendix \ref{appendix g}. \hfill$\blacksquare$
\begin{rmk}
	By Lemma \ref{lemma5}, the larger privacy noise parameters $\sigma_k^{(\zeta_i)},\sigma_k^{(\eta_i)}$ are, the smaller the cumulative differential privacy budget $\varepsilon_i$ is. While by Theorem \ref{thm2}, the larger privacy noise parameters $\sigma_k^{(\zeta_i)},\sigma_k^{(\eta_i)}$ are, the higher the oracle complexity is. For example, if the error $\varphi=0.02$, then the oracle complexity $O(10^{15})$ of Algorithm \ref{algorithm1} with \emph{Scheme (S1)} is higher than the oracle complexity $O(10^3)$ of the centralized SGD in \cite{lian2017can}. As a result, Algorithm~\ref{algorithm1} with \emph{Scheme (S1)} achieves privacy at the cost of increasing the oracle complexity.
\end{rmk}
\begin{thm0}\label{thm4}
	For step-sizes $\alpha_K,\beta_K,\gamma_K$, the sampling number $m_K$ satisfying \emph{Scheme (S1)}, and privacy noise parameters $\sigma_k^{(\zeta_i)}=(k+1)^{p_{\zeta_i}}$, $\sigma_k^{(\eta_i)}=(k+1)^{p_{\eta_i}}$, if Assumption \ref{asm2}(i) and the following conditions hold:
		\begin{align*}
			&p_m-p_\beta+\min\{\min_{i\in\mathcal{V}}\{p_{\eta_i}\}-1,0\}>0,\cr
			\noalign{\vskip -2pt}
			& p_m+\min\{0,p_\gamma-p_\alpha-p_\beta\}+\min\{\min_{i\in\mathcal{V}}\{p_{\zeta_i}\}-1,0\}>0,\cr
			\noalign{\vskip -5pt}
			&0\!<\!a_1\!<\!\min_{i\in\mathcal{V}}\{\frac{1}{\sum_{\hspace{-0.15em}j\in\mathcal{N}_{\mathcal{R},i}^{-}}\hspace{-0.65em}\mathcal{R}_{ij}}\},0\!<\!a_2\!<\!\min_{i\in\mathcal{V}}\{\frac{1}{\sum_{\hspace{-0.15em}j\in\mathcal{N}_{\mathcal{C},i}^{+}}\hspace{-0.4em}\mathcal{C}_{ji}}\},
	\end{align*}
	then the cumulative privacy budget $\varepsilon_i$ is finite for any $i\in\mathcal{V}$ even over infinite iterations.
\end{thm0}
{\bf Proof.} First, we compute $\sum_{k=0}^{K}\!\!\!\frac{\|\Delta y_{i,k}\|_1}{\sigma_k^{(\eta_i)}}$ for any $i\in\mathcal{V}$. Since $0<a_2<\min\{\min_{i\in\mathcal{V}}\{\frac{1}{\sum_{\hspace{-0.15em}j\in\mathcal{N}_{\mathcal{C},i}^{+}}\hspace{-0.65em}\mathcal{C}_{ji}}\}\}$, it can be seen that $0\!<\!\beta_K\sum_{j\in\mathcal{N}_{\mathcal{C},i}^{+}}\!\!\mathcal{C}_{ji}\!<\!1$. When $k=0,1$, $\|\Delta y_{i,k}\|_1=O(\frac{1}{(K+1)^{p_m}})$ by Lemma~\ref{lemma4}. When $2\leq k\leq K$, we have\vspace{-0.5em}
\begin{align}\label{73}
	\hspace{-1em}\|\Delta y_{i,k}\|_1=&O(\frac{|1\!-\!\beta_K\!\sum_{j\in\mathcal{N}_{\mathcal{C},i}^{+}}\!\!\mathcal{C}_{ji}|(1\!\!-\!\!|1\!\!-\!\!\beta_K\!\sum_{j\in\mathcal{N}_{\mathcal{C},i}^{+}}\!\!\mathcal{C}_{ji}|^k)}{m_K(1\!\!-\!\!|1\!\!-\!\!\beta_K\!\sum_{j\in\mathcal{N}_{\mathcal{C},i}^{+}}\!\!\mathcal{C}_{ji}|)})\notag\\
	\noalign{\vskip -4pt}
	=&O(\frac{1-\beta_K\!\sum_{j\in\mathcal{N}_{\mathcal{C},i}^{+}}\!\!\mathcal{C}_{ji}}{(\beta_K\!\sum_{j\in\mathcal{N}_{\mathcal{C},i}^{+}}\!\!\mathcal{C}_{ji})(K+1)^{p_m-p_\beta}})\cr
	\noalign{\vskip -2pt}
	=&O(\frac{1}{(K+1)^{p_m-p_\beta}}).
\end{align}
{\vskip -5pt}\noindent Then, for any $k=0,\dots,K$ and $i\in\mathcal{V}$, $\|\Delta y_{i,k}\|_1=O(\frac{1}{(K+1)^{p_m-p_\beta}})$, and $\sum_{k=0}^{K}\frac{\|\Delta y_{i,k}\|_1}{\sigma_k^{(\eta_i)}}$ can be rewritten as\vspace{-0.7em}
\begin{align*}
	\sum_{k=0}^{K}\!\frac{\|\Delta y_{i,k}\|_1}{\sigma_k^{(\eta_i)}}=&\frac{1}{(K\!+\!1)^{p_m\!-\!p_\beta}}O(\sum_{k=1}^{K}\!\frac{1}{k^{p_{\eta_i}}})\cr
	\noalign{\vskip -3pt}
	=&O(\frac{\ln (K\!+\!2)}{(K\!+\!1)^{p_m\!-p_\beta+\min\{p_{\eta_i}-1,0\}}}).
\end{align*}
{\vskip -5pt}\noindent Hence, if $p_m-p_\beta+\min\{\min_{i\in\mathcal{V}}\{p_{\eta_i}\}-1,0\}>0$ holds, then $\sum_{k=0}^{\infty}\frac{\|\Delta y_{i,k}\|_1}{\sigma_k^{(\eta_i)}}$ is finite for any $i\in\mathcal{V}$.

Next, we compute $\sum_{k\!=\!0}^{K}\!\!\frac{\|\Delta x_{i,k}\|_1}{\sigma_k^{(\zeta_i)}}$ for any $i\in\mathcal{V}$. Since $0<a_1<\min\{\min_{i\in\mathcal{V}}\{\frac{1}{\sum_{\hspace{-0.15em}j\in\mathcal{N}_{\mathcal{R},i}^{-}}\hspace{-0.65em}\mathcal{R}_{ij}}\}\}$, it can be seen that $0<\alpha_K\sum_{j\in\mathcal{N}_{\mathcal{R},i}^{-}}\!\!\!\!\mathcal{R}_{ij}$$<$$1$. When $k$$=$$0,1$, by Lemma \ref{lemma4}, $\|\Delta x_{i,k}\|_1$$=$ $O(\frac{1}{(K+1)^{p_m}})$. When $k=2,\dots,K$, by \eqref{73}, we have\vspace{-0.9em}
\begin{align}\label{74}
	\|\Delta x_{i,k}\|_1\leq&\sum\limits_{l=1}^{k-1}\!|1\!-\!\alpha_K\!\!\!\!\!\!\!\!\sum_{\hspace{1em}j\in\mathcal{N}_{\mathcal{R},i}^{-}}\!\!\!\!\!\!\!\!\mathcal{R}_{ij}|^{k\!-\!l}\gamma_K\|\Delta y_{l\!-\!1}\|_1\!+\!\gamma_K\|\Delta y_{k\!-\!1}\|_1\notag\\
	\noalign{\vskip -2pt}
	=&O(\frac{1-\alpha_K\hspace{-0.3em}\sum_{j\in\mathcal{N}_{\mathcal{R},i}^{-}}\hspace{-0.5em}\mathcal{R}_{i j}}{(\alpha_K\hspace{-0.3em}\sum_{j\in\mathcal{N}_{\mathcal{R},i}^{-}}\hspace{-0.5em}\mathcal{R}_{i j})(K+1)^{p_m+p_\gamma-p_\beta}})\notag\\
	\noalign{\vskip -2pt}
	=&O(\frac{1}{(K\!+\!1)^{p_m+p_\gamma-p_\alpha-p_\beta}}).
\end{align}
{\vskip -5pt}\noindent Then, for any $k=0,\dots,K$ and $i\in\mathcal{V}$, $\|\Delta x_{i,k}\|_1$$=$ $O(\frac{1}{(K\!+\!1)^{p_m\!+\!\min\{0,p_\gamma\!-\!p_\alpha\!-\!p_\beta\}}})$, and thus, $\sum_{k=0}^{K}\frac{\|\Delta x_{i,k}\|_1}{\sigma_k^{(\zeta_i)}}$ can be rewritten as\vspace{-0.4em}
\begin{align*}
	\sum_{k=0}^{K}\frac{\|\Delta x_{i,k}\|_1}{\sigma_k^{(\zeta_i)}}=&\frac{1}{(K\!+\!1)^{p_m+\min\{0,p_\gamma-p_\alpha-p_\beta\}}}O(\sum_{k=1}^{K}\frac{1}{k^{p_{\zeta_i}}})\cr
	\noalign{\vskip -4pt}
	=&O(\frac{\ln (K\!+\!2)}{(K\!+\!1)^{p_m+\min\{0,p_\gamma-p_\alpha-p_\beta\}+\min\{\zeta_i-1,0\}}}).
\end{align*}
{\vskip -5pt}\noindent If $p_m+\min\{0,p_\gamma-p_\alpha-p_\beta\}+\min\{\min_{i\in\mathcal{V}}\{p_{\zeta_i}\}-1,0\}>0$, then $\sum_{k=0}^{\infty}\frac{\|\Delta x_{i,k}\|_1}{\sigma_k^{(\zeta_i)}}$ is finite for any $i\in\mathcal{V}$. By Lemma \ref{lemma5}, this theorem is proved.~\hfill$\blacksquare$

\begin{thm0}\label{thm5}
	For step-sizes $\alpha_K,\beta_K,\gamma_K$, the sampling number $m_K$ satisfying \emph{Scheme (S2)}, and privacy noise parameters $\sigma_k^{(\zeta_i)}$$=$$p_{\zeta_i}^K$, $\sigma_k^{(\eta_i)}$$=$$p_{\eta_i}^K$, if Assumption \ref{asm2}(i) and the following conditions hold:
		\begin{align*}
			&0<p_{\zeta_i},p_{\eta_i}<1,p_m>\max_{i\in\mathcal{V}}\{\frac{1}{p_{\zeta_i}},\frac{1}{p_{\eta_i}}\},\cr
			&0\!<\!\alpha\!<\!\min_{i\in\mathcal{V}}\{\frac{1}{\sum_{\hspace{-0.15em}j\in\mathcal{N}_{\mathcal{R},i}^{-}}\hspace{-0.65em}\mathcal{R}_{ij}}\},0\!<\!\beta\!<\!\min_{i\in\mathcal{V}}\{\frac{1}{\sum_{\hspace{-0.15em}j\in\mathcal{N}_{\mathcal{C},i}^{+}}\hspace{-0.45em}\mathcal{C}_{ji}}\},
	\end{align*} then the cumulative privacy budget $\varepsilon_i$ is finite for any $i\in\mathcal{V}$ even over infinite iterations.
\end{thm0}
{\bf Proof.} By Lemma \ref{lemma4}, it can be seen that\vspace{0.3em}
\begin{align*}
	\smash{\sum_{k=0}^{K}\frac{\|\Delta x_{i,k}\|_1}{\sigma_k^{(\zeta_i)}}\!+\!\frac{\|\Delta y_{i,k}\|_1}{\sigma_k^{(\eta_i)}}=O(K(\frac{1}{p_mp_{\zeta_i}})^K\!+\!K(\frac{1}{p_mp_{\eta_i}})^K).}
\end{align*}
{\vskip 5pt}\noindent Hence, if $\frac{1}{p_m}<\min\{p_{\zeta_i},p_{\eta_i}\}$, then $\sum_{k=0}^{\infty}\frac{\|\Delta x_{i,k}\|_1}{\sigma_k^{(\zeta_i)}}+\frac{\|\Delta y_{i,k}\|_1}{\sigma_k^{(\eta_i)}}$ is finite. Therefore, this theorem is proved. \hfill$\blacksquare$
\begin{rmk}
	Theorems \ref{thm4} and \ref{thm5} establish the sufficient condition for Algorithm \ref{algorithm1} with \emph{Schemes (S1)}, \emph{(S2)} to achieve the finite cumulative differential privacy budget $\varepsilon_i$ even over infinite iterations, respectively. This is different from \cite{doan2021,xin2019distributed,pu2021distributed,koloskova2021improved,xin2020variance,lei2022distributed,wang2023gradient,lu2023convergence,chen2024accelerated,zhao2024asymptotic} that do not consider the privacy protection, and \cite{ding2021differentially,kang2021weighted,xu2022dp,wang2023decentralized,wang2023quantization,yan2024killing,liu2024distributed} that achieve the infinite cumulative differential privacy budget $\varepsilon_i$ over infinite iterations. Thus, compared to \cite{ding2021differentially,kang2021weighted,xu2022dp,wang2023decentralized,wang2023quantization,yan2024killing,liu2024distributed}, Algorithm \ref{algorithm1} with both \emph{Schemes (S1)} and \emph{(S2)} provides a higher differential privacy level.
\end{rmk}
\vspace{-1em}
\subsection{Trade-off between privacy and convergence}
Based on Theorems \ref{thm2}-\ref{thm5}, the trade-off between the privacy and the convergence  is given in the following corollary:
\begin{Cor}\label{cor1}
	(i) Under Assumptions \ref{asm1}-\ref{asm3}, \ref{asm4}, if $p_m-p_\beta+\min\{\min_{i\in\mathcal{V}}\{p_{\eta_i}\}-1,0\}>0,p_m+\min\{0,p_\gamma-p_\alpha-p_\beta\}+\min\{\min_{i\in\mathcal{V}}\{p_{\zeta_i}\}-1,0\}>0$ hold, then Algorithm \ref{algorithm1} with \emph{Scheme (S1)} achieves the polynomial mean square convergence rate and the finite cumulative differential privacy budget $\varepsilon_i$ for any $i\in\mathcal{V}$ even over infinite iterations simultaneously.
	
	\noindent(ii) Under Assumptions \ref{asm1}-\ref{asm3}, \ref{asm5}, if $0<p_{\zeta_i},p_{\eta_i}<1,p_m>\max_{i\in\mathcal{V}}\{\frac{1}{p_{\zeta_i}},\frac{1}{p_{\eta_i}}\}$ hold, then Algorithm \ref{algorithm1} with \emph{Scheme (S2)} achieves the exponential mean square convergence rate and the finite cumulative differential privacy budget $\varepsilon_i$ for any $i\in\mathcal{V}$ even over infinite iterations simultaneously.
\end{Cor}
{\bf Proof.} By Theorems \ref{thm2} and \ref{thm4}, Corollary \ref{cor1}(i) is proved. Then, by Theorems \ref{thm3} and \ref{thm5}, Corollary~\ref{cor1}(ii) is proved. \hfill$\blacksquare$
\begin{rmk}
	\emph{Scheme (S1)} achieves the polynomial mean square convergence rate and the finite cumulative differential privacy budget $\varepsilon_i$ over infinite iterations simultaneously under decreasing, constant and increasing privacy noises. For example, let $p_\alpha$$=$$0.987$, $p_\beta$$=$$0.69$, $p_\gamma=0.997$, $p_m$$=$$2$. Then, conditions in Corollary \ref{cor1}(i) are satisfied as long as $-0.3$$<$$p_{\zeta_i}$$<$$0.15$, $-0.3$$<$$p_{\eta_i}$$<$$0.15$. \emph{Scheme (S2)} achieves the exponential mean square convergence rate and the finite cumulative differential privacy budget $\varepsilon_i$ over infinite iterations simultaneously under decreasing privacy noises. For example, let $\alpha$$=$$0.1$, $\beta$$=$$0.1$, $\gamma$$=$$0.01$, $p_m$$=$$1.1$. Then,  conditions in Corollary \ref{cor1}(ii) are satisfied as long as $0.91$$<$$p_{\zeta_i}$$<$$0.95$, $0.91$$<$$p_{\eta_i}$$<$$0.95$.
\end{rmk}
\begin{rmk}
	Corollary \ref{cor1} shows the trade-off between privacy and the convergence rate in Algorithm \ref{algorithm1}. The smaller privacy noise parameters $\sigma_k^{(\zeta_i)}$, $\sigma_k^{(\eta_i)}$ are, the faster Algorithm \ref{algorithm1} converges, while the larger the cumulative differential privacy budget $\varepsilon_i$ is. Moreover, \emph{Scheme (S1)} achieves the polynomial mean square convergence rate and finite cumulative differential privacy budget $\varepsilon_i$ over infinite iterations under decreasing, constant, and increasing privacy noises, while \emph{Scheme (S2)} achieves the exponential mean square convergence rate and finite cumulative differential privacy budget $\varepsilon_i$ only for decreasing privacy noises. Then, the differential privacy level of \emph{Scheme (S1)} is higher than the one of \emph{Scheme (S2)}, while the convergence rate of \emph{Scheme (S2)} is faster than the one of \emph{Scheme (S1)}.
\end{rmk}
\begin{rmk}
	The parameter $a_4$ in the sampling number $m_K=\lfloor a_4 K^{p_m}\rfloor+1$ affects both convergence rate and the cumulative privacy budget. Since by \eqref{54c}, $\E\|\nabla F(x_{i,K\!+\!1})\|^2\!=\!O(\!\frac{a_4+1}{a_4(K\!+\!1)^{\theta-p_\gamma}})$ is decreasing with respect to $a_4$. Then, the larger the parameter $a_4$ is, the faster the convergence rate is. By Lemma \ref{lemma4}, the larger the parameter $a_4$ is, the smaller the sensitivity $\Delta_{i,k}^q$ is, and then by Theorem \ref{thm4}, the smaller the cumulative privacy budget $\varepsilon_i$ is.
\end{rmk}

Based on Corollary \ref{cor1}, we have the following corollary as the sampling number goes to infinity:
\begin{Cor}\label{cor2}
	Under the conditions of Corollary \ref{cor1}, Algorithm~\ref{algorithm1} with both \emph{Schemes (S1)}, \emph{(S2)} achieves the almost sure and mean square convergence and the finite cumulative differential privacy budget $\varepsilon_i$ for any $i$$\in$$\mathcal{V}$ over infinite iterations simultaneously as the sampling number goes to infinity.
\end{Cor}
\begin{rmk}
	The result of Corollary \ref{cor2} does not contradict the trade-off between privacy and utility. In fact, to achieve differential privacy, Algorithm \ref{algorithm1} incurs a compromise on the utility. However, different from \cite{kang2021weighted,yan2024killing,liu2024distributed} that compromise convergence accuracy to enable differential privacy, Algorithm~\ref{algorithm1} compromises the convergence rate and the sampling number (which are also utility metrics) instead. By Corollary~\ref{cor1}, the larger privacy noise parameters $\sigma_{k}^{(\zeta_i)}$, $\sigma_{k}^{(\eta_i)}$ are, the slower the convergence rate is. By Corollary \ref{cor2}, the sampling number $m_{K}$ is required to go to infinity when the convergence of Algorithm \ref{algorithm1} and the finite cumulative privacy budget $\varepsilon_i$ over infinite iterations are considered simultaneously. The ability to retain convergence accuracy makes our approach suitable for accuracy-critical scenarios.
\end{rmk}\vspace{-1em}
\section{Numerical examples}\label{section 4}
\begin{figure}[!htbp]
	\centering
	\subfloat[The directed graph $\mathcal{G}_{\mathcal{R}}$]{\includegraphics[width=0.168\textwidth]{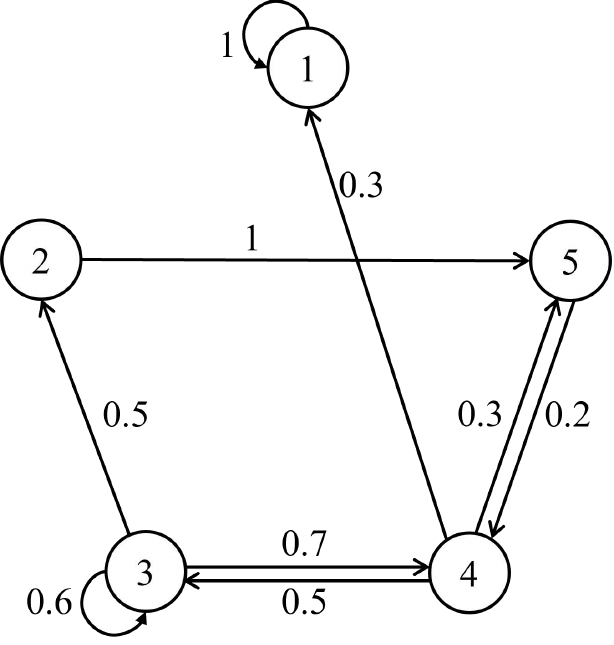}\label{fig1a}}\hspace{3em}
	\subfloat[The directed graph $\mathcal{G}_{\mathcal{C}}$]{\includegraphics[width=0.175\textwidth]{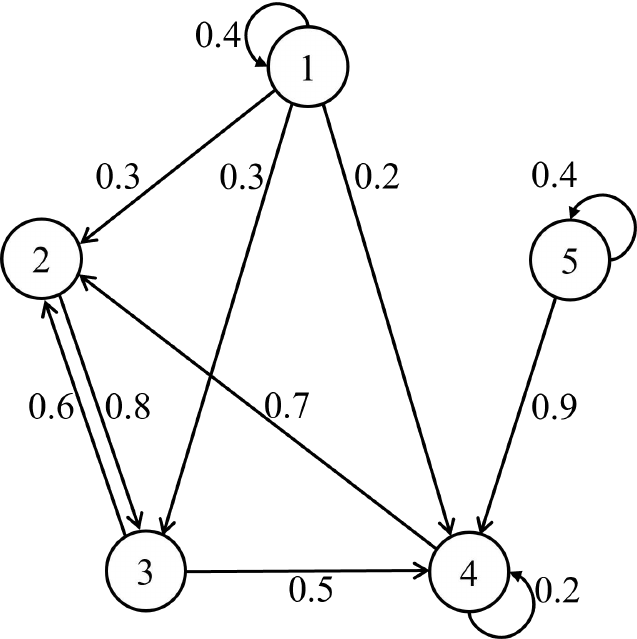}\label{fig1b}}
	\caption{\small Topology structures of directed graphs $\mathcal{G}_{\mathcal{R}},\mathcal{G}_{\mathcal{C}}$ induced by weight matrices $\mathcal{R},\mathcal{C}$ }
	\label{fig1}
\end{figure}

In this section, we train the machine learning model ResNet18 (\hspace{-0.05em}\cite{he2016deep}) in a distributed manner with the benchmark datasets ``MNIST'' (\hspace{-0.05em}\cite{mnist1998lecun}) and ``CIFAR-10'' (\hspace{-0.05em}\cite{cifar10torento,cifar}), respectively. Specifically, five agents cooperatively train ResNet18 over the directed graphs shown in Figs. \ref{fig1}\subref{fig1a} and \ref{fig1}\subref{fig1b}, which satisfy Assumption \ref{asm1}. Then, each benchmark dataset is divided into two subsets for training and testing, respectively. The training dataset of each benchmark dataset is uniformly divided into 5 subsets, each of which can only be accessed by one agent to update its model parameters.To ensure a fair comparison, we set the cumulative differential privacy budget $\varepsilon=\max_{i\in\mathcal{V}}\{\varepsilon_i\}$ for Algorithm \ref{algorithm1} with \emph{Schemes (S1)}, \emph{(S2)}. Then, the following three numerical experiments are given:
\begin{itemize}[leftmargin=1.5em]
	\item[(a)] the effect of privacy noises on Algorithm \ref{algorithm1}'s convergence rate and differential privacy level;
	\item[(b)] the comparison of Algorithm \ref{algorithm1} with \emph{Schemes (S1)}, \emph{(S2)} between the convergence rate and the differential \mbox{privacy level;}
	\item[(c)] the comparison between Algorithm \ref{algorithm1} with \emph{Schemes (S1)}, \emph{(S2)} and methods in \cite{kang2021weighted,wang2023quantization,yan2024killing,chen2024locallya,wang2024differentiallya,chen2024differentially} for the convergence rate and the differential privacy level.
\end{itemize}
\vspace{-1em}
\subsection{Effect of privacy noises}
\begin{figure}[!htbp]
	{\vskip -12pt}
	\centering
	\subfloat[Training accuracy on the ``MNIST'' dataset]{\includegraphics[width=0.165\textwidth]{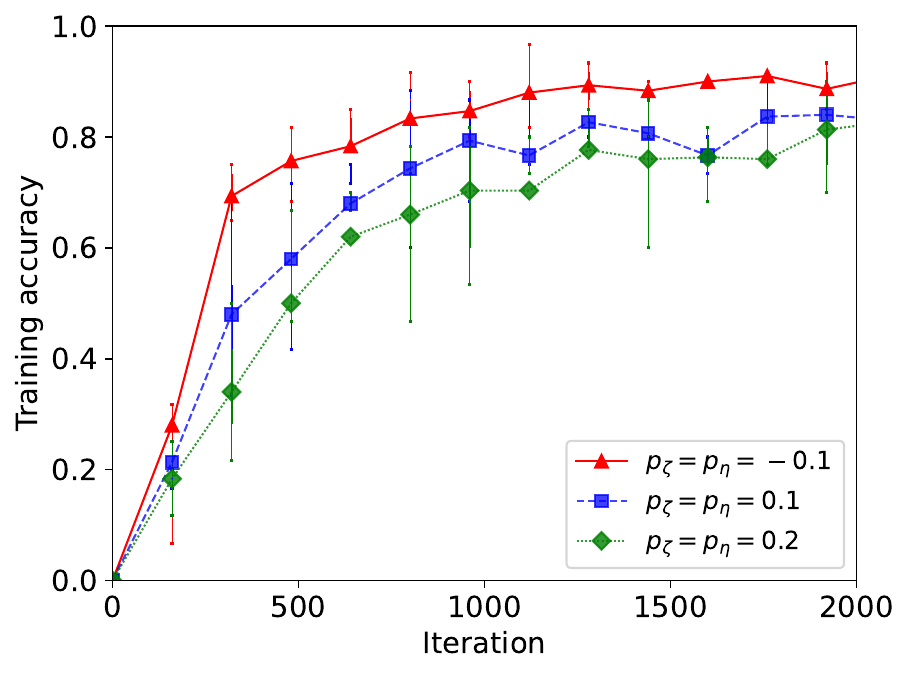}\label{fig2a}}
	\subfloat[Testing accuracy on the ``MNIST'' dataset]{\includegraphics[width=0.165\textwidth]{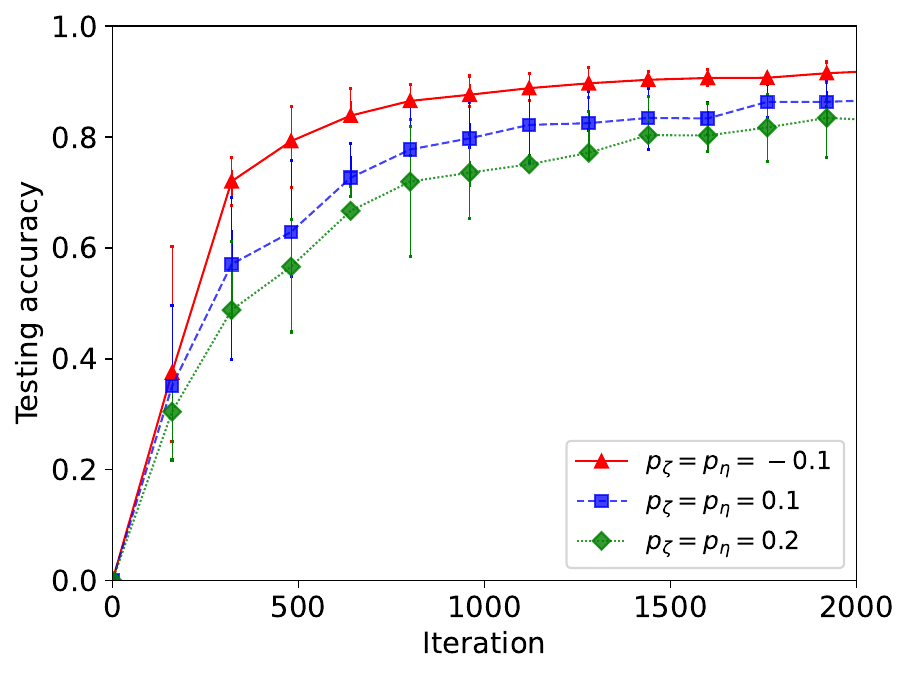}\label{fig2b}}
	\subfloat[Training accuracy on the ``CIFAR-10'' dataset]{\includegraphics[width=0.165\textwidth]{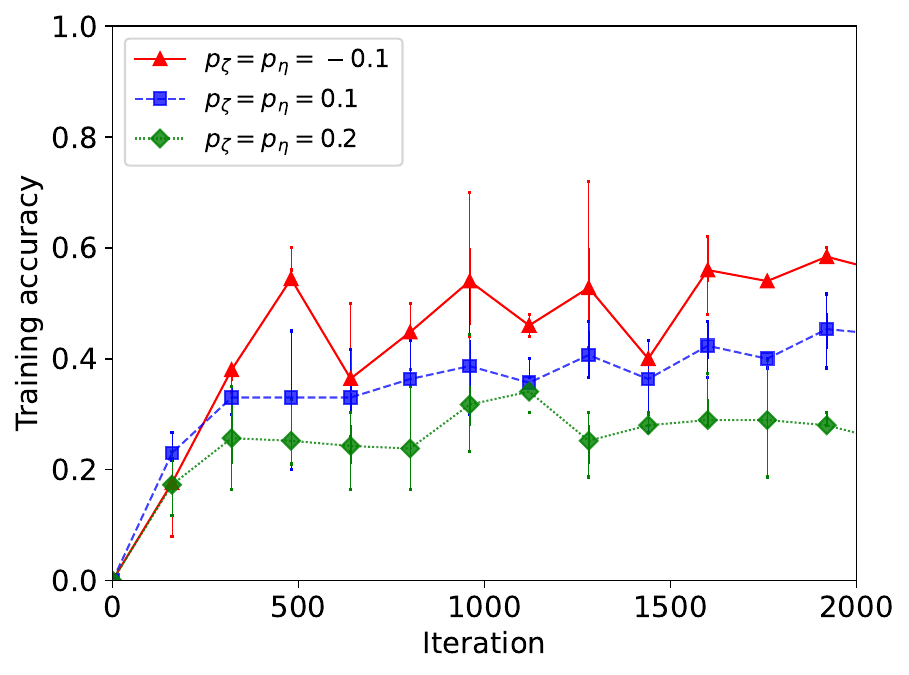}\label{fig2c}}\\
	{\vskip -1pt}
	\subfloat[Testing accuracy on the ``CIFAR-10'' dataset]{\includegraphics[width=0.165\textwidth]{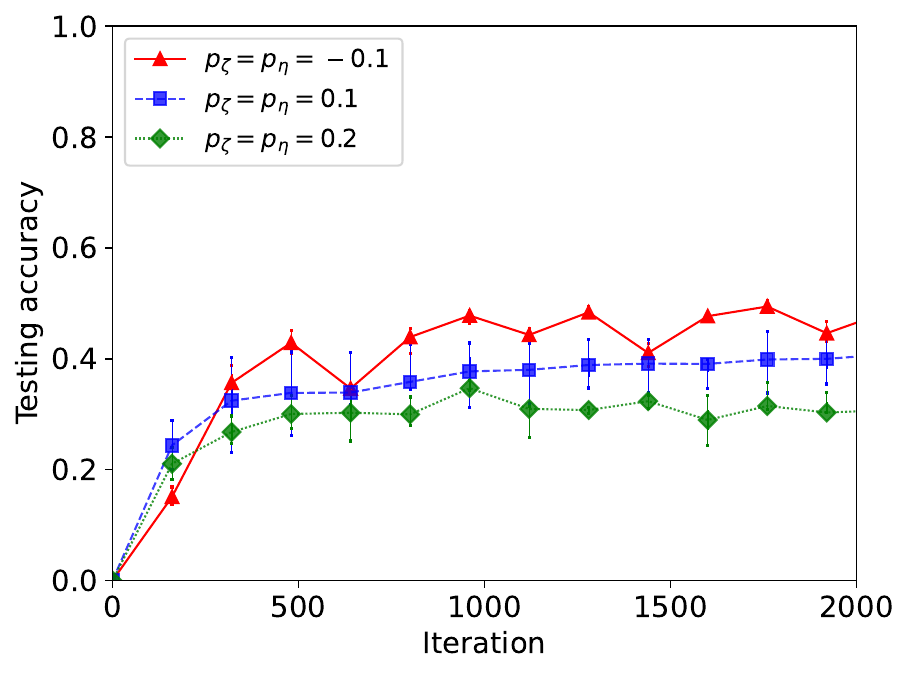}\label{fig2d}}\hspace{1em}
	\subfloat[Cumulative \mbox{differential privacy budget $\varepsilon$}]{\includegraphics[width=0.165\textwidth]{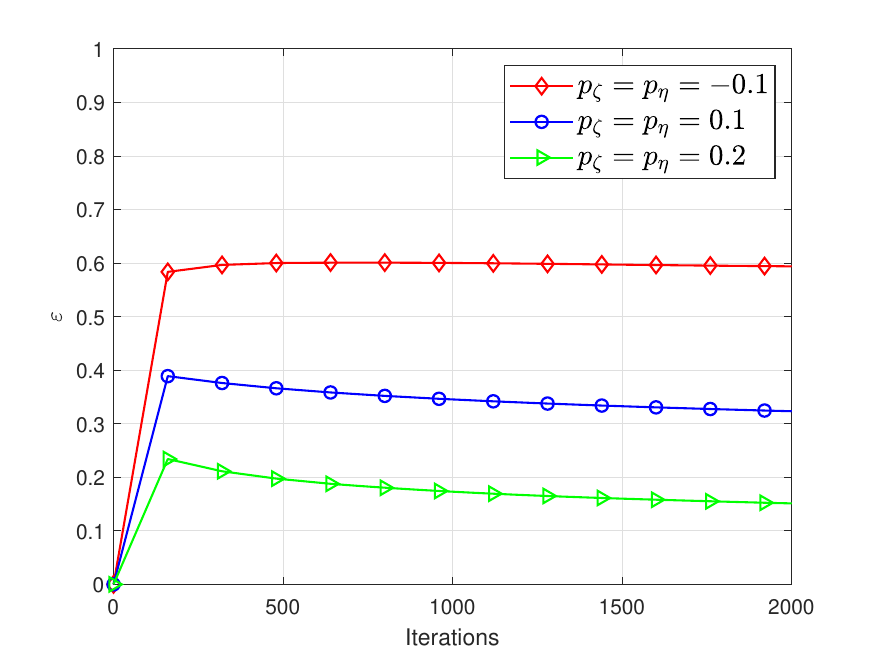}\label{fig2e}}\vspace{-0.5em}
	\caption{\small Accuracy and cumulative differential privacy budget $\varepsilon$ of Algorithm \ref{algorithm1} with \emph{Scheme (S1)} and $p_{\zeta_i},p_{\eta_i}=-0.1,0.1,0.2$}
	\label{fig2}
	{\vskip -14pt}
\end{figure}

First, let step-sizes $\alpha_K$$=$$\frac{72}{2000^{0.987}}$$=$$0.04$, $\beta_K$$=$$\frac{0.95}{2000^{0.69}}$$=$ $0.005$, $\gamma_K$$=$$\frac{98}{2000^{0.997}}$$=$$0.05$, the sampling number $m_K$$=$ $\lfloor0.00007\cdot2000^{1.78}\rfloor$$+$$1$$=$$53$, and privacy noise parameters $\sigma_k^{(\zeta_i)}=(k+1)^{p_{\zeta_i}}$, $\sigma_k^{(\eta_i)}=(k+1)^{p_{\eta_i}}$ with $p_{\zeta_i},p_{\eta_i}=-0.11+0.01i,0.09+0.01i,0.19+0.01i$ in \emph{Scheme (S1)} for $i=1,\dots,5$, respectively. Then, the training and testing accuracy on the benchmark datasets ``MNIST'' and ``CIFAR-10'' are given in Fig. \ref{fig2}\subref{fig2a}-\ref{fig2}\subref{fig2d}, from which one can see that the smaller privacy noise parameters $\sigma_k^{(\zeta_i)}$, $\sigma_k^{(\eta_i)}$ are, the faster Algorithm \ref{algorithm1} converges. This is consistent with the convergence rate analysis in Theorem~\ref{thm2}. Meanwhile, the cumulative differential privacy budget $\varepsilon$ of Algorithm \ref{algorithm1} is given in Fig. \ref{fig2}\subref{fig2e}, from which one can see that that the smaller privacy noise parameters $\sigma_k^{(\zeta_i)}$, $\sigma_k^{(\eta_i)}$ are, the smaller the cumulative differential privacy budget $\varepsilon$ is. This is consistent with the privacy analysis in Theorem \ref{thm4}, and thus consistent with the trade-off between the privacy and the convergence rate \mbox{in Corollary \ref{cor1}.}

\begin{figure}[!htbp]
	{\vskip -11pt}
	\centering
	\subfloat[Training accuracy on the ``MNIST'' dataset]{\includegraphics[width=0.165\textwidth]{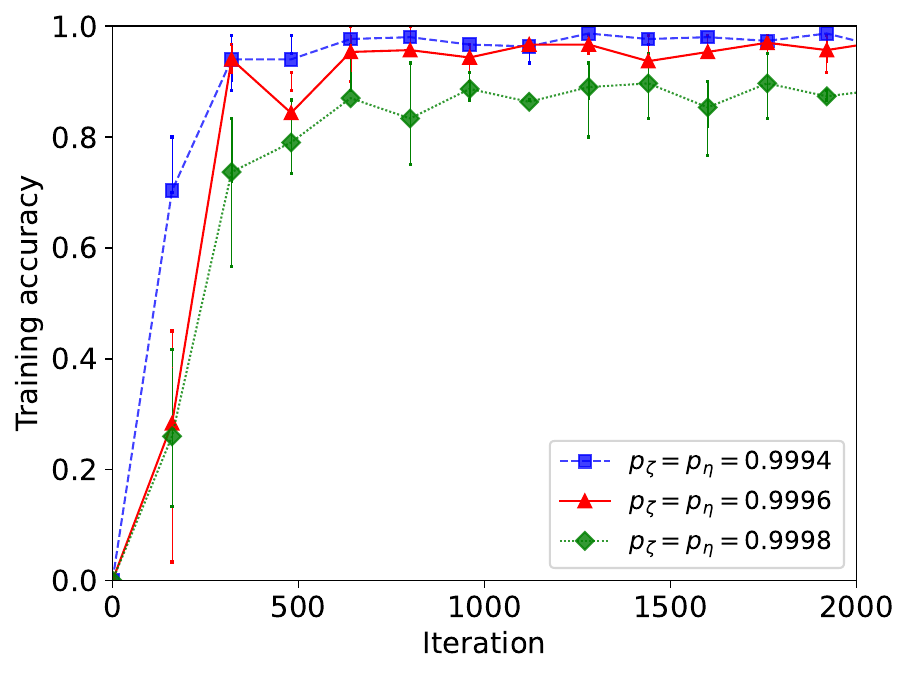}\label{fig3a}}
	\subfloat[Testing accuracy on the ``MNIST'' dataset]{\includegraphics[width=0.165\textwidth]{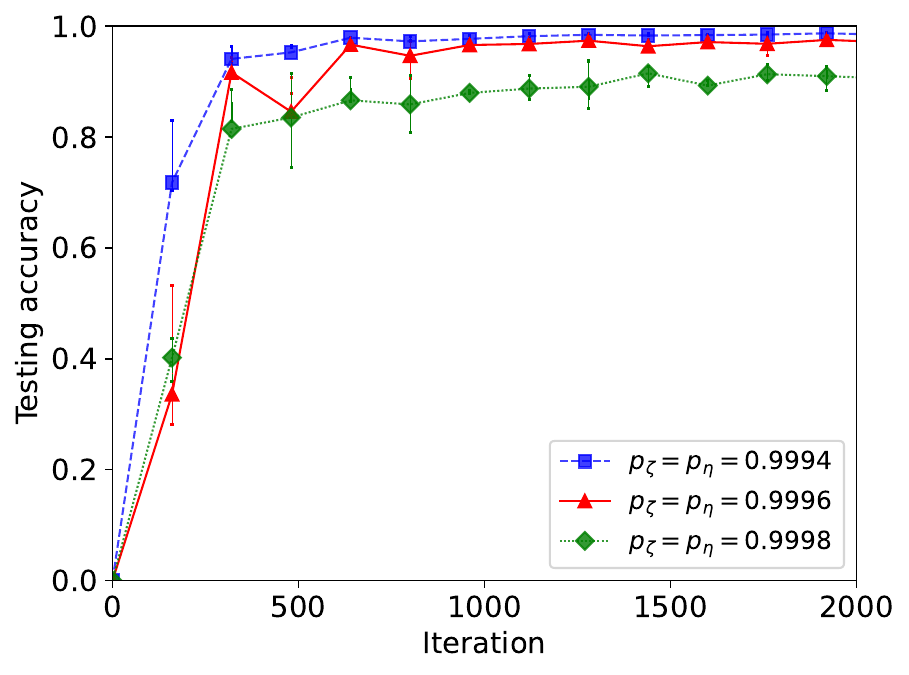}\label{fig3b}}
	\subfloat[Training accuracy on the ``CIFAR-10'' dataset]{\includegraphics[width=0.165\textwidth]{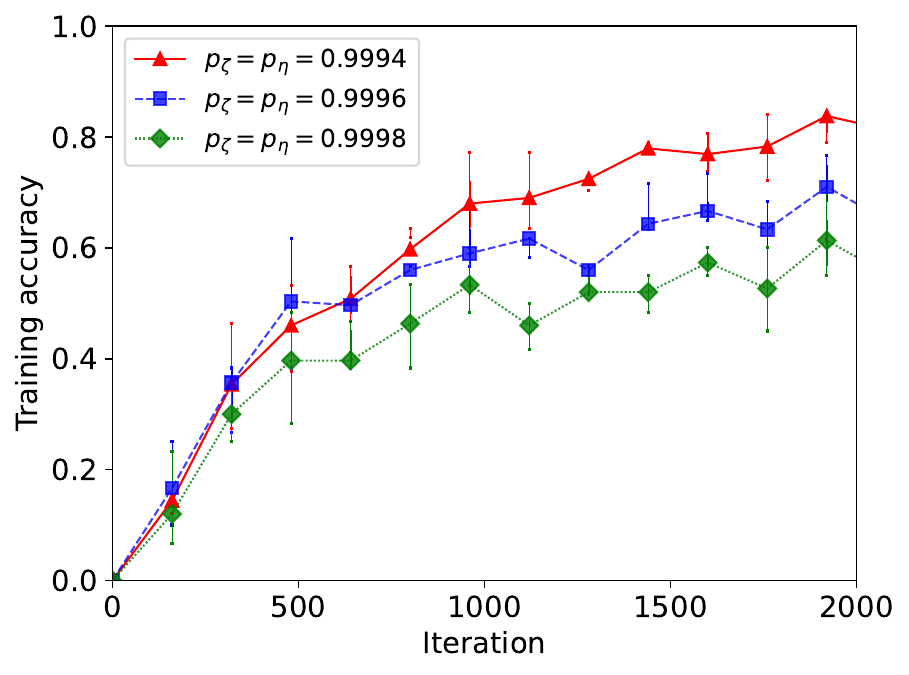}\label{fig3c}}\\
	{\vskip -1pt}
	\subfloat[Testing accuracy on the ``CIFAR-10'' dataset]{\includegraphics[width=0.165\textwidth]{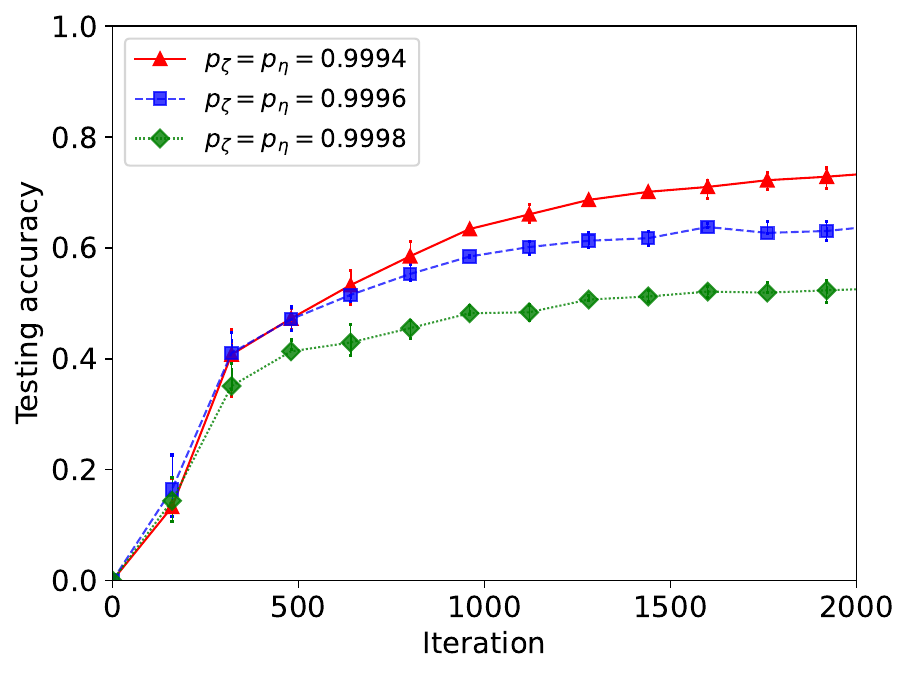}\label{fig3d}}\hspace{1em}
	\subfloat[Cumulative \mbox{differential privacy budget $\varepsilon$}]{\includegraphics[width=0.165\textwidth]{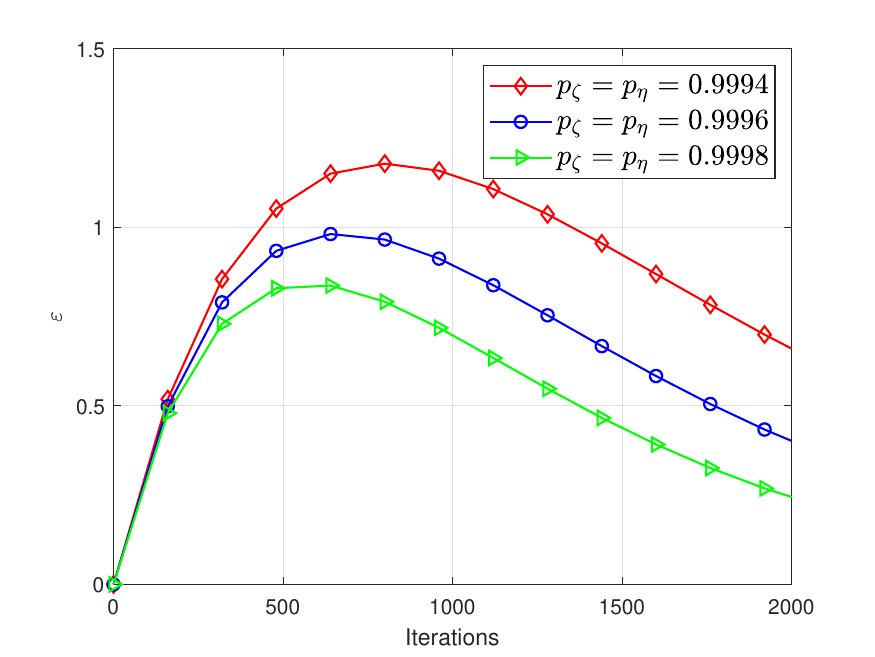}\label{fig3e}}\vspace{-0.5em}
	\caption{\small Accuracy and cumulative differential privacy budget $\varepsilon$ of Algorithm \ref{algorithm1} with \emph{Scheme (S2)} and $p_{\zeta_i},p_{\eta_i}=0.9994,0.9996,0.9998$}
	\label{fig3}
\end{figure}

Next, let step-sizes $\alpha_K=0.1$, $\beta_K=0.01$, $\gamma_K=0.1$, the sampling number $m_K=\lfloor1.002^{2000}\rfloor+1=55$, and privacy noise parameters $\sigma_k^{(\zeta_i)}={p_{\zeta_i}}^{2000}$, $\sigma_k^{(\eta_i)}={p_{\eta_i}}^{2000}$ with $p_{\zeta_i},p_{\eta_i}=0.99939+i\cdot 10^{-5},0.99959+i\cdot 10^{-5},0.99979+i\cdot 10^{-5}$ in \emph{Scheme (S2)} for $i=1,\dots,5$, respectively. Then, the training and testing accuracy on the benchmark datasets ``MNIST'' and ``CIFAR-10'' are given in Fig. \ref{fig3}\subref{fig3a}-\ref{fig3}\subref{fig3d}, from which one can see that the smaller privacy noise parameters $\sigma_k^{(\zeta_i)}$, $\sigma_k^{(\eta_i)}$ are, the faster Algorithm \ref{algorithm1} converges. This is consistent with the convergence rate analysis in Theorem \ref{thm3}. Meanwhile, the cumulative differential privacy budget $\varepsilon$ of Algorithm \ref{algorithm1} is given in Fig. \ref{fig3}\subref{fig3e}, from which one can see that that the smaller privacy noise parameters $\sigma_k^{(\zeta_i)}$, $\sigma_k^{(\eta_i)}$ are, the smaller the cumulative differential privacy budget $\varepsilon$ is. This is consistent with the privacy analysis in Theorem \ref{thm5}, and thus, consistent with the trade-off between the privacy and the convergence rate in Corollary \ref{cor1}.
\begin{rmk}
	Due to the increasing sample size $m_K$, the cumulative differential privacy budget $\varepsilon$ decreases in the later stages of the iterations in the numerical experiment. In \emph{Scheme (S1)}, the sampling number $m_K=\lfloor0.00007\cdot K^{1.78}\rfloor+1=O(K^{1.78})$. By Theorem \ref{thm4}, the cumulative differential privacy budget $\varepsilon=O(\frac{\ln (K+2)}{(K+1)^{0.22}})$. Denote the function $\psi_1(t)=\frac{\ln (t+2)}{(t+1)^{0.22}}$. Then, it can be seen that the function $\psi_1(t)$ decreases when $t$ satisfies $t+1\leq0.22(t+2)\ln (t+2)$, i.e., $t\geq87.54$. Thus, the cumulative differential privacy budget $\varepsilon$ decreases when the maximum iteration number $K\geq 88$. This result is consistent with Fig. \ref{fig2}\subref{fig2e}. Similarly, in \emph{Scheme (S2)}, the sampling number $m_K=\lfloor 1.002^K \rfloor+1=O(1.002^K)$. By Theorem \ref{thm5}, the cumulative differential privacy budget $\varepsilon=O(\frac{K}{1.0016^K})$. Denote the function $\psi_2(t)=\frac{t}{1.0016^t}$. Then, it can be seen that the function $\psi_2(t)$ decreases when  $t\geq 625.49$. Thus, the cumulative differential privacy budget $\varepsilon$ decreases when the maximum iteration number $K\geq 626$. This result is consistent with Fig. \ref{fig3}\subref{fig3e}.
\end{rmk}
\subsection{\hspace{-0.15em}Comparison  between \emph{Schemes \hspace{-0.09em}(S1)} and \hspace{-0.09em}\emph{(S2)}}
\begin{figure}[!htbp]
	\centering
	\subfloat[Training accuracy on the ``MNIST'' dataset]{\includegraphics[width=0.165\textwidth]{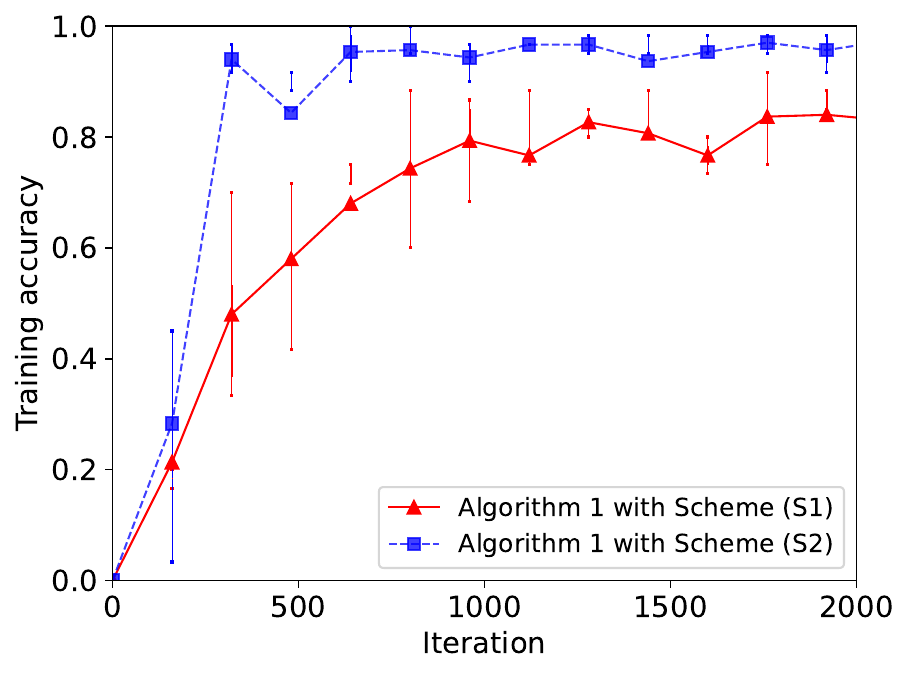}\label{fig4a}}
	\subfloat[Testing accuracy on the ``MNIST'' dataset]{\includegraphics[width=0.165\textwidth]{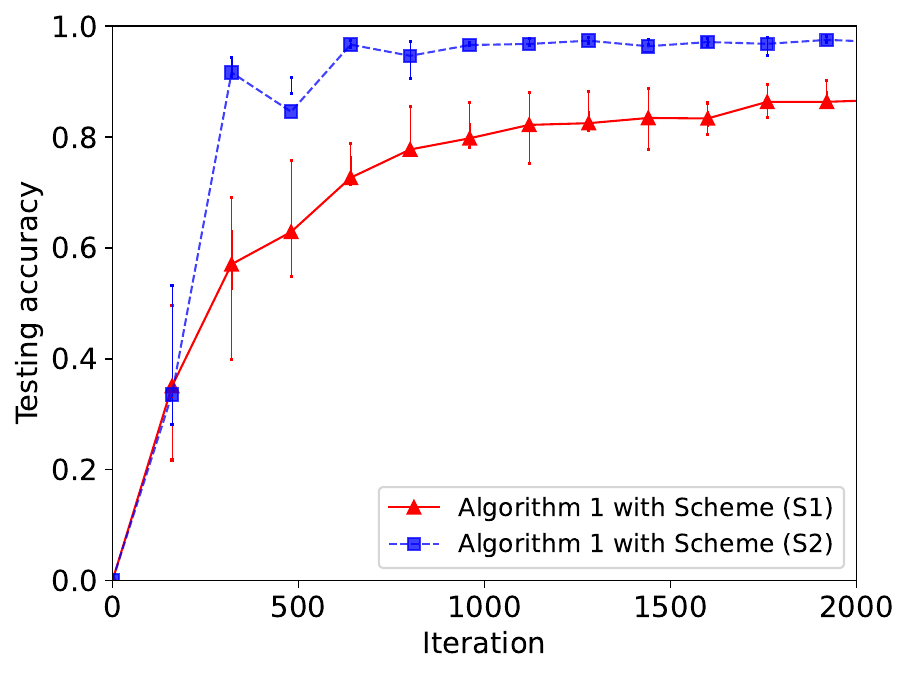}\label{fig4b}}
	\subfloat[Training accuracy on the ``CIFAR-10'' dataset]{\includegraphics[width=0.165\textwidth]{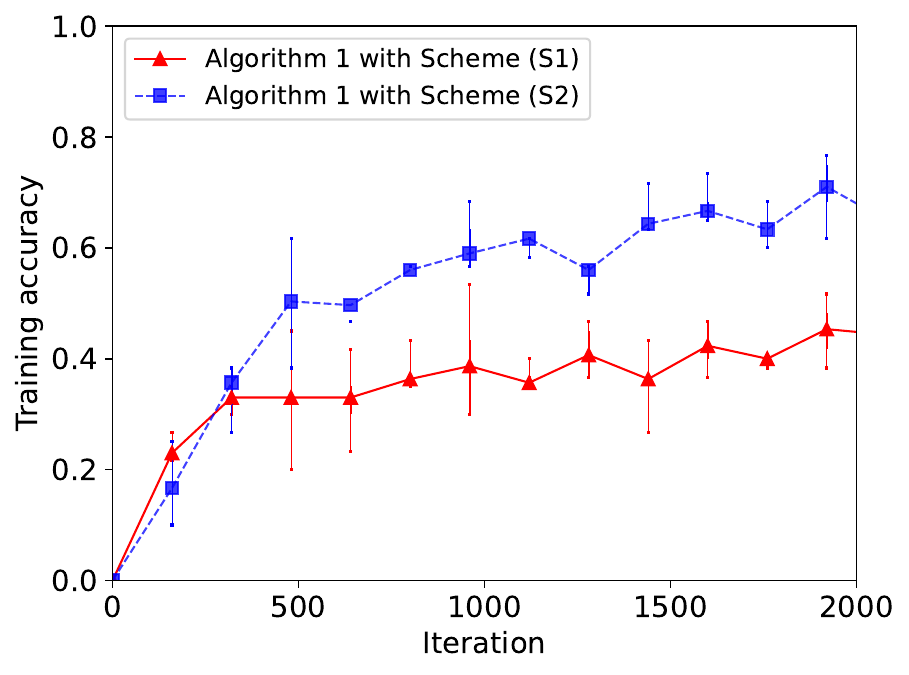}\label{fig4c}}\\
	{\vskip -1pt}
	\subfloat[Testing accuracy on the ``CIFAR-10'' dataset]{\includegraphics[width=0.165\textwidth]{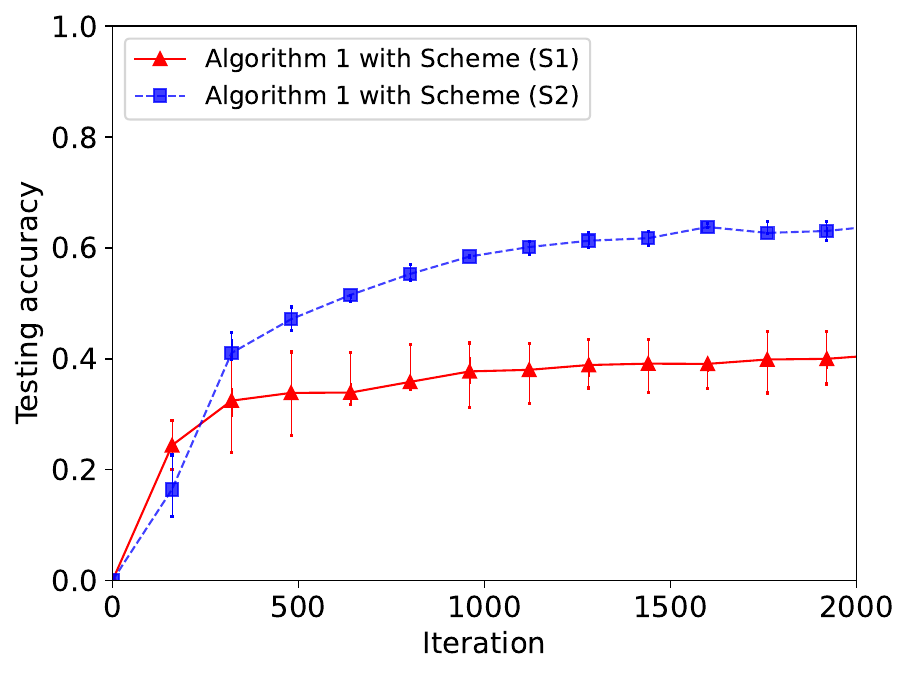}\label{fig4d}}\hspace{1em}
	\subfloat[Cumulative \mbox{differential privacy budget $\varepsilon$}]{\includegraphics[width=0.165\textwidth]{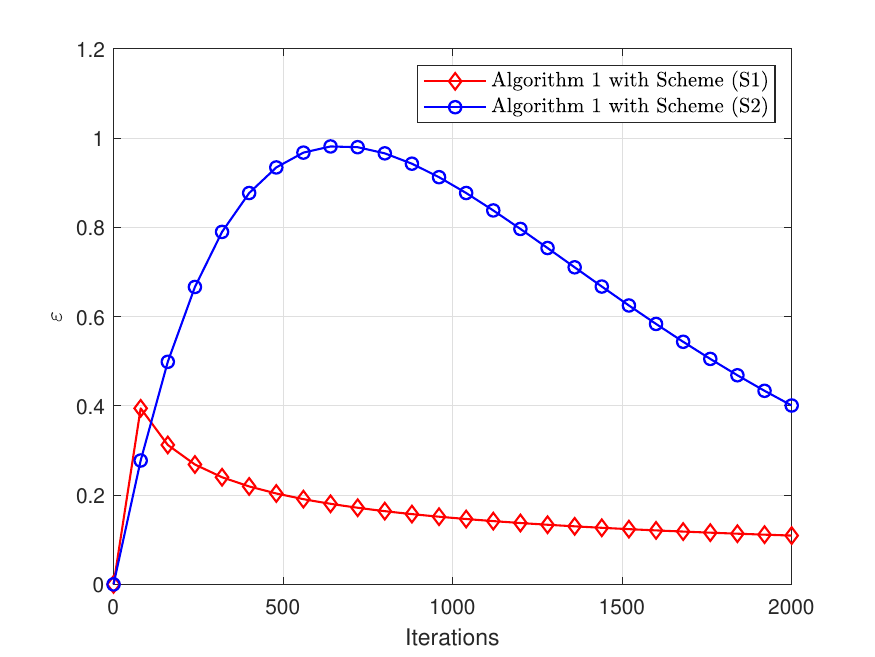}\label{fig4e}}
	\caption{\small Comparison of Algorithm \ref{algorithm1} with \emph{Schemes (S1)}, \emph{(S2)} on accuracy and cumulative differential privacy budget $\varepsilon$}
	\label{fig4}
\end{figure}

\begin{figure}[!h]
	\centering
	\subfloat[Training accuracy on the ``MNIST'' dataset]{\includegraphics[width=0.16\textwidth]{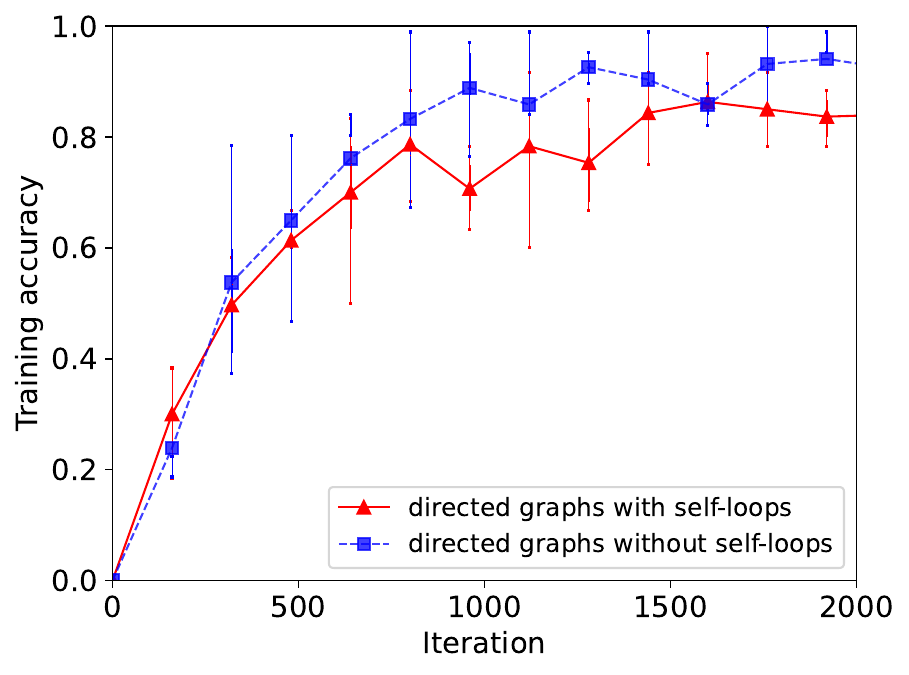}\label{fig7a}}
	\subfloat[Testing accuracy on the ``MNIST'' dataset]{\includegraphics[width=0.16\textwidth]{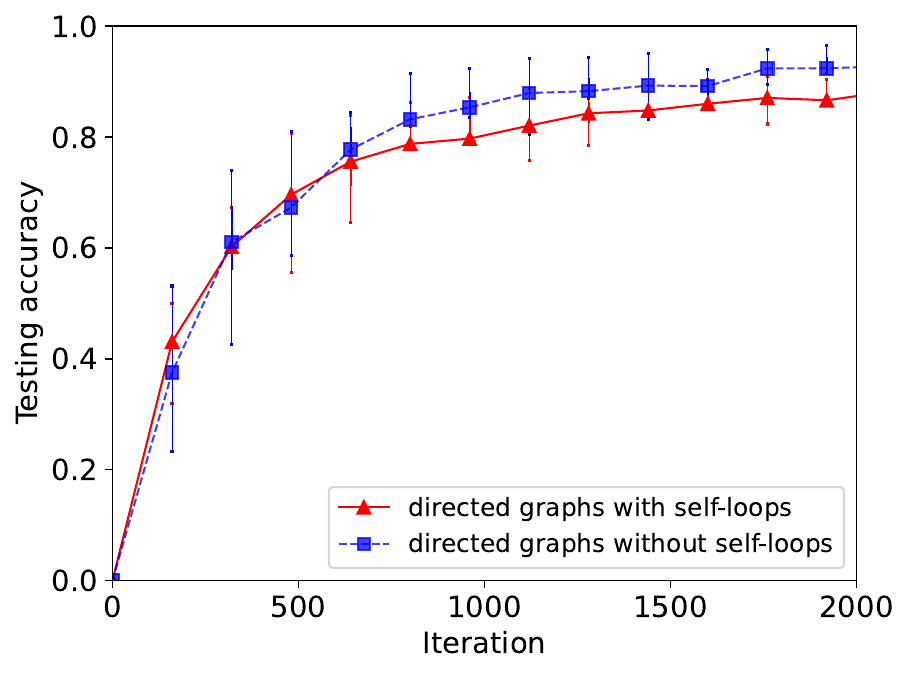}\label{fig7b}}
	\subfloat[Training accuracy on the ``CIFAR-10'' dataset]{\includegraphics[width=0.16\textwidth]{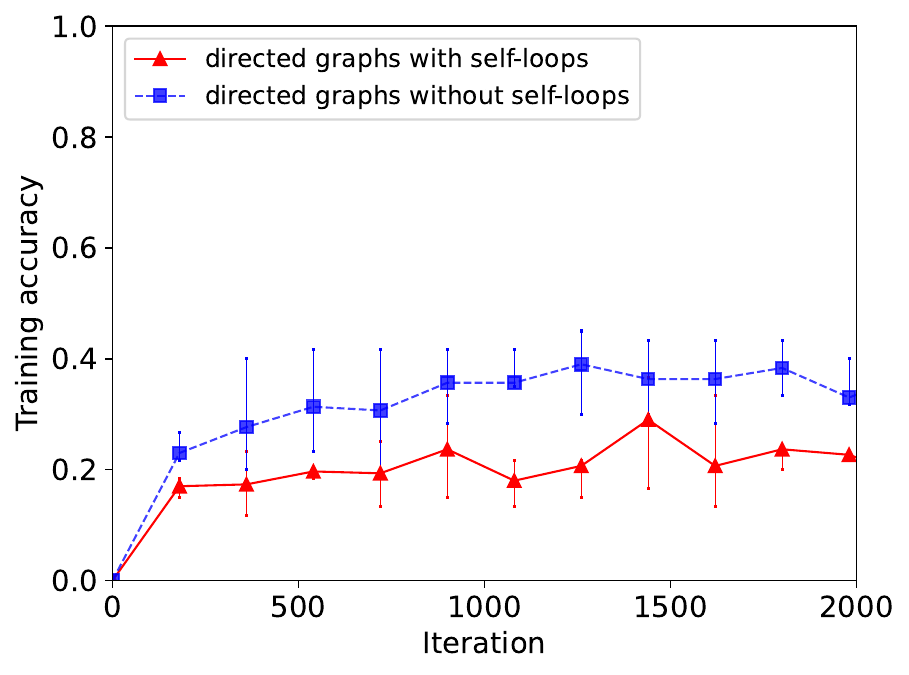}\label{fig7c}}\\\vspace{-1em}
	\subfloat[Testing accuracy on the ``CIFAR-10'' dataset]{\includegraphics[width=0.16\textwidth]{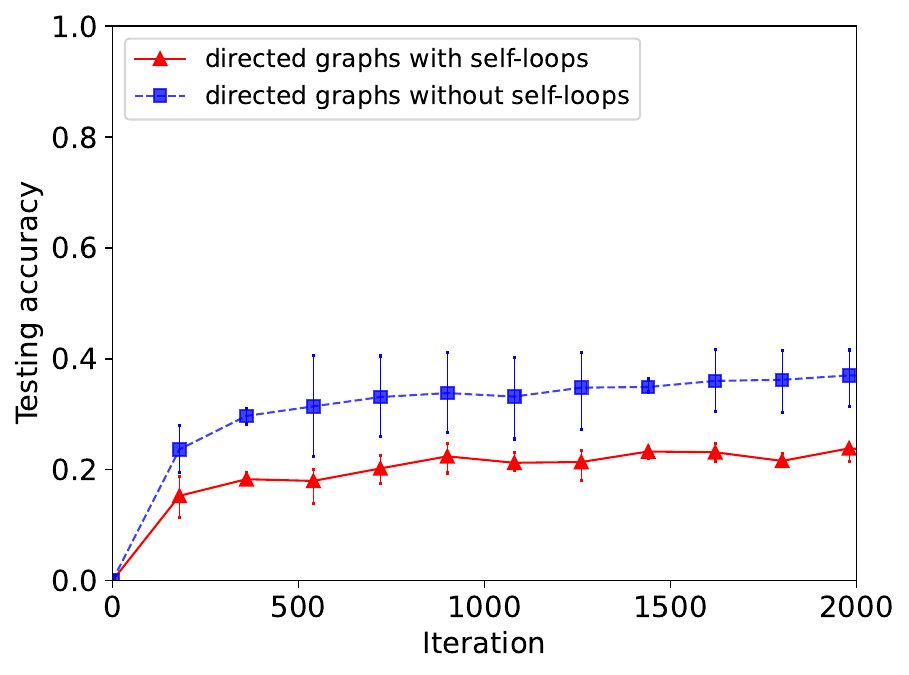}\label{fig7d}}\hspace{0.4em}
	\subfloat[Cumulative \mbox{differential privacy budget $\varepsilon$}]{\includegraphics[width=0.16\textwidth]{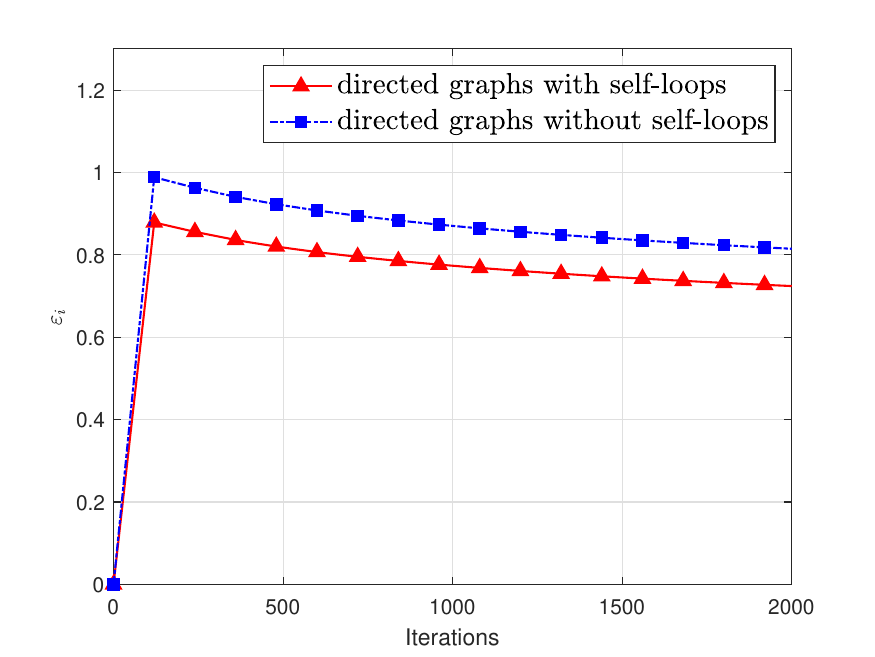}\label{fig7e}}\vspace{-0.5em}
	\caption{\small Accuracy and cumulative differential privacy budget $\varepsilon$ of Algorithm \ref{algorithm1} with \emph{Scheme (S1)} over directed graphs with and without self-loops}
	\label{fig7}
\end{figure}

\begin{figure}[!h]
	\centering\vspace{-2em}
	\subfloat[Training accuracy on the ``MNIST'' dataset]{\includegraphics[width=0.16\textwidth]{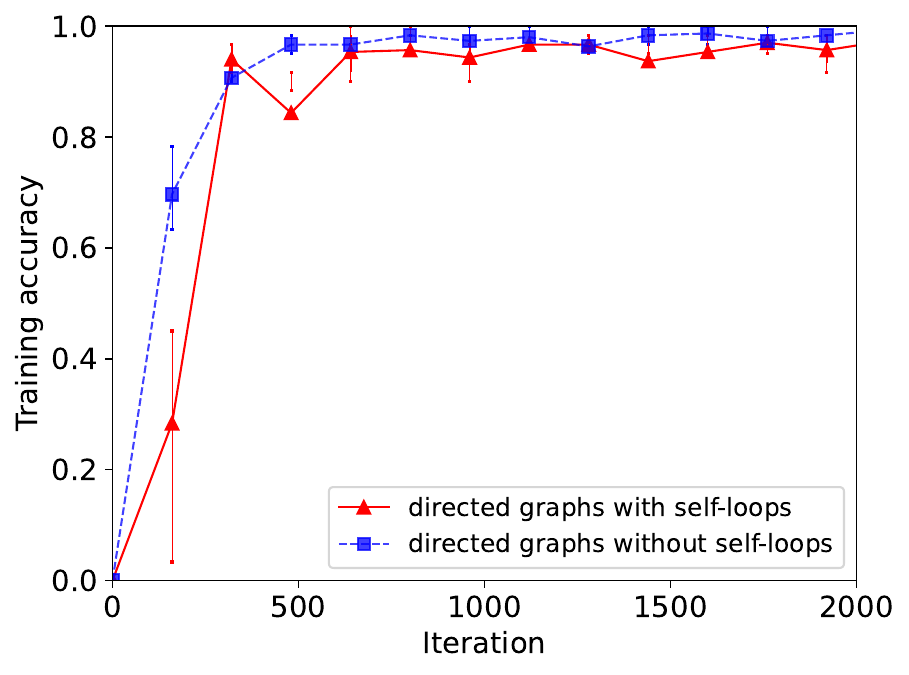}\label{fig8a}}
	\subfloat[Testing accuracy on the ``MNIST'' dataset]{\includegraphics[width=0.16\textwidth]{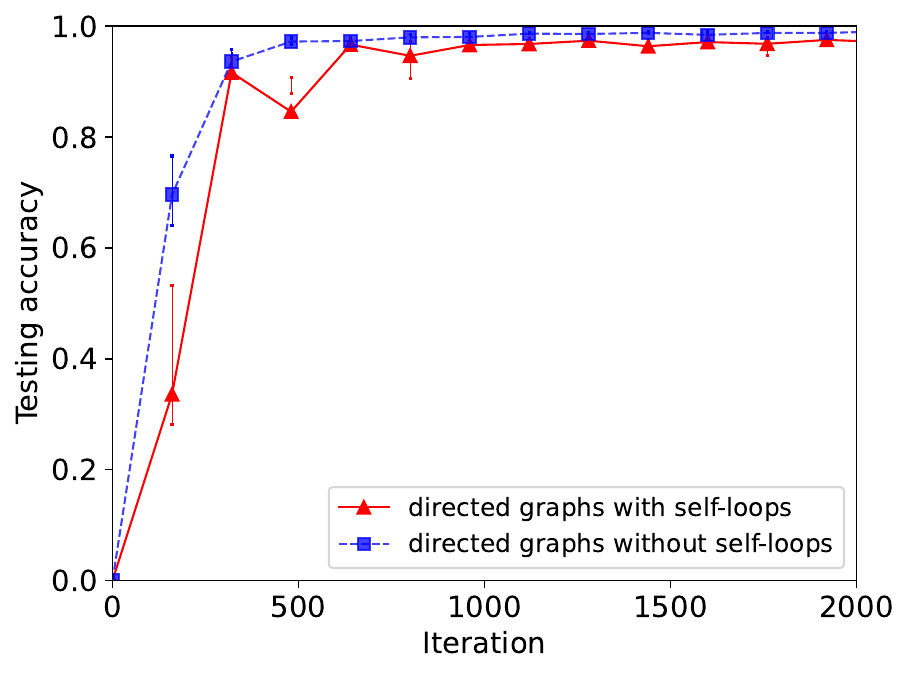}\label{fig8b}}
	\subfloat[Training accuracy on the ``CIFAR-10'' dataset]{\includegraphics[width=0.16\textwidth]{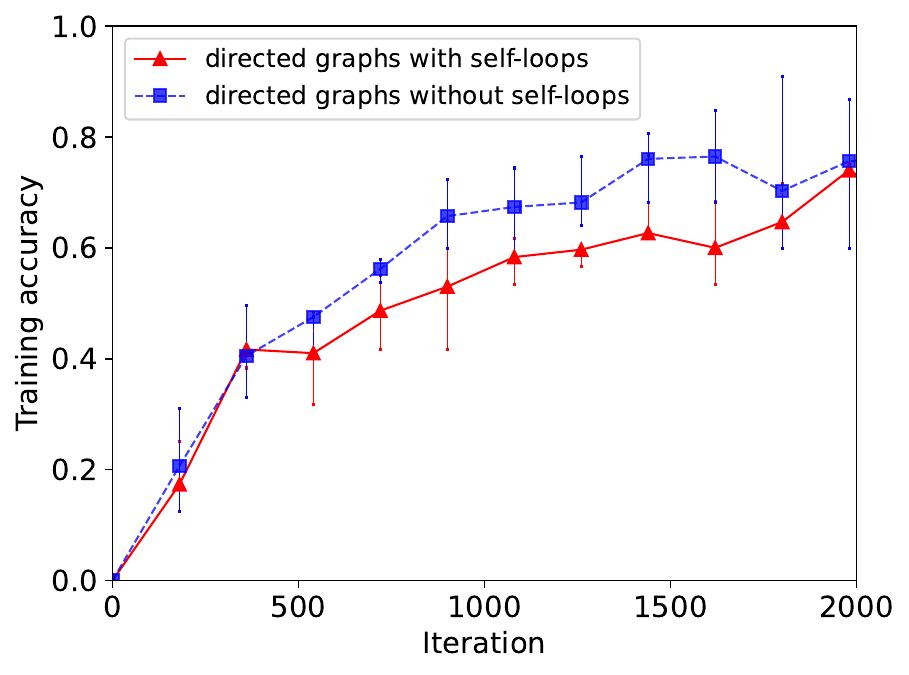}\label{fig8c}}\\\vspace{-1em}
	\subfloat[Testing accuracy on the ``CIFAR-10'' dataset]{\includegraphics[width=0.16\textwidth]{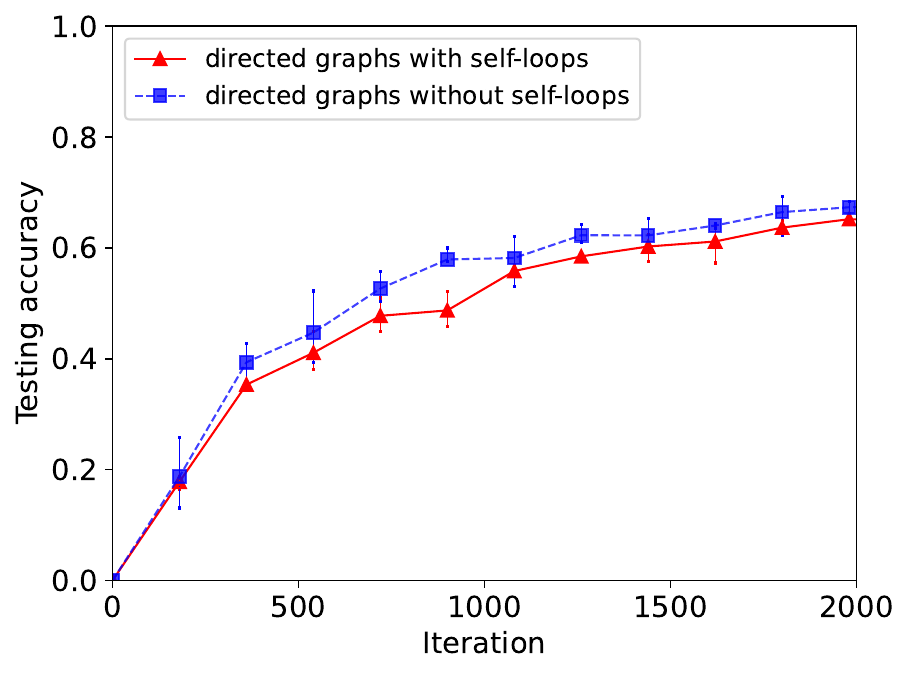}\label{fig8d}}\hspace{0.4em}
	\subfloat[Cumulative \mbox{differential privacy budget $\varepsilon$}]{\includegraphics[width=0.16\textwidth]{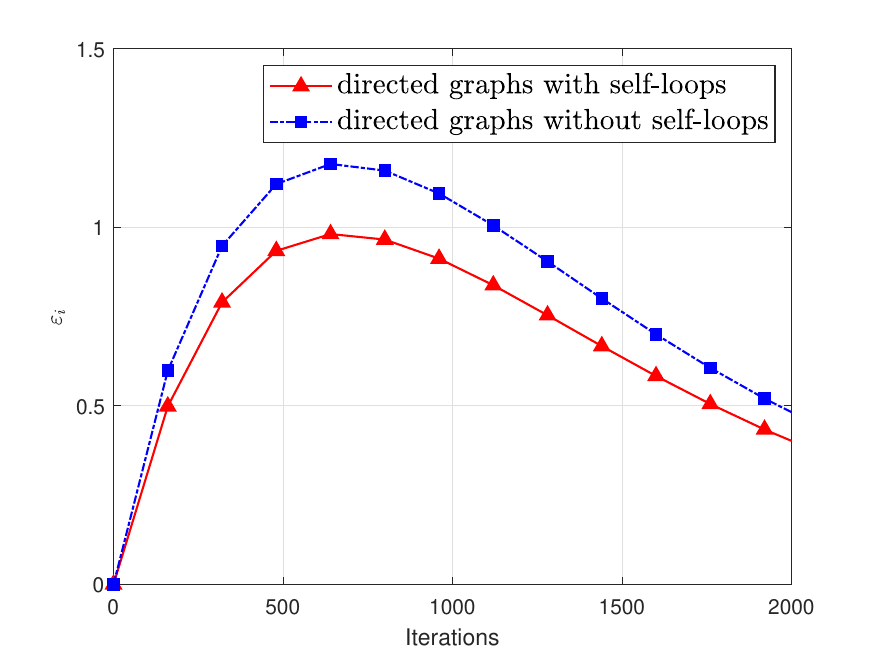}\label{fig8e}}\vspace{-0.5em}
	\caption{\small Accuracy and cumulative differential privacy budget $\varepsilon$ of Algorithm \ref{algorithm1} with \emph{Scheme (S2)} over directed graphs with and without self-loops}
	\label{fig8}
	\vspace{-2em}
\end{figure}

\vspace{-0.2em}
In this subsection, the comparison of Algorithm \ref{algorithm1} with \emph{Schemes (S1)}, \emph{(S2)} between the convergence rate and the differential privacy level is given. Let $p_{\zeta_i}=p_{\eta_i}=0.09+0.01i$ in \emph{Scheme (S1)}, and $p_{\zeta_i}=p_{\eta_i}=0.99959+i\cdot 10^{-5}$ in \emph{Scheme (S2)} for $i=1,\dots,5$, respectively. Then, from Fig. \ref{fig4}\subref{fig4a}-\ref{fig4}\subref{fig4d} one can see that Algorithm \ref{algorithm1} with \emph{Scheme (S2)} converges faster than Algorithm \ref{algorithm1} with \emph{Scheme (S1)}, while from Fig. \ref{fig4}\subref{fig4e} one can see that the cumulative differential privacy budget $\varepsilon$ of Algorithm \ref{algorithm1} with \emph{Scheme (S1)} is smaller than the cumulative differential privacy budget $\varepsilon$ of Algorithm~\ref{algorithm1} with \emph{Scheme (S1)}.
\begin{rmk}
		By \cite[Cor. 8.1.19]{horn2012matrix}, the spectral radius $\rho(\mathcal{R}), \rho(\mathcal{C})$ of directed graphs with self-loops are larger than those of directed graphs without self-loops. Then, by \eqref{54c} and \eqref{57}, both \emph{Schemes (S1)} and \emph{(S2)} converges slower over directed graphs with self-loops than over directed graphs without self-loops. This result is consistent with Figs. \ref{fig7}\subref{fig7a}-\ref{fig7}\subref{fig7d} and Figs. \ref{fig8}\subref{fig8a}-\ref{fig8}\subref{fig8d}. Note that $\sum_{j\in\mathcal{N}_{\mathcal{R},i}^{-}}\hspace{-0.5em}\mathcal{R}_{ij}$ and $\sum_{j\in\mathcal{N}_{\mathcal{C},i}^{+}}\hspace{-0.5em}\mathcal{C}_{ji}$ of directed graphs with self-loops are larger than those of directed graphs without self-loops. Then, by \eqref{73}, \eqref{74}, and Lemma \ref{lemma5}, the cumulative differential privacy budget $\varepsilon$ of both \emph{Schemes (S1)} and \emph{(S2)} over the directed graph with self-loops is smaller than the one over the directed graph without self-loops. This result is consistent with Figs. \ref{fig7}\subref{fig7e} and \ref{fig8}\subref{fig8e}. As a result, self-loops in directed graphs slow down the convergence rate of Algorithm~\ref{algorithm1} and enhance the differential privacy level of Algorithm \ref{algorithm1}.
\end{rmk}

\vspace{3em}
\subsection{Comparison with methods in \cite{kang2021weighted,wang2023quantization,yan2024killing,wang2024differentiallya,chen2024differentially}}
\begin{figure}[!htbp]
	{\vskip -16pt}
	\centering
	\subfloat[Training accuracy on \\the ``MNIST'' dataset]{\includegraphics[width=0.25\textwidth]{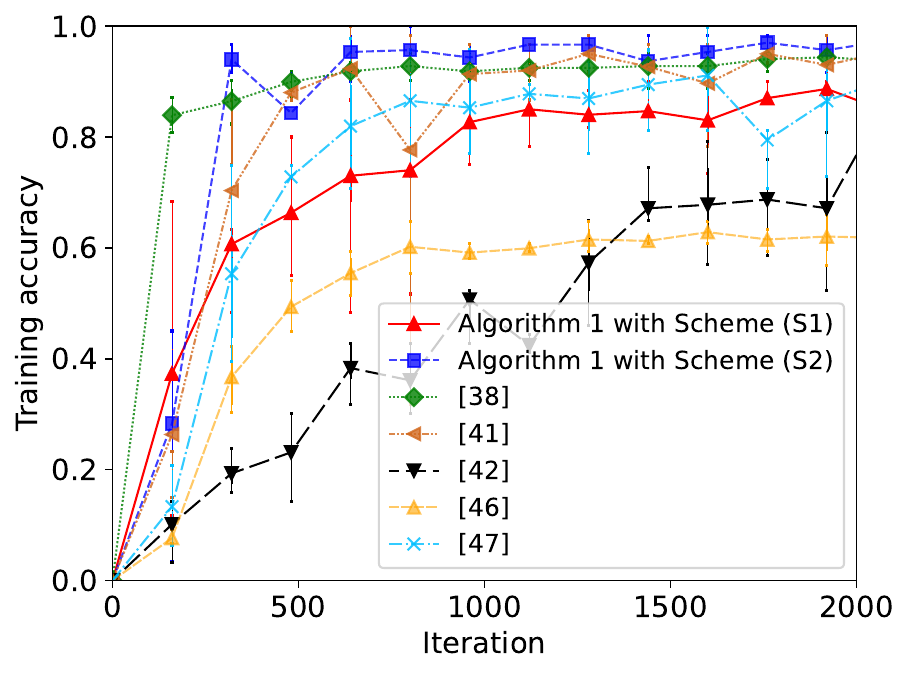}\label{fig5a}}
	\subfloat[Testing accuracy on \\the ``MNIST'' dataset]{\includegraphics[width=0.25\textwidth]{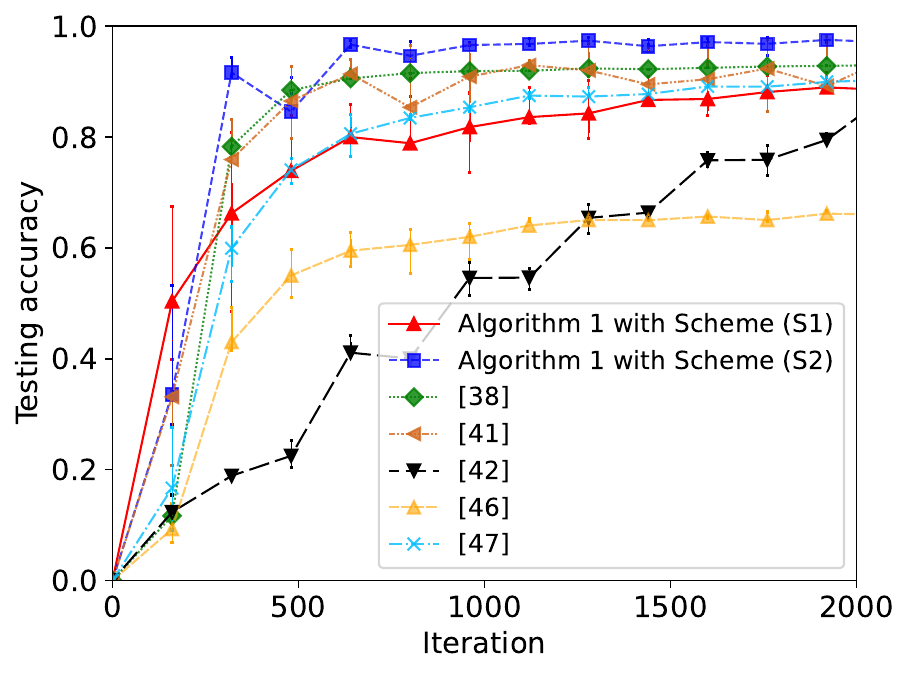}\label{fig5b}}\\
	{\vskip -12pt}
	\subfloat[Training accuracy on \\the ``CIFAR-10'' dataset]{\includegraphics[width=0.25\textwidth]{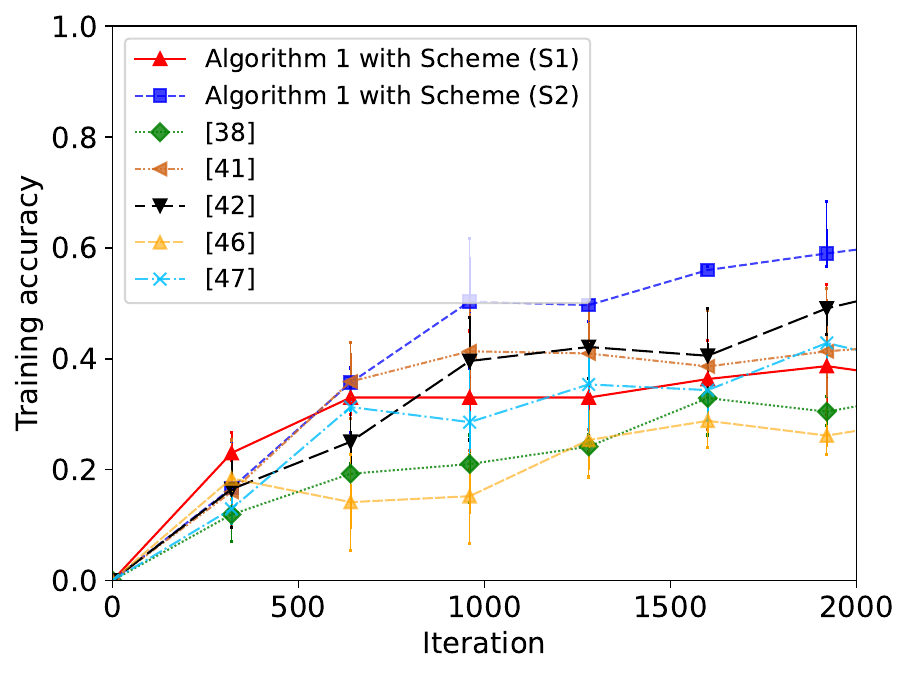}\label{fig5c}}
	\subfloat[Testing accuracy on \\the ``CIFAR-10'' dataset]{\includegraphics[width=0.25\textwidth]{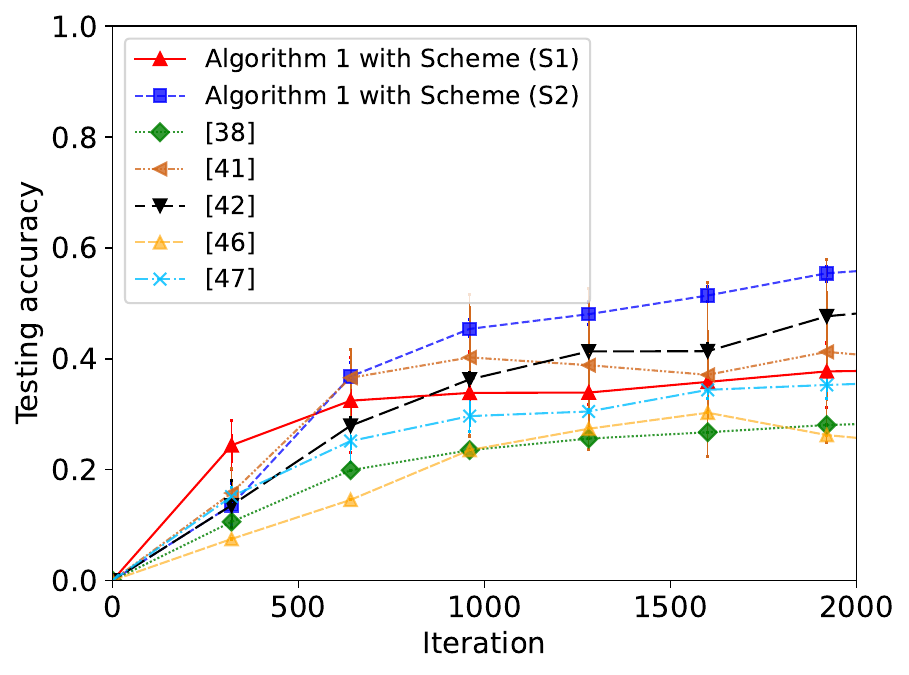}\label{fig5d}}
	\caption{\small Comparison of accuracy on the benchmark datasets ``MNIST'' and ``CIFAR-10''}
	\label{fig5}
	{\vskip -10pt}
\end{figure}

\vspace{-0.2em}
Let $p_{\zeta_i}=p_{\eta_i}=0.99959+i\cdot 10^{-5}$ in \emph{Scheme (S2)} for $i=1,\dots,5$, and iterations step-sizes $\alpha_K, \beta_K, \gamma_K$, the sampling number $m_K$, and privacy noise parameters $\sigma_k^{(\zeta_i)}, \sigma_k^{(\eta_i)}$ in \emph{Scheme (S1)} and \cite{kang2021weighted,wang2023quantization,yan2024killing,wang2024differentiallya,chen2024differentially} be the same as \emph{Scheme (S2)} to ensure a fair comparison. Then, the comparison of the convergence rate and the differential privacy level between Algorithm \ref{algorithm1} and the methods in \cite{kang2021weighted,wang2023quantization,yan2024killing,wang2024differentiallya,chen2024differentially} is given in Fig. \ref{fig5}. From Fig. \ref{fig5}, one can see that Algorithm \ref{algorithm1} with \emph{Scheme (S2)} converges faster than methods in \cite{kang2021weighted,wang2023quantization,yan2024killing,wang2024differentiallya,chen2024differentially}.

\begin{figure}[!htbp]
	{\vskip -15pt}
	\centering
	\subfloat[$\varepsilon$ under same $\sigma_k^{(\zeta_i)},\sigma_k^{(\eta_i)}$]{\includegraphics[width=0.25\textwidth]{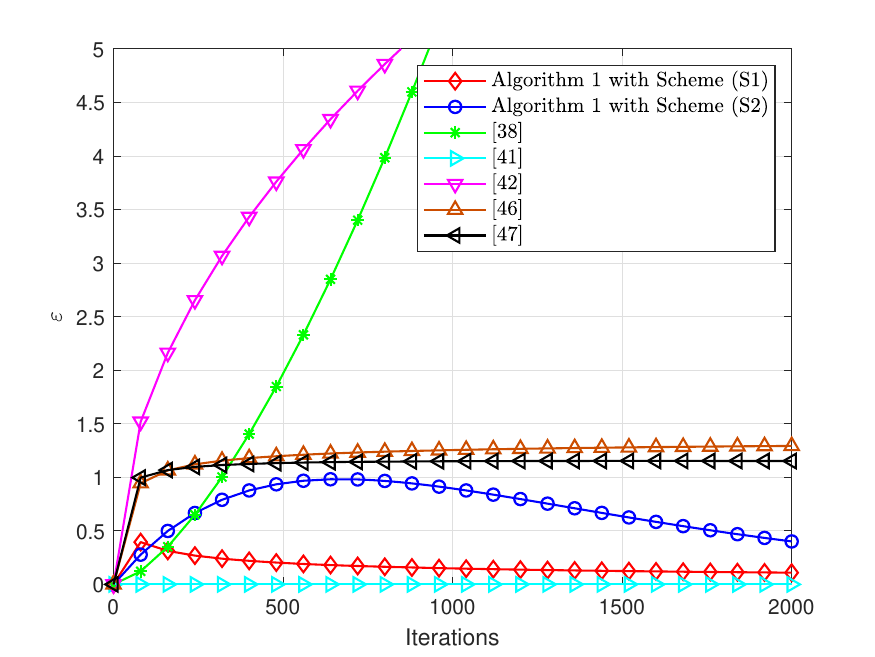}\label{fig6a}}
	\subfloat[$\delta$ under same $\sigma_k^{(\zeta_i)},\sigma_k^{(\eta_i)}$]{\includegraphics[width=0.25\textwidth]{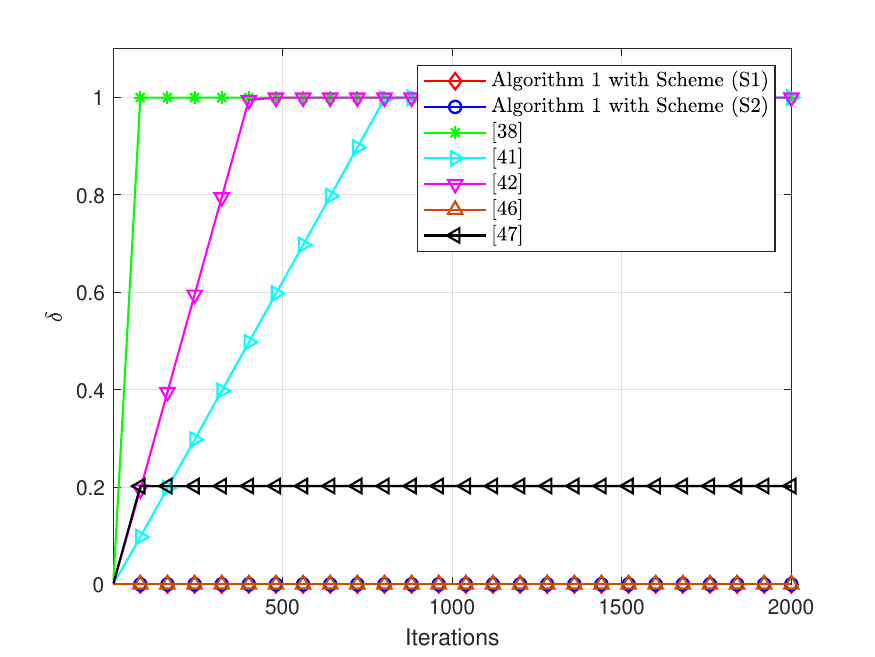}\label{fig6b}}
	\caption{\small Comparison of cumulative differential privacy \\budgets $\varepsilon$ and $\delta$}
	\label{fig6}
	{\vskip -15pt}
\end{figure}

A comparison of the differential privacy level between Algorithm \ref{algorithm1} and the methods in \cite{kang2021weighted,wang2023quantization,yan2024killing,wang2024differentiallya,chen2024differentially} is given in Fig. \ref{fig6}. By Fig. \ref{fig6}\subref{fig6a}, the cumulative differential privacy budget $\varepsilon$ of Algorithm \ref{algorithm1} with both \emph{Schemes (S1)} and \emph{(S2)} is smaller than the ones in \cite{kang2021weighted,yan2024killing,wang2024differentiallya,chen2024differentially}. By Fig. \ref{fig6}\subref{fig6b}, \cite{wang2023quantization} achieves the cumulative differential privacy budget $\delta = 1$ after 800 iterations, and thus, the one therein cannot protect sampled gradients after 800 iterations. Thus, Algorithm \ref{algorithm1} with both \emph{Schemes (S1)} and \emph{(S2)} provides a higher differential privacy level than methods in \cite{kang2021weighted,wang2023quantization,yan2024killing,wang2024differentiallya,chen2024differentially}.
\section{Conclusion}\label{section 5}
In this paper, we have proposed a new differentially private gradient-tracking-based distributed stochastic optimization algorithm over directed graphs. Two novel schemes of step-sizes and the sampling number are given: \emph{Scheme (S1)} uses polynomially decreasing step-sizes and the increasing sampling number with the maximum iteration number. \emph{Scheme (S2)} uses constant step-sizes and the exponentially increasing sampling number with the maximum iteration number. By using the sampling parameter-controlled subsampling method, both schemes achieve the finite cumulative privacy budget even over infinite iterations, and thus, enhance the differential privacy level compared to the existing ones. By using the gradient-tracking method, the almost sure and mean square convergence of the algorithm is shown for nonconvex objectives over directed graphs with spanning trees.  Further, when  nonconvex objectives satisfy the Polyak-{\L}ojasiewicz condition, the polynomial mean square convergence rate (\emph{Scheme (S1)}) and the exponential mean square convergence rate (\emph{Scheme (S2)}) are given, respectively. Furthermore, the oracle complexity of the algorithm, the trade-off between the privacy and the convergence are shown. Finally, numerical examples of distributed training on the benchmark datasets ``MNIST'' and ``CIFAR-10'' are given to show the effectiveness of the algorithm.
\appendices
\section{Useful lemmas}\label{appendix a}
\setcounter{lemma}{0}
\renewcommand{\thelemma}{A.\arabic{lemma}}
\begin{lemma}\label{lemma a1}
	\!(\!\!\!\cite[Lemma A.1]{chen2024differentially}) If a function $h:\mathbb{R}^d\rightarrow \mathbb{R}$ has a global minimum $h(x^*)$ and satisfies $\|\nabla h(x)-\nabla h(y)\|\leq L_1\|x-y\|,\forall x,y\in\mathbb{R}^d$, then following statements holds:
	
	\noindent(i) $h(y)\leq h(x)\!+\!\langle \nabla h(x),y\!-\!x \rangle\!+\!\frac{L_1}{2}\|y\!-\!x\|^2$, $\forall x,y\in \mathbb{R}^d$.
	
	\noindent(ii) $\|\nabla h(x)\|^2\leq 2L_1\left(h(x)-h(x^*)\right)$, $\forall x\in\mathbb{R}^d$.
\end{lemma}
\begin{lemma}\label{lemma a2}
	\!(\!\!\!\cite[Cor. 8.1.29, Th. 8.4.4]{horn2012matrix}) For any $n=1,2,\dots$, let $A\in\mathbb{R}^{n\times n}$ be a nonnegative matrix and $x\in\mathbb{R}^n$ be a positive vector. Then, following statements hold: 

	\noindent(i) If there exists $\rho>0$ such that $Ax\leq\rho x$, then $\rho(A)\leq\rho$.
	
	\noindent(ii) If $A$ is irreducible, then $\rho(A)>0$ and there exists a positive vector $y=[y_1,\dots,y_n]^\top\in\mathbb{R}^n$ such that $y^\top A=\rho(A) y^\top$.
\end{lemma}

\section{Proof of Lemma \ref{lemma1}}\label{appendix b}
The proof of Lemma \ref{lemma1} is given in the following three steps.

{\bf Step 1.} First, we prove Lemma \ref{lemma1}(i). Let $\mathcal{G}_{-\mathcal{L}_1}$ be the directed graph induced by the matrix $-\mathcal{L}_1$. Then, by Assumption~\ref{asm1}, $\mathcal{G}_{-\mathcal{L}_1}$ has the same spanning trees as $\mathcal{G}_{\mathcal{R}}$. By \cite[Lemma 3.3]{ren2005consensus}, we have $\varpi_1^{(1)}=0$ and $\text{Re}(\varpi_l^{(1)})>0$ for any $l=2,\dots,n$. Similarly,  we have $\varpi_1^{(2)}=0$ and $\text{Re}(\varpi_l^{(2)})>0$ for any $l=2,\dots,n$.

{\bf Step 2.} In this step, we prove that there exist unique nonnegative vectors $v_1,v_2$$\in$$\mathbb{R}^n$ such that $v_1^\top(I_n-\alpha_K\mathcal{L}_1)$$=$$v_1^\top$, $(I_n-\beta_K\mathcal{L}_2)v_2$$=$$v_2$, $v_1^\top\mathbf{1}_n$$=$$n$, $v_2^\top\mathbf{1}_n$$=$$n$, $v_1^\top v_2$$>$$0$. Since $0$$<$$\alpha_K$$<$$\min_{i\in\mathcal{V}}\{\!\frac{1}{\sum_{j\in\mathcal{N}_{\mathcal{R},i}^{-}}\hspace{-0.65em}\mathcal{R}_{ij}}\!\}$ and $0$$<$$\beta_K$$<$$\min_{i\in\mathcal{V}}\{\!\frac{1}{\sum_{j\in\mathcal{N}_{\mathcal{C},i}^{+}}\hspace{-0.65em}\mathcal{C}_{ji}}\!\}$, matrices $I_n-\alpha_K\mathcal{L}_1$ and $ I_n-\beta_K\mathcal{L}_2^\top$ are nonnegative. Let $\mathcal{G}_{I_n-\alpha_K\mathcal{L}_1},\mathcal{G}_{I_n-\beta_K\mathcal{L}_2^\top}$ be directed graphs induced by matrices $I_n-\alpha_K\mathcal{L}_1$ and $I_n-\beta_K\mathcal{L}_2^\top$, respectively. Then, by Assumption \ref{asm1}, $\mathcal{G}_{I_n-\alpha_K\mathcal{L}_1}$ has the same spanning trees as $\mathcal{G}_{\mathcal{R}}$, and $\mathcal{G}_{I_n-\beta_K\mathcal{L}_2^\top}$ has the same spanning trees as $\mathcal{G}_{\mathcal{C}^\top}$. 

\indent Note that $(I_n-\alpha_K\mathcal{L}_1)\mathbf{1}_n=\mathbf{1}_n,(I_n-\beta_K\mathcal{L}_2^\top)\mathbf{1}_n=\mathbf{1}_n$. Then, by \cite[Lemma~1]{pu2020push}, there exist unique nonnegative vectors $v_1,v_2\in\mathbb{R}^n$ such that $v_1^\top(I_n-\alpha_K\mathcal{L}_1)=v_1^\top$, $v_2^\top(I_n-\beta_K\mathcal{L}_2^\top)=v_2^\top$, $v_1^\top\mathbf{1}_n=n$, $v_2^\top\mathbf{1}_n=n$. Thus, by $v_2^\top(I_n-\beta_K\mathcal{L}_2^\top)=v_2^\top$, we have $(I_n-\beta_K\mathcal{L}_2)v_2=v_2$.

By Assumption \ref{asm1}, $\mathcal{G}_{\mathcal{R}}$ and $\mathcal{G}_{\mathcal{C}^\top}$ contain at least one spanning tree. Then $\mathcal{G}_{I_n-\alpha_K\mathcal{L}_1}$ and $\mathcal{G}_{I_n-\beta_K\mathcal{L}_2^\top}$ contain at least one spanning tree. Thus, by \cite[Lemma~1]{pu2020push}, we have $v_1^\top v_2>0$.

{\bf Step 3.} In this step, we prove that there exist $r_1, r_2>0$ such that $\rho(W_1-\alpha_K\mathcal{L}_1)\leq 1-r_1\alpha_K$, $\rho(W_2-\beta_K\mathcal{L}_2)\leq 1-r_2\beta_K$. By {\bf Step 1}, the eigenvalues of the matrix $I_n-\alpha_K\mathcal{L}_1$ are $\{1,1-\alpha_K\varpi_2^{(1)},\dots,1-\alpha_K\varpi_n^{(1)}\}$. Note that the matrix $I_n-\alpha_K\mathcal{L}_1$ is nonnegative and row-stochastic. Then, by \cite[Cor. 8.1.29]{horn2012matrix}, $\rho(I_n-\alpha_K\mathcal{L}_1)=1$ is the algebraically simple eigenvalue, and $|1-\alpha_K\varpi_l^{(1)}|<1$ holds for any $l=2,\dots,n$.

Let $r_1$$=$$\min_{l=2,\dots,n}\{\frac{(2+|\varpi_l^{(1)}|^2)\text{Re}(\varpi_l^{(1)})}{2+2|\varpi_l^{(1)}|^2}\}$. Then, for any $l=2,\dots,n$, by $0$$<$$\alpha_K$$<$ $\min_{l=2,\dots,n}\{\!\frac{\text{Re}(\varpi_l^{(1)})}{1+|\varpi_l^{(1)}|^2}\!\}$ and Bernoulli's inequality (\!\!\cite[Ex. 5.4.7]{zorich2015analysis}), we have
\begin{align*}
	|1-\alpha_K\varpi_l^{(1)}|=&\sqrt{1-2\alpha_K\text{Re}(\varpi_l^{(1)})+\alpha_K^2|\varpi_l^{(1)}|^2}\cr
	\noalign{\vskip -3pt}
	\leq&1-\alpha_K\text{Re}(\varpi_l^{(1)})+\frac{\alpha_K^2|\varpi_l^{(1)}|^2}{2}
\end{align*}
	\begin{align}\label{b2,1}
		\leq&1-\alpha_K(\text{Re}(\varpi_l^{(1)})-\frac{|\varpi_l^{(1)}|^2}{2}\min_{l=2,\dots,n}\{\frac{\text{Re}(\varpi_l^{(1)})}{1+|\varpi_l^{(1)}|^2}\})\cr
		\noalign{\vskip -3pt}
		\leq&1-\alpha_K(\text{Re}(\varpi_l^{(1)})-\frac{\text{Re}(\varpi_l^{(1)})|\varpi_l^{(1)}|^2}{2+2|\varpi_l^{(1)}|^2})\cr
		\noalign{\vskip -6pt}
		\leq&1-r_1\alpha_K.
\end{align}

\noindent Note that $0,1$ are the eigenvalues of the matrix $\frac{1}{n}\mathbf{1}_nv_1^\top$ and 1 is the algebraically simple eigenvalue. Then, by $(W_1-\alpha_K\mathcal{L}_1)\mathbf{1}_n=0$, the eigenvalues of $W_1-\alpha_K\mathcal{L}_1$ are $\{0,1-\alpha_K\varpi_2^{(1)},\dots,1-\alpha_K\varpi_n^{(1)}\}$. Thus, by \eqref{b2,1} we have
	\begin{align*}
		\hspace{-0.3em}\rho(W_1\!\!-\!\alpha_K\mathcal{L}_1)=&\max\{|1\!-\!\alpha_K\varpi_2^{(1)}|,\dots,|1\!-\!\alpha_K\varpi_n^{(1)}|\}\cr
		\leq&1\!-\!r_1\alpha_K.
	\end{align*}
	\noindent Similarly, let $r_2=\min_{l=2,\dots,n}\{\frac{(2+|\varpi_l^{(2)}|^2)\text{Re}(\varpi_l^{(2)})}{2+2|\varpi_l^{(2)}|^2}\}$. Then, we have $\rho(W_2-\beta_K\mathcal{L}_2)\leq 1-r_2\beta_K$. Therefore, this lemma is proved. $\hfill\blacksquare$

\section{Proof of Lemma \ref{lemma a10}}\label{appendix c0}
The following four steps are given to prove Lemma \ref{lemma a10}.

{\bf Step 1:} First, we prove the following inequality holds for any $k=0,\dots,K$, $K=0,1,\dots$:
\begin{align}\label{a6}
	&\E\|(W_1\otimes I_d) x_{k+1}\|^2\cr
	\leq& A_K^{(11)}\E\|(W_{\!1}\!\!\otimes\!\! I_d)x_k\|^2+ A_K^{(12)}\E\|(W_2\otimes I_d)y_k\|^2\notag\\
	&+\frac{A_K^{(13)}}{2L_1}\E\|\nabla F(\bar{x}_k)\|^2+u_k^{(1)}.
\end{align}
By Assumption \ref{asm1}, Lemma \ref{lemma1} holds. Note that by Lemma \ref{lemma1}(ii), $\mathcal{L}_1 W_1=W_1\mathcal{L}_1=\mathcal{L}_1$. Then, multiplying $W_1\otimes I_d$ on both sides of \eqref{eq7} implies\vspace{-0.6em}
\begin{align}\label{2}
	&(W_1\otimes I_d) x_{k+1}\cr
	\noalign{\vskip -4pt}
	=&\left((I_n-\alpha_K\mathcal{L}_1)\otimes I_d\right)\!(W_1\otimes I_d)x_k+\alpha_K(W_1\mathcal{R}\otimes I_d)\zeta_k\cr
	\noalign{\vskip -4pt}
	&-\gamma_K(W_1\otimes I_d)y_k,\cr
	\noalign{\vskip -4pt}
	=&\left((I_n-\alpha_K\mathcal{L}_1)\otimes I_d\right)\!(W_1\otimes I_d)x_k+\alpha_K(W_1\mathcal{R}\otimes I_d)\zeta_k\cr
	\noalign{\vskip -4pt}
	&-\gamma_K(W_1W_2\otimes I_d)y_k-\frac{\gamma_K}{n}(W_1v_2\mathbf{1}_n^\top\otimes I_d)y_k.
\end{align}
Let $\bar{y}_k$$=$$\frac{1}{n}(\mathbf{1}_n^\top\otimes I_d)y_k$. Then, by \eqref{2}, taking the mathematical expectation of $\|(W_1\otimes I_d) x_{k+1}\|^2$ implies
\begin{align}\label{3}
	&\E\|(W_1\otimes I_d) x_{k+1}\|^2\cr
	=&\E\|\left((I_n-\alpha_K\mathcal{L}_1)\otimes I_d\right) (W_1\otimes I_d)x_k+\alpha_K(\mathcal{R}\otimes I_d)\zeta_k\cr
	&-\gamma_K(W_1W_2\otimes I_d)y_k-\gamma_K(W_1v_2\otimes I_d)\bar{y}_k\|^2.
\end{align}
For any $k=0,\dots,K$, let $\mathcal{F}_k=\sigma(\{x_k,y_k\})$. Then, since $\zeta_k$ is independent of $\mathcal{F}_k$ and has the Laplacian distribution $\text{Lap}(\sigma_k^{(\zeta_i)})$, we have
\begin{align}\label{2d}
	&\E(\zeta_k|\mathcal{F}_k)=\E\zeta_k=0,\cr
	&\E(\|\zeta_k\|^2|\mathcal{F}_k)=\E\|\zeta_k\|^2\leq2nd\max_{i\in\mathcal{V}}\{(\sigma_k^{(\zeta_i)})^2\}.
\end{align}
Then, substituting \eqref{2d} into \eqref{3} implies
\begin{align}\label{4}
	&\E\|(W_1\otimes I_d) x_{k+1}\|^2\cr
	=&\E(\E(\|((I_n\!-\!\alpha_K\mathcal{L}_1)\!\otimes\! I_d)(W_1\!\otimes\! I_d)x_k\!-\!\gamma_K(W_1W_2\!\otimes\! I_d)y_k\cr
	&-(W_1v_2\otimes I_d)\bar{y}_k+\alpha_K(\mathcal{R}\otimes I_d)\zeta_k\|^2|\mathcal{F}_k))\cr
	=&\E(\E(\|((I_n\!-\!\alpha_K\mathcal{L}_1)\!\otimes\! I_d)(W_1\!\otimes\! I_d)x_k\!-\!\gamma_K(W_1W_2\!\otimes\! I_d)y_k\cr
	&-(W_1v_2\otimes I_d)\bar{y}_k\|^2|\mathcal{F}_k)+\E(\|\alpha_K(\mathcal{R}\otimes I_d)\zeta_k\|^2|\mathcal{F}_k))\cr
	=&\E\|((I_n\!-\!\alpha_K\mathcal{L}_1)\!\otimes\! I_d)(W_1\!\otimes\! I_d)x_k\!-\!\gamma_K(W_1W_2\!\otimes\! I_d)y_k\cr
	&-(W_1v_2\otimes I_d)\bar{y}_k\|^2+\E\|\alpha_K(\mathcal{R}\otimes I_d)\zeta_k\|^2\cr
	\leq&\E\left(\|((I_n\!-\!\alpha_K\mathcal{L}_1)\!\otimes\! I_d)(W_1\otimes I_d)x_k\!-\!\gamma_K(W_1W_2\!\otimes\! I_d)y_k\right.\cr
	\noalign{\vskip -3pt}
	&\left.-\gamma_K(W_1v_2\!\otimes\! I_d)\bar{y}_k\|^2\right)\!+\!2nd\rho(\mathcal{R})^2\alpha_K^2\max_{i\in\mathcal{V}}\{(\sigma_k^{(\zeta_i)})^2\}.
\end{align}
Note that for any $\mathbf{a},\mathbf{b}\in\mathbb{R}^{d},r>0$, the following Cauchy-Schwarz inequality (\!\!\cite[Ex. 4(b)]{zorich2015analysis}) holds:
\begin{align}\label{5}
	\smash{\|\mathbf{a}+\mathbf{b}\|^2\leq(1+r)\|\mathbf{a}\|^2+\left(1+\frac{1}{r}\right)\|\mathbf{b}\|^2.}
\end{align} 
Then, by Lemma \ref{lemma1}(ii), setting $r$$=$$r_1\alpha_K$ in \eqref{5} and substituting \eqref{5} into \mbox{\eqref{4} imply}\vspace{-0.5em}
\begin{align}\label{6}
	&\E\|(W_1\otimes I_d) x_{k+1}\|^2\cr
	\leq&(1\!+\!r_1\alpha_K)\E\|\left((I_n-\alpha_K\mathcal{L}_1)\otimes I_d\right) (W_1\otimes I_d)x_k\|^2\cr
	&+\left(1\!+\!\frac{1}{r_1\alpha_K}\right)\E\|\gamma_K(W_{\!1}W_{\!2}\!\otimes\! I_d)y_k\!+\!\gamma_K(W_{\!1}\!v_{2}\!\otimes\! I_d)\bar{y}_k\|^2\cr
	&+2nd\rho(\mathcal{R})^2\alpha_K^2\max_{i\in\mathcal{V}}\{(\sigma_k^{(\zeta_i)})^2\}.
\end{align}
{\vskip -5pt}\noindent Since $v_1^\top \mathbf{1}_n=n$ holds by Lemma \ref{lemma1}(ii), $W_1^2=W_1$. Then, $((I_n\!-\!\alpha_K\mathcal{L}_1)\!\otimes\! I_d)(W_1\!\otimes\! I_d)x_k$$=$$((W_1\!-\!\alpha_K \mathcal{L}_1)\!\otimes\! I_d)(W_1\!\otimes\! I_d)x_k$. Thus, by $\rho(W_1\!-\!\alpha_K \mathcal{L}_1)\leq 1-r_1\alpha_K$ in Lemma \ref{lemma1}(ii), we have\vspace{-0.5em}
\begin{align}\label{7}
	&(1+r_1\alpha_K)\E\|((I_n-\alpha_K \mathcal{L}_1)\otimes I_d) (W_1\otimes I_d)x_k\|^2\cr
	\leq&(1+r_1\alpha_K)(1-r_1\alpha_K)^2\E\|(W_1\otimes I_d)x_k\|^2\cr
	\leq&(1-r_1\alpha_K)\E\|(W_1\otimes I_d)x_k\|^2.
\end{align}
{\vskip -3pt}\noindent Substituting \eqref{7} into \eqref{6} implies\vspace{-0.5em}
\begin{align}\label{8}
	&\E\|(W_1\otimes I_d) x_{k+1}\|^2\cr
	\leq&\smash{(1-r_1\alpha_K)\E\|(W_1\otimes I_d)x_k\|^2+2nd\rho(\mathcal{R})^2\alpha_K^2\max_{i\in\mathcal{V}}\{(\sigma_k^{(\zeta_i)})^2\}}~~~~~~\cr
	&+\!\!\frac{(1\!\!+\!\!r_1\alpha_K)\gamma_K^2}{r_1\alpha_K}\!\E(\|(W_1\!W_2\!\otimes\!I_d)y_k\!+\!(W_1\!v_2\!\otimes\! I_d)\bar{y}_k\|^2).
\end{align}
Since for any $\mathbf{a}_1$, $\mathbf{a}_2$, $\dots$, $\mathbf{a}_m\in\mathbb{R}^{d}$, the following \mbox{inequality holds:}
\begin{align}\label{9}
	\|\sum_{i=1}^{m}\mathbf{a}_i\|^2\leq m\sum_{i=1}^{m}\|\mathbf{a}_i\|^2.
\end{align}
Then, setting $m=2$ in \eqref{9} and substituting \eqref{9} into \eqref{8} imply\vspace{-0.5em}
\begin{align}\label{10}
	&\smash{\E\|(W_1\otimes I_d) x_{k+1}\|^2}\cr
	\leq&\smash{(1-r_1\alpha_K)\E\|(W_1\otimes I_d)x_k\|^2+2nd\rho(\mathcal{R})^2\alpha_K^2\max_{i\in\mathcal{V}}\{(\sigma_k^{(\zeta_i)})^2\}}\cr
	&+\frac{2(1+r_1\alpha_K)\gamma_K^2}{r_1\alpha_K}\E\|(W_1W_2\otimes I_d)y_k\|^2\cr
	&+\frac{2(1+r_1\alpha_K)\gamma_K^2}{r_1\alpha_K}\E\|(W_1v_2\otimes I_d)\bar{y}_k\|^2\cr
	\leq&\smash{(1-r_1\alpha_K)\E\|(W_1\otimes I_d)x_k\|^2+2nd\rho(\mathcal{R})^2\alpha_K^2\max_{i\in\mathcal{V}}\{(\sigma_k^{(\zeta_i)})^2\}}\cr
	&+\frac{2(1+r_1\alpha_K)\rho(W_1)^2\gamma_K^2}{r_1\alpha_K}\E\|(W_2\otimes I_d)y_k\|^2\cr
	&+\frac{2(1+r_1\alpha_K)\rho(W_1)^2\|v_2\|^2\gamma_K^2}{r_1\alpha_K}\E\|\bar{y}_k\|^2.
\end{align}Note that by $W_1=I_n-\frac{1}{n}\mathbf{1}_nv_1^\top$, we have $\rho(W_1)=1$. Then, \eqref{10} can be rewritten as
\begin{align}\label{10.1}
	&\smash{\E\|(W_1\otimes I_d) x_{k+1}\|^2}\cr
	\leq&\smash{(1-r_1\alpha_K)\E\|(W_1\otimes I_d)x_k\|^2+2nd\rho(\mathcal{R})^2\alpha_K^2\max_{i\in\mathcal{V}}\{(\sigma_k^{(\zeta_i)})^2\}}\cr
	&+\frac{2(1+r_1\alpha_K)\gamma_K^2}{r_1\alpha_K}\E\|(W_2\otimes I_d)y_k\|^2\cr
	&+\frac{2(1+r_1\alpha_K)\|v_2\|^2\gamma_K^2}{r_1\alpha_K}\E\|\bar{y}_k\|^2.
\end{align}

Multiplying $\mathbf{1}_n^\top\otimes I_d$ on both sides of \eqref{eq8} and using $y_0=g_0$ result in\vspace{-0.5em}
\begin{align}\label{11}
	\bar{y}_k=&\bar{y}_{k-1}+\frac{1}{n}(\mathbf{1}_n^\top\otimes I_d)(g_k-g_{k-1})+\frac{\beta_K}{n}(\mathbf{1}_n^\top\mathcal{C}\otimes I_d)\eta_{k-1}\notag\\
	\noalign{\vskip -3pt}
	=&\frac{1}{n}(\mathbf{1}_n^\top\otimes I_d)g_0+\sum_{l=0}^{k-1}\frac{1}{n}(\mathbf{1}_n^\top\otimes I_d)(g_{l+1}-g_l)\notag\\
	\noalign{\vskip -5pt}
	&+\frac{\beta_K}{n}\sum_{l=0}^{k-1}(\mathbf{1}_n^\top\mathcal{C}\otimes I_d)\eta_{l}\notag\\
	\noalign{\vskip -5pt}
	=&\frac{1}{n}(\mathbf{1}_n^\top\otimes I_d)g_k+\frac{\beta_K}{n}\sum_{l=0}^{k-1}(\mathbf{1}_n^\top\mathcal{C}\otimes I_d)\eta_{l}.
\end{align}
By Assumption \ref{asm2}(ii), we have
\begin{align}\label{2g}
	&\E((g_k\!-\!\nabla f(x_k))|\mathcal{F}_k)=\E(g_k\!-\!\nabla f(x_k))=0,\cr
	&\E(\|g_k\!-\!\nabla f(x_k)\|^2|\mathcal{F}_k)=\E\|g_k\!-\!\nabla f(x_k)\|^2\leq\frac{n\sigma_g^2}{m_K}.
\end{align}
Let $\nabla f(x_k)$$=$$[\nabla f_1(x_{1,k})^\top\!,\dots,\nabla f_n(x_{n,k})^\top]^\top$. Then, by \eqref{2d} and \eqref{2g}, taking the mathematical expectation of $\|\bar{y}_k\|^2$ implies
\begin{align}\label{12}
	&\E\|\bar{y}_k\|^2\cr
	=&\E\|\bar{y}_k-\E\bar{y}_k\|^2+\|\E\bar{y}_k\|^2\cr
	\noalign{\vskip -3pt}
	=&\E\|\frac{1}{n}(\mathbf{1}_n^\top\!\otimes\! I_d)(g_k\!-\!\nabla f(x_k))\!+\!\frac{\beta_K}{n}\!\sum_{l=0}^{k-1}(\mathbf{1}_n^\top\mathcal{C}\!\otimes\! I_d)\eta_{l}\|^2\cr
	\noalign{\vskip -3pt}
	&+\|\E(\frac{1}{n}(\mathbf{1}_n^\top\!\otimes\! I_d)\nabla f(x_k))\|^2\cr
	\noalign{\vskip -3pt}
	=&\E\|\frac{1}{n}(\mathbf{1}_n^\top\!\otimes\! I_d)(g_k\!-\!\nabla f(x_k))\|^2\!+\!\frac{\beta_K^2}{n^2}\sum_{l=0}^{k-1}\E\|(\mathbf{1}_n^\top\mathcal{C}\!\otimes\! I_d)\eta_{l}\|^2\cr
	\noalign{\vskip -3pt}
	&+\|\E(\frac{1}{n}(\mathbf{1}_n^\top\!\otimes\! I_d)\nabla f(x_k))\|^2\cr
	\leq&\frac{\sigma_g^2}{m_K}+\frac{2d\rho(\mathcal{C})^2\beta_K^2}{n}\sum_{l=0}^{k-1}\max_{i\in\mathcal{V}}\{(\sigma_l^{(\eta_i)})^2\}\cr
	&+\|\E(\frac{1}{n}(\mathbf{1}_n^\top\!\otimes\! I_d)\nabla f(x_k))\|^2.
\end{align}
By Jensen's inequality (\!\!\cite[Cor. 4.3.1]{chow2012probability}), $\|\E(\frac{1}{n}(\mathbf{1}_n^\top\!\otimes\! I_d)\nabla f(x_k))\|^2\leq \E\|\frac{1}{n}(\mathbf{1}_n^\top\!\otimes\! I_d)\nabla f(x_k)\|^2$. Then, \eqref{12} can be rewritten as
\begin{align}\label{12.2}
	\E\|\bar{y}_k\|^2\leq&\frac{\sigma_g^2}{m_K}+\frac{2d\rho(\mathcal{C})^2\beta_K^2}{n}\sum_{l=0}^{k-1}\max_{i\in\mathcal{V}}\{(\sigma_l^{(\eta_i)})^2\}\cr
	&+\E\|\frac{1}{n}(\mathbf{1}_n^\top\!\otimes\! I_d)\nabla f(x_k)\|^2.
\end{align}
Since $\nabla f(x_k)=(\nabla f(x_k)-\nabla f((\mathbf{1}_n\otimes I_d)\bar{x}_k))+\nabla f((\mathbf{1}_n\otimes I_d)\bar{x}_k)$, setting $m=2$ in \eqref{9} and substituting \eqref{9} into $\E\|\frac{1}{n}(\mathbf{1}_n^\top\otimes I_d)\nabla f(x_k)\|^2$ imply
\begin{align}\label{13}
	&\smash{\E\|\frac{1}{n}(\mathbf{1}_n^\top\otimes I_d)\nabla f(x_k)\|^2}\cr
	\leq&2\E\|\frac{1}{n}\sum_{i=1}^{n}(\nabla f_i(x_{i,k})-\nabla f_i(\bar{x}_k))\|^2+2\E\|\nabla F(\bar{x}_k)\|^2\cr
	\noalign{\vskip -3pt}
	\leq&\frac{2}{n}\sum_{i=1}^{n}\E\|\nabla f_i(x_{i,k})-\nabla f_i(\bar{x}_k)\|^2+2\E\|\nabla F(\bar{x}_k)\|^2.
\end{align}
{\vskip -5pt}\noindent By Assumption \ref{asm2}(i) and Jensen's inequality, for any $x,y\in\mathbb{R}^d$ we have
\begin{align}\label{13.5}
	\|\nabla f_i(x)-\nabla f_i(y)\|^2=&\|\E(g_i(x,\lambda_i)-g_i(y,\lambda_i))\|^2\cr
	\leq&\E\|g_i(x,\lambda_i)-g_i(y,\lambda_i)\|^2\cr
	\leq&L_1^2\|x-y\|^2.
\end{align}Then by \eqref{13.5}, \eqref{13} can be rewritten as
\begin{align}\label{14}
	&\E\|\frac{1}{n}(\mathbf{1}_n^\top\otimes I_d)\nabla f(x_k)\|^2\cr
	\leq&\frac{2L_1^2}{n}\sum_{i=1}^{n}\E\|x_{i,k}-\bar{x}_k\|^2+2\E\|\nabla F(\bar{x}_k)\|^2\cr
	=&\frac{2L_1^2}{n}\|(W_1 \otimes I_d)x_k\|^2+2\E\|\nabla F(\bar{x}_k)\|^2.
\end{align}
Substituting \eqref{14} into \eqref{12.2} implies
\begin{align}\label{16}
	\E\|\bar{y}_k\|^2\leq&\frac{2L_1^2}{n}\|(W_1 \otimes I_d)x_k\|^2+2\E\|\nabla F(\bar{x}_k)\|^2\cr
	&+\frac{\sigma_g^2}{m_K}+\frac{2d\rho(\mathcal{C})^2\beta_K^2}{n}\sum_{l=0}^{k-1}\max_{i\in\mathcal{V}}\{(\sigma_l^{(\eta_i)})^2\}.
\end{align}
Thus, substituting \eqref{16} into \eqref{10.1} implies \eqref{a6}.

{\bf Step 2:} In this step, we prove the following inequality holds for any $k=0,\dots,K$:
\begin{align}\label{a7}
	&\E\|(W_2\otimes I_d) y_{k+1}\|^2\notag\\
	\leq& A_K^{(21)}\E\|(W_{\!1}\!\!\otimes\!\! I_d)x_k\|^2+ A_K^{(22)}\E\|(W_2\otimes I_d)y_k\|^2\notag\\
	&+\frac{A_K^{(23)}}{2L_1}\E\|\nabla F(\bar{x}_k)\|^2+u_k^{(2)}.
\end{align}
By Assumption \ref{asm1}, Lemma \ref{lemma1} holds. Note that by Lemma \ref{lemma1}(ii), $\mathcal{L}_2 W_2=W_2\mathcal{L}_2=\mathcal{L}_2$. Then, multiplying $W_2\otimes I_d$ on both sides of \eqref{eq8} leads to\vspace{-0.5em}
\begin{align}\label{18}
	\hspace{-0.75em}(W_2\!\otimes\!I_d)y_{k+1}\!=&(\!(I_n\!\!-\!\!\beta_K\mathcal{L}_2)\!\otimes\!I_d)(W_2\!\otimes\!I_d)y_k\cr
	&+\!\beta_K(\mathcal{C}\!\otimes\!I_d)\eta_k\!+\!(W_2\!\otimes\!I_d)\!(g_{k\!+\!1}\!-\!g_{k}).
\end{align}
By \eqref{18}, taking the mathematical expectation of $\|(W_2\otimes I_d)y_{k+1}\|^2$ implies
\begin{align}\label{19}
	&\E\|(W_2\otimes I_d)y_{k+1}\|^2\cr
	=&\E\|((I_n-\beta_K\mathcal{L}_2)\otimes I_d)(W_2\otimes I_d)y_k+\beta_K(W_2\mathcal{C}\otimes I_d)\eta_k\cr
	&+(W_2\otimes I_d)\left(g_{k+1}-g_k\right)\|^2.
\end{align} For any $k=0,\dots,K$, let $\mathcal{H}_k=\sigma(\{x_{k+1},y_k\})$. Then, since $\eta_k$ is independent of $\mathcal{H}_k$ and has the Laplacian distribution $\text{Lap}(\sigma_k^{(\eta_i)})$, we have
\begin{align}\label{5a}
	&\E(\eta_k|\mathcal{H}_k)=\E\eta_k=0,\cr
	&\E(\|\eta_k\|^2|\mathcal{H}_k)=\E\|\eta_k\|^2\leq2nd\max_{i\in\mathcal{V}}\{(\sigma_k^{(\eta_i)})^2\}.
\end{align}Moreover, since $g_{k+1}-\nabla f(x_{k+1})$ is independent of $\mathcal{H}_k$, by Assumption \ref{asm2}(ii) we have
\begin{align}\label{5b}
	&\E((g_{k\!+\!1}\!\!-\!\!\nabla f(x_{k\!+\!1})|\mathcal{F}_k)\!=\!\E(g_{k\!+\!1}\!\!-\!\!\nabla f(x_{k\!+\!1}))\!=\!0,\cr
	&\E(\|g_{k\!+\!1}\!\!-\!\!\nabla f(x_{k\!+\!1})\|^2|\mathcal{F}_k)\!\leq\!\E\|g_{k\!+\!1}\!\!-\!\!\nabla f(x_{k\!+\!1})\|^2\!=\!\frac{n\sigma_g^2}{m_K}.~~~~~
\end{align}Then, by \eqref{5a}, \eqref{5b} and the law of total expectation (\!\!\cite[Th. 7.1.1(ii)]{chow2012probability}), \eqref{19} can be rewritten as
\begin{align}\label{5c}
	&\E\|(W_2\otimes I_d)y_{k+1}\|^2\cr
	=&\E(\E(\|((I_n-\beta_K\mathcal{L}_2)\otimes I_d)(W_2\otimes I_d)y_k\cr
	&+(W_2\otimes I_d)(\nabla f(x_{k+1})-g_k)\|^2|\mathcal{H}_k)\cr
	&+\E(\|\beta_K(\mathcal{C}\otimes I_d)\eta_k\|^2|\mathcal{H}_k)\cr
	&+\E(\|(W_2\otimes I_d)(g_{k+1}-\nabla f(x_{k+1})\|^2|\mathcal{H}_k))\cr
	\leq&\E\|((I_n-\beta_K\mathcal{L}_2)\otimes I_d)(W_2\otimes I_d)y_k\cr
	&+(W_2\otimes I_d)(\nabla f(x_{k+1})-g_k)\|^2+\frac{\rho(W_2)^2n\sigma_g^2}{m_K}\cr
	&+2nd\rho(W_2)^2\rho(\mathcal{C})^2\beta_K^2\max_{i\in\mathcal{V}}\{(\sigma_k^{(\eta_i)})^2\}.
\end{align}Note that by $W_2=I_n-\frac{1}{n}v_2\mathbf{1}_n^\top$, we have $\rho(W_2)=1$. Then, \eqref{5c} can be rewritten as
\begin{align}\label{5c1}
	&\E\|(W_2\otimes I_d)y_{k+1}\|^2\cr
	\leq&\E\|((I_n-\beta_K\mathcal{L}_2)\otimes I_d)(W_2\otimes I_d)y_k\cr
	&+(W_2\otimes I_d)(\nabla f(x_{k+1})-g_k)\|^2\cr
	&+2nd\rho(\mathcal{C})^2\beta_K^2\max_{i\in\mathcal{V}}\{(\sigma_k^{(\eta_i)})^2\}+\frac{n\sigma_g^2}{m_K}.
\end{align}Then, setting $r=r_2\beta_K$ in \eqref{5} and substituting \eqref{5} into \eqref{5c1} result in\vspace{-0.2em}
\begin{align}\label{22}
	&\E\|(W_2\otimes I_d)y_{k+1}\|^2\cr
	\leq&(1+r_2\beta_K)\E\|((I_n-\beta_K\mathcal{L}_2)\otimes I_d)(W_2\otimes I_d)y_k\|^2\cr
	&+\left(1\!+\!\frac{1}{r_2\beta_K}\right)\E\|(W_2\otimes I_d)(\nabla f(x_{k+1})-g_k)\|^2\cr
	&+2nd\rho(\mathcal{C})^2\beta_K^2\max_{i\in\mathcal{V}}\{(\sigma_k^{(\eta_i)})^2\}+\frac{n\sigma_g^2}{m_K}.
\end{align}
Note that $\nabla f(x_{k+1})-g_k=\nabla f(x_{k+1})-\nabla f(x_k)+\nabla f(x_k)-g_k$. Then, setting $m=2$ in \eqref{9} and substituting \eqref{9}, \eqref{2g} into \eqref{22}~imply
\begin{align*}
	&\E\|(W_2\otimes I_d)y_{k+1}\|^2\cr
	\leq&(1+r_2\beta_K)\E\|((I_n-\beta_K\mathcal{L}_2)\otimes I_d)(W_2\otimes I_d)y_k\|^2
\end{align*}
\begin{align}\label{5d}
	&+\frac{2(1+r_2\beta_K)}{r_2\beta_K}\E\|(W_2\otimes I_d)(\nabla f(x_{k+1})-\nabla f(x_k))\|^2\notag\\
	&+\frac{2(1+r_2\beta_K)}{r_2\beta_K}\E\|(W_2\otimes I_d)(\nabla f(x_k)-g_k)\|^2\notag\\
	&+2nd\rho(\mathcal{C})^2\beta_K^2\max_{i\in\mathcal{V}}\{(\sigma_k^{(\eta_i)})^2\}+\frac{n\sigma_g^2}{m_K}\notag\\
	\leq&(1+r_2\beta_K)\E\|((I_n-\beta_K\mathcal{L}_2)\otimes I_d)(W_2\otimes I_d)y_k\|^2\notag\\
	&+\frac{2(1+r_2\beta_K)}{r_2\beta_K}\E\|\nabla f(x_{k+1})-\nabla f(x_k)\|^2\notag\\
	&+2nd\rho(\mathcal{C})^2\beta_K^2\max_{i\in\mathcal{V}}\{(\sigma_k^{(\eta_i)})^2\}+\frac{n(2+3r_2\beta_K)\sigma_g^2}{r_2\beta_Km_K}.
\end{align}

By \eqref{13.5}, it can be seen that\vspace{-0.5em}
\begin{align*}
	\|\nabla f(x_{k+1})-\nabla f(x_k)\|^2=&\sum_{i=1}^{n}\|\nabla f_i(x_{i,k+1})-\nabla f_i(x_{i,k})\|^2\cr
	\noalign{\vskip -5pt}
	\leq&L_1^2\sum_{i=1}^{n}\|x_{i,k+1}-x_{i,k}\|^2\cr
	=&L_1^2\|x_{k+1}-x_k\|^2.
\end{align*}
{\vskip -5pt}\noindent Thus, we have
\begin{align}\label{23}
	\E\left\|\nabla\!f(x_{k+1})\!-\!\nabla\!f(x_k)\right\|^2\leq L_1^2\E\left\|x_{k+1}\!-\!x_k\right\|^2.
\end{align}
{\vskip -2pt}\noindent Rearranging \eqref{eq7} gives\vspace{-0.4em}
\begin{align}\label{24}
	&x_{k+1}-x_k\cr
	=&-\alpha_K(\mathcal{L}_1\otimes I_d)(W_1\otimes I_d)x_k+\alpha_K(\mathcal{R}\otimes I_d)\zeta_k\cr
	&-\gamma_K(W_2\otimes I_d)y_k-\gamma_K(v_2\otimes I_d)\bar{y}_k.
\end{align}By \eqref{2d} and the law of total expectation, taking the mathematical expectation of $\|x_{k+1}-x_k\|^2$ gives
\begin{align}\label{24.1}
	&\E\|x_{k+1}-x_k\|^2\cr
	=&\E\|\alpha_K(\mathcal{L}_1\otimes I_d)(W_1\otimes I_d)x_k-\gamma_K(W_2\otimes I_d)y_k\cr
	&-\gamma_K(v_2\otimes I_d)\bar{y}_k\|^2+\E\|\alpha_K(\mathcal{R}\otimes I_d)\zeta_k\|^2\cr
	\leq&\E\|\alpha_K(\mathcal{L}_1\otimes I_d)(W_1\otimes I_d)x_k+\gamma_K(W_2\otimes I_d)y_k\cr
	&+\gamma_K(v_2\otimes I_d)\bar{y}_k\|^2+2nd\rho(\mathcal{R})^2\alpha_K^2\max_{i\in\mathcal{V}}\{(\sigma_k^{(\zeta_i)})^2\}.
\end{align}
Setting $m=3$ in \eqref{9} and substituting \eqref{9} into \eqref{24.1} implies
\begin{align}\label{25}
	&\E\|x_{k+1}-x_k\|^2\cr
	\leq&3\E\|\alpha_K(\mathcal{L}_1\otimes I_d)(W_1\otimes I_d)x_k\|^2+3\E\|\gamma_K\left(W_2\otimes I_d\right)y_k\|^2\cr
	\noalign{\vskip -3pt}
	&+3\E\|\gamma_K(v_2\otimes I_d)\bar{y}_k\|^2+2nd\rho(\mathcal{R})^2\alpha_K^2\max_{i\in\mathcal{V}}\{(\sigma_k^{(\zeta_i)})^2\}\cr
	\noalign{\vskip -3pt}
	\leq&3\alpha_K^2\rho(\mathcal{L}_1)^2\E\|(W_1\otimes I_d)x_k\|^2+3\gamma_K^2\E\|\left(W_2\otimes I_d\right)y_k\|^2\cr
	\noalign{\vskip -3pt}
	&+3\|v_2\|^2\gamma_K^2\E\|\bar{y}_k\|^2+2nd\rho(\mathcal{R})^2\alpha_K^2\max_{i\in\mathcal{V}}\{(\sigma_k^{(\zeta_i)})^2\}.
\end{align}
Substituting \eqref{16} into \eqref{25} results in
\begin{align}\label{25.1}
	&\E\|x_{k+1}-x_k\|^2\cr
	\leq&(3\alpha_K^2\rho(\mathcal{L}_1)^2+\frac{6\|v_2\|^2\gamma_K^2L_1^2}{n})\E\|(W_1\otimes I_d)x_k\|^2\cr
	&+3\gamma_K^2\E\|\left(W_2\otimes I_d\right)y_k\|^2+6\|v_2\|^2\gamma_K^2\E\|\nabla F(\bar{x}_k)\|^2\cr
	&+\frac{3\|v_2\|^2\sigma_g^2\gamma_K^2}{m_K}+2nd\rho(\mathcal{R})^2\alpha_K^2\max_{i\in\mathcal{V}}\{(\sigma_k^{(\zeta_i)})^2\}\cr
	&+\frac{6d\|v_2\|^2\rho(\mathcal{C})^2\beta_K^2\gamma_K^2}{n}\sum_{l=0}^{k-1}\max_{i\in\mathcal{V}}\{(\sigma_l^{(\eta_i)})^2\}.
\end{align}

Substituting \eqref{25.1} into \eqref{23} leads to\vspace{-0.3em}
\begin{align}\label{26}
	&\E\|\nabla f(x_{k+1})-\nabla f(x_k)\|^2\cr
	\leq&\left(3\alpha_K^2\rho(\mathcal{L}_1)^2+\frac{6\|v_2\|^2\gamma_K^2L_1^2}{n}\right)L_1^2\E\|(W_1\otimes I_d)x_k\|^2\cr
	&+\!3\gamma_K^2L_1^2\E\|\!(W_2\!\otimes\! I_d)y_k\|^2\!\!+\!6\|v_2\|^2\gamma_K^2L_1^2\E\|\nabla F(\bar{x}_k)\|^2\cr
	&+\frac{3\|v_2\|^2\gamma_K^2\sigma_g^2L_1^2}{m_K}+2nd\rho(\mathcal{R})^2L_1^2\alpha_K^2\max_{i\in\mathcal{V}}\{(\sigma_k^{(\zeta_i)})^2\}\cr
	&+\frac{6d\|v_2\|^2\rho(\mathcal{C})^2\beta_K^2\gamma_K^2L_1^2}{n}\sum_{l=0}^{k-1}\max_{i\in\mathcal{V}}\{(\sigma_l^{(\eta_i)})^2\}.
\end{align}
{\vskip -3pt}\noindent Note that by Lemma \ref{lemma1}(ii), $W_2^2=W_2$ holds. Then, $((I_n\!-\!\beta_K\mathcal{L}_2)\otimes I_d)(W_2\otimes I_d)y_k=((W_2\!-\!\beta_K\mathcal{L}_2)\otimes I_d)(W_2\otimes I_d)y_k$. Thus, by $\rho(W_2\!-\!\beta_K \mathcal{L}_2)\leq 1-r_2\beta_K$ in Lemma~\ref{lemma1}(ii), we have\vspace{-0.5em}
\begin{align}
	\label{27}
	&(1+r_2\beta_K)\E\|((I_n-\beta_K\mathcal{L}_2)\otimes I_d)(W_2\otimes I_d)y_k\|^2\cr
	\leq&(1+r_2\beta_K)(1-\beta_K r_2)^2\E\|(W_2\otimes I_d)y_k\|^2\cr
	\leq&(1-r_2\beta_K)\E\|(W_2\otimes I_d)y_k\|^2.
\end{align}
{\vskip -3pt}\noindent Then, substituting \eqref{26} and \eqref{27} into \eqref{5d} implies \eqref{a7}.

{\bf Step 3:} In this step,  we prove the following inequality holds for any $k=0,\dots,K$:
\begin{align}\label{a8}
	&(\frac{(v_1^{\!\top} v_2)\gamma_K}{2n}-\frac{2(v_1^{\!\top} v_2)^2\gamma_K^2L_1}{n^2})\sum_{k=0}^K\!\E\|\nabla F(\bar{x}_k)\|^2\cr
	\leq&A_K^{(31)}\sum_{k=0}^K\E\|(W_1\otimes I_d)x_k\|^2+A_K^{(32)}\sum_{k=0}^K\E\|(W_2\otimes I_d)y_k\|^2\cr
	&+F(\bar{x}_0)\!-\!F(x^*)+\sum_{k=0}^{K}u_k^{(3)}.
\end{align}
By Assumption \ref{asm1}, Lemma \ref{lemma1} holds. Then, multiplying $\frac{1}{n}(v_1^\top\otimes I_d)$ on both sides of \eqref{eq7} results in\vspace{-0.3em}
\begin{align}\label{29}
	\bar{x}_{k+1}=\bar{x}_k+\frac{\alpha_K}{n}(v_1^\top\mathcal{R}\otimes I_d)\zeta_k-\frac{\gamma_K}{n}(v_1^\top\otimes I_d)y_k.
\end{align}
{\vskip -4pt}\noindent Thus, setting $y=\bar{x}_{k+1}$, $x=\bar{x}_k$ in Lemma \ref{lemma a1}(i) and substituting \eqref{29} into Lemma \ref{lemma a1}(i) gives\vspace{-0.4em}
\begin{align}\label{30}
	&F(\bar{x}_{k\!+\!1})\notag\\
	\leq&F(\bar{x}_k)\!+\!\langle\nabla\!F(\bar{x}_k),\bar{x}_{k\!+\!1}\!-\!\bar{x}_k\rangle\!+\!\frac{L_1}{2}\!\|\bar{x}_{k\!+\!1}\!-\!\bar{x}_k\|^2\notag\\
	\noalign{\vskip 4pt}
	=&\smash{F(\bar{x}_k)+\langle\nabla F(\bar{x}_k),\frac{\alpha_K}{n}(v_1^\top\!\mathcal{R}\!\otimes\! I_d)\zeta_k\!-\!\frac{\gamma_K}{n}(v_1^\top\!\!\otimes\! I_d)y_k\rangle}\notag\\
	&+\frac{L_1}{2}\|\frac{\alpha_K}{n}(v_1^\top\!\mathcal{R}\otimes I_d)\zeta_k\!-\!\frac{\gamma_K}{n}(v_1^\top\!\!\otimes\! I_d)y_k\|^2.
\end{align}
Taking the mathematical expectation on both sides of \eqref{30} implies
\begin{align}\label{30.1}
	&\E F(\bar{x}_{k\!+\!1})\notag\\
	\hspace{-1em}\leq&\E F(\bar{x}_k)+\E \langle\nabla F(\bar{x}_k),\frac{\alpha_K}{n}(v_1^\top\!\mathcal{R}\!\otimes\! I_d)\zeta_k\!-\!\frac{\gamma_K}{n}(v_1^\top\!\!\otimes\! I_d)y_k\rangle\notag\\
	&+\frac{L_1}{2}\E\|\frac{\alpha_K}{n}(v_1^\top\!\mathcal{R}\otimes I_d)\zeta_k\!-\!\frac{\gamma_K}{n}(v_1^\top\!\!\otimes\! I_d)y_k\|^2\notag\\
	\hspace{-1em}=&\E F(\bar{x}_k)\!+\!\frac{L_1}{2}\E(\E(\|\frac{\alpha_K}{n}(v_1^\top\!\mathcal{R}\!\otimes\! I_d)\zeta_k\!-\!\frac{\gamma_K}{n}(v_1^\top\!\!\otimes\! I_d)y_k\|^2|\mathcal{F}_k))\notag\\
	&\E(\langle\nabla F(\bar{x}_k),\E(\frac{\alpha_K}{n}(v_1^\top\mathcal{R}\!\otimes\! I_d)\zeta_k|\mathcal{F}_k)\!-\!\frac{\gamma_K}{n}(v_1^\top\!\otimes\! I_d)y_k\rangle)\notag\\
	\hspace{-1em}\leq&\E F(\bar{x}_k)\!-\!\gamma_K\E \langle\nabla F(\bar{x}_k),\frac{1}{n}(v_1^\top\!\!\otimes\! I_d)y_k\rangle\!+\!\frac{\gamma_K^2L_1}{2n^2}\E\|(v_1^\top\!\!\otimes\! I_d)y_k\|^2\notag\\
	&+\frac{d\rho(\mathcal{R})^2\|v_1\|^2\alpha_K^2L_1}{n}\max_{i\in\mathcal{V}}\{(\sigma_k^{(\zeta_i)})^2\}.
\end{align}

Note that $\frac{1}{n}(v_1^\top\otimes I_d)y_k=\frac{1}{n}((v_1^\top W_2)\otimes I_d)y_k+\frac{(v_1^\top v_2)}{n}\bar{y}_k$. Then, we have
\begin{align}\label{30.2}
	&-\gamma_K\E\langle\nabla F(\bar{x}_k),\frac{1}{n}(v_1^\top\!\otimes\! I_d)y_k\rangle\cr
	\hspace{-1em}=&-\frac{(v_1^\top v_2)\gamma_K}{n}\E\langle\nabla F(\bar{x}_k),\bar{y}_k\!+\!\frac{1}{v_1^\top v_2}((v_1^\top W_2)\!\otimes\! I_d)y_k\rangle.
\end{align}
Since $-\langle\mathbf{a},\mathbf{b}\rangle$$=$$\frac{\|\mathbf{a}-\mathbf{b}\|^2\!-\|\mathbf{a}\|^2\!-\|\mathbf{b}\|^2}{2}$ for any $\mathbf{a},\mathbf{b}\in\mathbb{R}^d$, it can be seen that\vspace{-0.5em}
\begin{align}\label{31}
	&-\gamma_K\E\langle\nabla F(\bar{x}_k),\frac{1}{n}(v_1^\top\otimes I_d)y_k\rangle\cr
	\noalign{\vskip -3pt}
	=&\frac{(v_1^\top v_2)\gamma_K}{2n}(\E\|\nabla F(\bar{x}_k)\!-\!\frac{1}{v_1^\top v_2}((v_1^\top W_2)\!\otimes\! I_d)y_k\!-\!\bar{y}_k\|^2\cr
	\noalign{\vskip -5pt}
	&-\E\|\nabla F(\bar{x}_k)\|^2-\E\|\frac{1}{v_1^\top v_2}((v_1^\top W_2)\!\otimes\! I_d)y_k\!+\!\bar{y}_k\|^2)\cr
	\noalign{\vskip -4pt}
	\leq&\frac{(v_1^\top v_2)\gamma_K}{2n}\E\|\nabla F(\bar{x}_k)\!-\!\frac{1}{v_1^\top v_2}((v_1^\top W_2)\!\otimes\! I_d)y_k\!-\!\bar{y}_k\|^2\cr
	\noalign{\vskip -3pt}
	&-\frac{(v_1^\top v_2)\gamma_K}{2n}\E\|\nabla F(\bar{x}_k)\|^2.
\end{align}

Note that\vspace{-0.5em}
\begin{align}\label{32}
	&\E\|\nabla F(\bar{x}_k)-\frac{1}{v_1^\top v_2}((v_1^\top W_2)\otimes I_d)y_k-\bar{y}_k\|^2\cr
	=&\E\|\nabla F(\bar{x}_k)-\frac{1}{n}(\mathbf{1}_n^\top\otimes I_d)g_k+\frac{1}{n}(\mathbf{1}_n^\top\otimes I_d)g_k-\bar{y}_k\cr
	&-\!\frac{1}{v_1^\top v_2}((v_1^\top W_2)\!\otimes\! I_d)y_k\|^2.
\end{align}
Then, setting $m=3$ in \eqref{9} and substituting \eqref{9} into \eqref{32} imply\vspace{-0.5em}
\begin{align}\label{33}
	&\E\|\nabla F(\bar{x}_k)-\frac{1}{v_1^\top v_2}((v_1^\top W_2)\otimes I_d)y_k-\bar{y}_k\|^2\cr
	\leq&3\E\|\nabla F(\bar{x}_k)-\frac{1}{n}(\mathbf{1}_n^\top\otimes I_d)g_k\|^2\cr
	&+3\E\|\frac{1}{n}(\mathbf{1}_n^\top\otimes I_d)g_k-\bar{y}_k\|^2+3\E\|\frac{1}{v_1^\top v_2}((v_1^\top W_2)\!\otimes\! I_d)y_k\|^2\cr
	=&3\E\|\nabla F(\bar{x}_k)-\frac{1}{n}(\mathbf{1}_n^\top\otimes I_d)\nabla f(x_k)+\frac{1}{n}(\mathbf{1}_n^\top\otimes I_d)\nabla f(x_k)\cr
	&-\frac{1}{n}(\mathbf{1}_n^\top\otimes I_d)g_k\|^2+3\E\|\frac{1}{n}(\mathbf{1}_n^\top\otimes I_d)g_k-\bar{y}_k\|^2\cr
	&+3\E\|\frac{1}{v_1^\top v_2}((v_1^\top W_2)\!\otimes\! I_d)y_k\|^2.
\end{align}
{\vskip -5pt}\noindent Thus, substituting \eqref{11}, \eqref{2g}, and \eqref{5a} into \eqref{33} implies\vspace{-0.5em}
\begin{align}\label{33.1}
	&\E\|\nabla F(\bar{x}_k)-\frac{1}{v_1^\top v_2}((v_1^\top W_2)\otimes I_d)y_k-\bar{y}_k\|^2\cr
	\leq&3\E\|\nabla F(\bar{x}_k)-\frac{1}{n}(\mathbf{1}_n^\top\otimes I_d)\nabla f(x_k)\|^2\cr
	&+3\E\|\frac{1}{n}(\mathbf{1}_n^\top\otimes I_d)\nabla f(x_k)-\frac{1}{n}(\mathbf{1}_n^\top\otimes I_d)g_k\|^2\cr
	&+3\E\|\frac{\beta_K}{n}\sum_{l=0}^{k-1}(\mathbf{1}_n^\top\mathcal{C}\otimes I_d)\eta_{l}\|^2\cr
	&+3\E\|\frac{1}{v_1^\top v_2}((v_1^\top W_2)\!\otimes\! I_d)y_k\|^2\cr
	\leq&3\E\|\nabla F(\bar{x}_k)-\frac{1}{n}(\mathbf{1}_n^\top\otimes I_d)\nabla f(x_k)\|^2\cr
	&+\frac{3\|v_1\|^2}{(v_1^\top v_2)^2}\E\|(W_2\otimes I_d)y_k\|^2\cr
	&+\frac{3\sigma_g^2}{m_K}+\frac{6d\rho(\mathcal{C})^2\beta_K^2}{n}\sum_{l=0}^{k-1}\max_{i\in\mathcal{V}}\{(\sigma_l^{(\eta_i)})^2\}.
\end{align}
By \eqref{13.5}, we have \vspace{-0.5em}
\begin{align}\label{33.2}
	&\E\|\nabla F(\bar{x}_k)-\frac{1}{n}(\mathbf{1}_n^\top\otimes I_d)\nabla f(x_k)\|^2\cr
	=&\E\|\frac{1}{n}\sum_{i=1}^{n}(\nabla f_i(x_{i,k})-\nabla f_i(\bar{x}_k))\|^2\cr
	\leq&\frac{L_1^2}{n}\E\|(W_1\otimes I_d)x_k\|^2.
\end{align}
Then, substituting \eqref{33.2} into \eqref{33.1} implies\vspace{-0.5em}
\begin{align}\label{34}
	&\E\|\nabla F(\bar{x}_k)-\frac{1}{v_1^\top v_2}((v_1^\top W_2)\otimes I_d)y_k-\bar{y}_k\|^2\cr
	\leq&\frac{3L_1^2}{n}\E\|(W_1\otimes I_d)x_k\|^2+\frac{3\|v_1\|^2}{(v_1^\top v_2)^2}\E\|(W_2\otimes I_d)y_k\|^2\cr
	&+\frac{3\sigma_g^2}{m_K}+\frac{6d\rho(\mathcal{C})^2\beta_K^2}{n}\sum_{l=0}^{k-1}\max_{i\in\mathcal{V}}\{(\sigma_l^{(\eta_i)})^2\}.
\end{align}
Substituting \eqref{34} into \eqref{31} implies
\begin{align}\label{35}
	&-\gamma_K\E\langle\nabla F(\bar{x}_k),\frac{1}{n}(v_1^\top\otimes I_d)y_k\rangle\cr
	\hspace{-2em}\leq&-\frac{(v_1^\top\! v_2)\gamma_K}{2n}\E\|\nabla F(\bar{x}_k)\|^2\!+\!\frac{3(v_1^\top\! v_2)\gamma_KL_1^2}{2n^2}\E\|(W_1\!\otimes\! I_d)x_k\|^2\cr
	&+\frac{3\|v_1\|^2\gamma_K}{2n(v_1^\top v_2)}\E\|(W_2\otimes I_d)y_k\|^2+\frac{3(v_1^\top v_2)\sigma_g^2\gamma_K}{2nm_K}\cr	
	&+\frac{3(v_1^\top v_2)d\rho(\mathcal{C})^2\beta_K^2\gamma_K}{n^2}\sum_{l=0}^{k-1}\max_{i\in\mathcal{V}}\{(\sigma_l^{(\eta_i)})^2\}.
\end{align}
Then, substituting \eqref{35} into \eqref{30.1} result in
\begin{align}\label{36}
	&\E F(\bar{x}_{k+1})\cr
	\hspace{-2em}\leq&\E F(\bar{x}_k)-\frac{(v_1^\top\! v_2)\gamma_K}{2n}\E\|\nabla F(\bar{x}_k)\|^2\cr
	&+\frac{3(v_1^\top\! v_2)\gamma_KL_1^2}{2n^2}\E\|(W_1\!\otimes\! I_d)x_k\|^2\!+\!\frac{3\|v_1\|^2\gamma_K}{2n(v_1^\top v_2)}\E\|(W_2\!\otimes\! I_d)y_k\|^2\notag\\
	&+\frac{\gamma_K^2L_1}{2n^2}\E\|(v_1^\top\!\otimes\! I_d)y_k\|^2+\frac{3(v_1^\top v_2)\sigma_g^2\gamma_K}{2nm_K}\cr
	&+\frac{3(v_1^\top v_2)d\rho(\mathcal{C})^2\beta_K^2\gamma_K}{n^2}\sum_{l=0}^{k-1}\max_{i\in\mathcal{V}}\{(\sigma_l^{(\eta_i)})^2\}\cr
	&+\frac{d\rho(\mathcal{R})^2\|v_1\|^2\alpha_K^2L_1}{n}\max_{i\in\mathcal{V}}\{(\sigma_k^{(\zeta_i)})^2\}.
\end{align}

Note that by setting $m=2$ in \eqref{9}, we have
\begin{align}\label{37}
	&\frac{\gamma_K^2L_1}{2n^2}\E\|(v_1^\top\!\otimes\! I_d)y_k\|^2\cr
	=&\frac{\gamma_K^2L_1}{2n^2}\E\|(v_1^\top W_2\otimes I_d)y_k+(v_1^\top v_2) \bar{y}_k\|^2\cr
	\leq&\frac{\gamma_K^2L_1}{n^2}\E\|(v_1^\top W_2\otimes I_d)y_k\|^2+\frac{\gamma_K^2L_1}{n^2}\E\|(v_1^\top v_2)\bar{y}_k\|^2\cr
	\leq&\frac{\|v_1\|^2\gamma_K^2L_1}{n^2}\E\|(W_2\otimes I_d)y_k\|^2+\frac{(v_1^\top v_2)^2\gamma_K^2L_1}{n^2}\E\|\bar{y}_k\|^2.~~~~~~~~~~~~
\end{align}
Then, substituting \eqref{16} into \eqref{37} implies
\begin{align}\label{37.1}
	&\frac{\gamma_K^2L_1}{2n^2}\E\|(v_1^\top\!\otimes\! I_d)y_k\|^2\cr
	\leq&\frac{2(v_1^\top v_2)^2\gamma_K^2L_1^3}{n^3}\E\|(W_1\otimes I_d)x_k\|^2\cr
	&+\frac{\|v_1\|^2\gamma_K^2L_1}{n^2}\E\|(W_2\otimes I_d)y_k\|^2\cr
	&+\frac{2(v_1^\top v_2)^2\gamma_K^2L_1}{n^2}\E\|\nabla F(\bar{x}_k)\|^2+\frac{(v_1^\top v_2)^2\sigma_g^2\gamma_K^2L_1}{n^2m_K}\cr
	&+\frac{2d(v_1^\top v_2)^2\rho(\mathcal{C})^2\beta_K^2\gamma_K^2L_1}{n^3}\sum_{l=0}^{k-1}\max_{i\in\mathcal{V}}\{(\sigma_l^{(\eta_i)})^2\}.
\end{align}
Thus, substituting \eqref{37.1} into \eqref{36} gives
\begin{align}\label{39}
	&\E F(\bar{x}_{k+1})\notag\\
	\hspace{-2em}\leq&\E F(\bar{x}_k)\!+\!A_K^{(31)}\E\|(W_1\!\otimes\! I_d)x_k\|^2\!+\!A_K^{(32)}\E\|(W_2\!\otimes\! I_d)y_k\|^2\notag\\
	&+\!(-\frac{(v_1^\top v_2)\gamma_K}{2n}\!+\!\frac{2(v_1^\top v_2)^2\gamma_K^2L_1}{n^2})\E\|\nabla F(\bar{x}_k)\|^2\!+\!u_k^{(3)}\!\!.
\end{align}Rearranging \eqref{39} gives
\begin{align}\label{8a}
	&(\frac{(v_1^\top v_2)\gamma_K}{2n}\!-\!\frac{2(v_1^\top v_2)^2\gamma_K^2L_1}{n^2})\E\|\nabla F(\bar{x}_k)\|^2\notag\\
	\leq&\E (F(\bar{x}_k)-F(\bar{x}_{k+1}))+A_K^{(31)}\E\|(W_1\!\otimes\! I_d)x_k\|^2\notag\\
	&+A_K^{(32)}\E\|(W_2\!\otimes\! I_d)y_k\|^2+u_k^{(3)}.
\end{align}
Then, summing \eqref{8a} from 0 to $K$ and using $F(x_{K+1})\geq F(x^*)$ result in \eqref{a8}.

{\bf Step 4:} In this step, we prove \eqref{41b} holds for any $k=0,\dots,K$. By Lemma \ref{lemma a1}(ii), \eqref{a6} and \eqref{a7} can be rewritten as
\begin{align}
	&\E\|(W_1\otimes I_d) x_{k+1}\|^2\notag\\
	\leq& A_K^{(11)}\E\|(W_{\!1}\!\!\otimes\!\! I_d)x_k\|^2+ A_K^{(12)}\E\|(W_2\otimes I_d)y_k\|^2\notag\\
	&+A_K^{(13)}\E(F(\bar{x}_k)-F(x^*))+u_k^{(1)},\label{b1,1}\\
	&\E\|(W_2\otimes I_d) y_{k+1}\|^2\notag\\
	\leq& A_K^{(21)}\E\|(W_{\!1}\!\!\otimes\!\! I_d)x_k\|^2+ A_K^{(22)}\E\|(W_2\otimes I_d)y_k\|^2\notag\\
	&+A_K^{(23)}\E(F(\bar{x}_k)-F(x^*))+u_k^{(2)}.\label{b1,2}
\end{align}Moreover, by Assumption \ref{asm3} and $\gamma_K<\frac{n}{4(v_1^\top v_2)L_1}$, we have \vspace{0.1em}$(-\frac{(v_1^\top v_2)\gamma_K}{2n}+\frac{2(v_1^\top v_2)^2\gamma_K^2L_1}{n^2})\|\nabla F(\bar{x}_k)\|^2\leq(-\frac{(v_1^\top v_2)\mu\gamma_K}{n}+$ $\frac{4(v_1^\top v_2)^2\mu\gamma_K^2L_1}{n^2})$$(F(\bar{x}_k)$$-$$F(x^*))$. Then, \eqref{39} can be \mbox{rewritten as}
\begin{align}\label{40}
	&\E F(\bar{x}_{k+1})\notag\\
	\noalign{\vskip -5pt}
	\hspace{-3em}\leq&\E F(\bar{x}_k)\!+\!(-\frac{(v_1^\top v_2)\mu\gamma_K}{n}\!+\!\frac{4(v_1^\top v_2)^2\mu\gamma_K^2L_1}{n^2})(F(\bar{x}_k)\!-\!F(x^*))\notag\\
	&+\!A_K^{(31)}\E\|(W_1\!\otimes\! I_d)x_k\|^2\!+\!A_K^{(32)}\E\|(W_2\!\otimes\! I_d)y_k\|^2\!+\!u_k^{(3)}\!\!.
\end{align}
Thus, subtracting $F(x^*)$ from both sides of \eqref{40} implies
\begin{align}\label{b1,3}
	&\E (F(\bar{x}_{k+1})-F(x^*))\notag\\
	\leq&A_K^{(33)}\E (F(\bar{x}_k)-F(x^*))\!+\!A_K^{(31)}\E\|(W_1\!\otimes\! I_d)x_k\|^2\notag\\
	&+A_K^{(32)}\E\|(W_2\!\otimes\! I_d)y_k\|^2+u_k^{(3)}.
\end{align}
Hence, combining \eqref{b1,1}, \eqref{b1,2}, and \eqref{b1,3} results in \eqref{41b}. Therefore, this lemma is proved. $\hfill\blacksquare$

\section{Proof of Theorem \ref{thm1}}\label{appendix c1}
We proceed with the following two cases for \emph{Scheme (S1)} and \emph{Scheme (S2)}.

\indent{\bf Case 1.} If Assumptions \ref{asm1}, \ref{asm2}, \ref{asm4} holds under \emph{Scheme (S1)}, then the proof of the almost sure and mean square convergence of Algorithm \ref{algorithm1} is given in the following four steps:
	
\indent{\bf Step 1.} First, we prove that there exists $G_3>0$ such that for any $K=0,1,\dots$, $\E(\mathbf{1}_3^\top V_K)\leq G_3$. Let $\tilde{v}=[\tilde{v}_1,\tilde{v}_2,\tilde{v}_3]^\top$ be a positive vector, and the matrix $D_K$ defined as follows:
\begin{align*}
	D_K=\left[\begin{array}{cccc}
		\!\!A_K^{(11)}&\!\!\!\!A_K^{(12)}\!\!&\!\!\!\!A_K^{(13)}\!\!&\!\!\!\!0\!\!\\
		\!\!A_K^{(21)}&\!\!\!\!A_K^{(22)}\!\!&\!\!\!\!A_K^{(23)}\!\!&\!\!\!\!0\!\!\\
		\!\!A_K^{(31)}&\!\!\!\!A_K^{(32)}\!\!&\!\!\!\!1\!\!&\!\!\!\!{\scriptstyle  -\!\frac{(v_1^{\!\top}\! v_2)\gamma_K}{2n}\!+\!\frac{2(v_1^{\!\top}\! v_2)^2\gamma_K^2L_1}{n^2}}\!\!\\
		0&0&0&0
	\end{array}\right].
\end{align*}Then, by $p_\beta<p_\alpha< p_\gamma$ in Assumption \ref{asm4}, there exists a positive integer $K_0$ such that for any $K=K_0,K_0+1,\dots$, the following inequality holds:
\begin{align}\label{44}
	[\tilde{v}_1,\tilde{v}_2,\tilde{v}_3,0]D_K\leq(1\!+\!\frac{16\|v_2\|^2\gamma_K^2L_1}{n^2r_1\alpha_K})[\tilde{v}_1,\tilde{v}_2,\tilde{v}_3,0].
\end{align}Thus, by \eqref{39}, \eqref{b1,1}, \eqref{b1,2}, and \eqref{44}, we have
\begin{align}\label{46}
	\E(\tilde{v}^\top V_{k\!+\!1})\leq&[\tilde{v}_1,\tilde{v}_2,\tilde{v}_3,0]D_K\left[\begin{array}{c}
		\!\!\E V_k\!\!\\
		\!\!\E\|\nabla F(\bar{x}_k)\|^2\!\!
	\end{array}\right]\!+\!\tilde{v}^\top u_k\notag\\
	\leq&(1\!+\!\frac{16\|v_2\|^2\gamma_K^2L_1}{n^2r_1\alpha_K})\E(\tilde{v}^\top V_k)\!+\!\tilde{v}^\top u_k.
\end{align}
{\vskip -3pt}\noindent Let $\theta=\min\{p_m-p_\beta,2p_\alpha-p_\beta-2\max\{\max_{i\in\mathcal{V}}\{p_{\zeta_i}\},0\}$, $2p_\beta-2\max\{\max_{i\in\mathcal{V}}\{p_{\eta_i}\},0\}\}$. Then, by Assumption \ref{asm4}, $\tilde{v}^\top u_k=O(\frac{1}{(K+1)^\theta})$ holds for any $k=0,\dots,K$. Thus, iteratively computing \eqref{46} results in\vspace{-0.5em}
\begin{align}\label{48}
	&\E(\tilde{v}^\top V_{K\!+\!1})\cr
	=&(1\!+\!\frac{16\|v_2\|^2\gamma_K^2L_1}{n^2r_1\alpha_K})^{K\!+\!1}\E(\tilde{v}^{\!\top}\!V_0)\notag\\
	\noalign{\vskip -3pt}
	&+O\!\left(\sum_{k=0}^{K}(1\!+\!\frac{16\|v_2\|^2\gamma_K^2L_1}{n^2r_1\alpha_K})^k\frac{1}{(K\!+\!1)^\theta}\right)\!.
\end{align}
{\vskip -3pt}\noindent Since $2p_\gamma-p_\alpha\geq1$ in Assumption \ref{asm4}, $\lim_{K\to\infty}(1\!+\!\frac{16\|v_2\|^2\gamma_K^2L_1}{n^2r_1\alpha_K})^{K\!+\!1}$ $<\infty$. Then, there exists $G_1>0$ such that for any $K=0,1,\dots$, $(1\!+\!\frac{16\|v_2\|^2\gamma_K^2L_1}{n^2r_1\alpha_K})^{K\!+\!1}\leq G_1$. Thus, \eqref{48} can be rewritten as\vspace{-0.3em}
\begin{align}\label{49}
	\hspace{-1em}\E(\tilde{v}^\top V_{K\!+\!1})\leq&G_1\E(\tilde{v}^{\!\top}\!V_0)+\!O\!\!\left(\!\sum_{k=0}^{K}\!\frac{1}{(K\!+\!1)^\theta}\!\right)\cr
	=&G_1\E(\tilde{v}^{\!\top}\!V_0)+O\left(\frac{1}{(K+1)^{\theta-1}}\right).
\end{align}
{\vskip -4pt}\noindent By $2p_\gamma-p_\alpha\geq1$, $2p_\alpha-p_\beta-2\max\{\max_{i\in\mathcal{V}}\{p_{\zeta_i}\},0\}$$\geq$$1$, $2p_\beta-2\max\{\max_{i\in\mathcal{V}}\{p_{\eta_i}\},0\}\geq1$, $p_m-p_\beta\geq1$ in Assumption~\ref{asm4}, we have $\theta\geq1$. Thus, there exists $G_2>0$ such that for any for any $K=K_0,K_0\!+\!1,\dots$, $\E(\tilde{v}^\top V_{K\!+\!1})\leq G_2$. 

\noindent Let $G_3$$=$$(\frac{1}{\tilde{v}_1}$$+$$\frac{1}{\tilde{v}_2}$$+$$\frac{1}{\tilde{v}_3})$$\max\{\E(\tilde{v}^\top\! V_0),\! \E(\tilde{v}^\top\! V_1),\dots,\E(\tilde{v}^\top\! V_{K_0})$, $G_2\}$. Then, for any $K=0,1,\dots$, we have
\begin{align*}
	\E(\mathbf{1}_3^\top V_K)\leq&\max\{\frac{1}{\tilde{v}_1},\frac{1}{\tilde{v}_2},\frac{1}{\tilde{v}_3}\}\E(\tilde{v}^\top V_K)\cr
	\leq&(\frac{1}{\tilde{v}_1}+\frac{1}{\tilde{v}_2}+\frac{1}{\tilde{v}_3})\E(\tilde{v}^\top V_K)\leq G_3.
\end{align*}

\indent{\bf Step 2:} In this step, we prove that for any $i\in\mathcal{V}$, $\liminf_{K\to\infty}\|(W_1\otimes I_d)$$ x_{K+1}\|^2=0$ a.s., $\lim_{K\to\infty}\E\|(W_1\otimes I_d)$ $x_{K+1}\|^2=0$. By {\bf Step 1}, there exists $G_3>0$ such that for any $K=0,1,\dots$, $\E$$\|(W_1$$\otimes$$ I_d)x_K\|^2$$\leq$$ G_3$, $\E$$\|(W_2$$\otimes$$ I_d)y_K\|^2$ $\leq$$ G_3$, $\E(F(\bar{x}_K)-F(x^*))$$\leq$$ G_3$. Then, substituting these inequalities into \eqref{b1,1} gives
\begin{align}\label{50}
	&\E\|(W_1\!\otimes\! I_d) x_{k+1}\|^2\cr
	\leq&(1\!-\!r_1\alpha_K)\E\|(W_1\!\otimes \!I_d)x_k\|^2+u_k^{(1)}\cr
	&+\!\frac{2(1\!+\!r_1\alpha_K)\gamma_K^2G_3}{r_1\alpha_K}(1\!+\!\frac{4\|v_2\|^2L_1}{n^2}\!+\!\frac{2\|v_2\|^2L_1^2}{n^3})\cr
	=&(1\!-\!r_1\alpha_K)\E\|(W_1\!\otimes \!I_d)x_k\|^2\cr
	&+O(\frac{1}{(K+1)^{2p_\gamma-p_\alpha,2p_\alpha-2\max\{\max_{i\in\mathcal{V}}\{p_{\zeta_i}\},0\}}}).
\end{align}
Iteratively computing \eqref{50} gives
\begin{align}\label{51}
	&\E\|(W_1\otimes I_d) x_{K+1}\|^2\cr
	\leq&(1\!-\!r_1\alpha_K)^{K+1}\|(W_1\!\otimes \!I_d)x_0\|^2\cr
	&+O(\frac{1}{(K+1)^{2p_\gamma-p_\alpha,2p_\alpha-2\max\{\max_{i\in\mathcal{V}}\{p_{\zeta_i}\},0\}}}\sum_{k=0}^{K}(1\!-\!r_1\alpha_K)^k)\cr
	=&O(\frac{1}{(K+1)^{2p_\gamma-2p_\alpha,p_\alpha-2\max\{\max_{i\in\mathcal{V}}\{p_{\zeta_i}\},0\}}}).
\end{align}Then by \eqref{51}, $\lim_{K\to\infty}\E\|(W_1\otimes I_d) x_{K+1}\|^2=0$. By \cite[Th. 4.2.3]{chow2012probability}, $\|(W_1\otimes I_d) x_{K+1}\|^2$ converges in probability to 0, and thus, there exists a sequence $\{\|(W_1\otimes I_d) x_{l_K}\|^2,K=0,1,\dots\}$ such that $\lim_{K\to\infty}\|(W_1\otimes I_d) x_{l_K}\|^2=0$, a.s.. Hence, we have
\begin{align*}
	\liminf_{K\to\infty}\|(W_1\!\otimes\! I_d) x_{K+1}\|^2=\liminf_{K\to\infty}\|(W_1\!\otimes\! I_d) x_K\|^2=0\text{ a.s.}.
\end{align*}

\indent{\bf Step 3:} In this step, we prove that $\liminf_{K\to\infty}$ $\|\nabla F(\bar{x}_{K+1})\|^2=0$ a.s., $\liminf_{K\to\infty}\E\|\nabla F(\bar{x}_{K+1})\|^2=0$. By {\bf Step 1}, there exists $G_3>0$ such that for any $K=0,1,\dots$, $\E$$\|(W_1$$\otimes$$ I_d)x_K\|^2$$\leq$$ G_3$, $\E$$\|(W_2$$\otimes$$ I_d)y_K\|^2$ $\leq$$ G_3$, $\E(F(\bar{x}_K)-F(x^*))$$\leq$$ G_3$. Then, substituting these inequalities into \eqref{b1,2} gives
\begin{align}\label{eq52}
	&\E\|(W_2\otimes I_d) y_{k+1}\|^2\notag\\
	\leq& (1-r_2\beta_K)\E\|(W_2\otimes I_d)y_k\|^2+u_k^{(2)}\notag\\
	&+\frac{6(1+r_2\beta_K)\gamma_K^2L_1^2G_3}{r_2\beta_K}(1+4\|v_2\|^2L_1+\frac{2\|v_2\|^2L_1^2}{n})\notag\\
	&+\frac{6(1+r_2\beta_K)\rho(\mathcal{L}_1)^2\alpha_K^2L_1^2G_3}{r_2\beta_K}.
\end{align}
Similar to {\bf Step 2}, by \eqref{eq52} we have
\begin{align}\label{eq53}
	\E\|(W_2\otimes I_d) y_{K\!+\!1}\|^2\!=\!O(\frac{1}{(K\!+\!1)^{1-p_\beta}}).
\end{align}
Substituting \eqref{eq52} and \eqref{eq53} into \eqref{a8} implies
\begin{align*}
	&(\frac{(v_1^{\!\top} v_2)\gamma_K}{2n}-\frac{2(v_1^{\!\top} v_2)^2\gamma_K^2L_1}{n^2})\sum_{k=0}^K\!\E\|\nabla F(\bar{x}_k)\|^2\cr
	\leq&\E (F(\bar{x}_0)\!-\!F(x^*))+\sum_{k=0}^{K}u_k^{(3)}\cr
	&+O(\frac{1}{(K+1)^{\min\{3p_\gamma-2p_\alpha-1,1+p_\gamma-p_\beta\}}}).
\end{align*}
By $p_\beta$$<$$p_\alpha$$<$$p_\gamma$, $2p_\gamma$$-$$p_\alpha\geq1$ in Assumption \ref{asm4}, we have
\begin{align}\label{eq53.5}
	(\frac{(v_1^{\!\top} v_2)\gamma_K}{2n}-\frac{2(v_1^{\!\top} v_2)^2\gamma_K^2L_1}{n^2})\sum_{k=0}^K\!\E\|\nabla F(\bar{x}_k)\|^2<\infty.
\end{align}

Next, we prove $\liminf_{K\to\infty}$ $\E\|\nabla F(\bar{x}_{K+1})\|^2=0$ by contradiction. Suppose there exists $G_4>0$ such that $\liminf_{K\to\infty}$ $\E\|\nabla F(\bar{x}_{K+1})\|^2=G_4$. Then, there exists a positive integer $K_1$ such that $\E\|\nabla F(\bar{x}_{K_1})\|^2\geq \frac{G_4}{2}$ for any $K=K_1,K_1+1,\dots$. Thus, we have
\begin{align}\label{eq54}
	&(\frac{(v_1^{\!\top} v_2)\gamma_K}{2n}-\frac{2(v_1^{\!\top} v_2)^2\gamma_K^2L_1}{n^2})\sum_{k=0}^K\!\E\|\nabla F(\bar{x}_k)\|^2\cr
	\geq&(\frac{(v_1^{\!\top} v_2)\gamma_K}{2n}-\frac{2(v_1^{\!\top} v_2)^2\gamma_K^2L_1}{n^2})\sum_{k=K_1}^K\!\E\|\nabla F(\bar{x}_k)\|^2\cr
	\geq&(\frac{(v_1^{\!\top} v_2)\gamma_K}{2n}-\frac{2(v_1^{\!\top} v_2)^2\gamma_K^2L_1}{n^2})\frac{(K-K_1+1)G_4}{2}.
\end{align}
Note that when $K$ goes to infinity, the right hand side of \eqref{eq54} goes to infinity. Then, $(\frac{(v_1^{\!\top} v_2)\gamma_K}{2n}-\frac{2(v_1^{\!\top} v_2)^2\gamma_K^2L_1}{n^2})\sum_{k=0}^K\!\E\|\nabla F(\bar{x}_k)\|^2$ goes to infinity, and thus, contradicts \eqref{eq53.5}. Hence, $\liminf_{K\to\infty}$ $\E\|\nabla F(\bar{x}_{K+1})\|^2=0$. 

Then by $\liminf_{K\to\infty}\E\|\nabla F(\bar{x}_{K+1})\|^2=0$, there exists a sequence $\{\E\|\nabla F(\bar{x}_{l_K})\|^2,K=0,1,\dots\}$ such that $\lim_{K\to\infty}\E\|\nabla F(\bar{x}_{l_K})\|^2=0$. By \cite[Th. 4.2.3]{chow2012probability}, $\|\nabla F(\bar{x}_{l_K})\|^2$ converges in probability to 0, and then, there exists a sequence $\{\E\|\nabla F(\bar{x}_{s_K})\|^2,K=0,1,\dots\}$ such that $\lim_{K\to\infty}\E\|\nabla F(\bar{x}_{s_K})\|^2=0$ a.s.. Thus, $\liminf_{K\to\infty}\|\nabla F(\bar{x}_{K\!+\!1})\|^2=0$ a.s..

\indent{\bf Step 4:} In this step, we prove that $\liminf_{K\to\infty}$ $\|\nabla F(x_{i,K\!+\!1})\|^2$$=$$0\text{ a.s.}$, $\liminf_{K\to\infty}$ $\E\|\nabla F(x_{i,K\!+\!1})\|^2$ $=$ $0$ for any $i\in\mathcal{V}$. By \eqref{13.5}, the following inequality holds for any $i\in\mathcal{V}$:
\begin{align}\label{s4,1}
	&\|\nabla F(x_{i,K\!+\!1})\|^2\cr
	=&\|\nabla F(\bar{x}_{K\!+\!1})\!+\!\nabla F(x_{i,K\!+\!1})\!-\!\nabla F(\bar{x}_{K\!+\!1})\|^2\cr
	\leq& 2\|\nabla F(\bar{x}_{K\!+\!1})\|^2\!+\!2\|\nabla F(x_{i,K\!+\!1})\!-\!\nabla F(\bar{x}_{K\!+\!1})\|^2\cr
	\leq& 2\|\nabla F(\bar{x}_{K\!+\!1})\|^2\!+\!2L_1^2\|x_{i,K\!+\!1}-\bar{x}_{K\!+\!1}\|^2\cr
	\leq& 2\|\nabla F(\bar{x}_{K\!+\!1})\|^2\!+\!2L_1^2\|(W_1\!\otimes\! I_d)x_{K\!+\!1}\|^2.
\end{align}Then, by {\bf Steps 2} and {\bf 3}, we have $\liminf_{K\to\infty}\!\|\nabla F(x_{i,K\!+\!1})\|^2$ $=$$0\text{ a.s.}$, $\liminf_{K\to\infty}$ $\E\|\nabla F(x_{i,K\!+\!1})\|^2$$=$$0$ for any $i\in\mathcal{V}$. Therefore, the almost sure and mean square convergence of Algorithm \ref{algorithm1} with \emph{Scheme (S1)} is proved. 

\indent{\bf Case 2.} If Assumptions \ref{asm1}, \ref{asm2}, \ref{asm5} holds under \emph{Scheme (S2)}, then the proof of the almost sure and mean square convergence of Algorithm \ref{algorithm1} is given in the following three steps:

{\bf Step 1:} First, for any $k=0,\dots,K$, $K=0,1,\dots$, let vectors $\textbf{V}_k,\textbf{u}_k,\textbf{b}$ and the matrix $\textbf{M}_K$ defined as follows:
\begin{align*}
	&\textbf{u}_k=[u_k^{(1)},u_k^{(2)}]^\top,\textbf{b}\!=\![\textbf{b}_1,\textbf{b}_2]^\top\!=\![\frac{A_K^{(13)}}{2L},\frac{A_K^{(23)}}{2L}]^\top,\cr
	&\textbf{V}_k=\left[\begin{matrix}
		\E\|(W_1\!\otimes\! I_d) x_k\|^2\\
		\E\|(W_2\!\otimes\! I_d)y_k\|^2
	\end{matrix}\right],\textbf{M}_K=\left[\begin{matrix}
		A_K^{(11)}\!\!\!&A_K^{(12)}\\
		A_K^{(21)}\!\!\!&A_K^{(22)}
	\end{matrix}\right].
\end{align*}Then, in this step, we give the upper bound of $\sum_{k=0}^{K+1}\textbf{V}_k$. By \eqref{a6} and \eqref{a7}, we have
\begin{align}\label{t1c2,3}
	\textbf{V}_{k+1}\leq \textbf{M}_K\textbf{V}_k+\textbf{b}\E\|\nabla F(\bar{x}_k)\|^2\!+\!\textbf{u}_k.
\end{align}Iteratively computing \eqref{t1c2,3} results in $\textbf{V}_{k+1}\leq\textbf{M}_K^{k+1}\textbf{V}_0+\sum_{l=0}^k\textbf{M}_K^{k-l}(\textbf{b}\E\|\nabla F(\bar{x}_l)\|^2\!+\!\textbf{u}_l)$. Thus, summing the inequality above from 0 to $K+1$ gives
\begin{align}\label{t1c2,4}
	\sum_{k=0}^{K+1}\textbf{V}_{k}\leq&(\sum_{k=0}^{K+1}\textbf{M}_K^k)\textbf{V}_0+\sum_{k=0}^K\sum_{l=0}^k\textbf{M}_K^{k-l}(\textbf{b}\E\|\nabla F(\bar{x}_l)\|^2\!+\!\textbf{u}_l)\notag\\
	\noalign{\vskip -3pt}
	\leq&(\sum_{k=0}^{\infty}\textbf{M}_K^k)(\textbf{V}_0+\sum_{k=0}^K(\textbf{b}\E\|\nabla F(\bar{x}_k)\|^2\!+\!\textbf{u}_k)).
\end{align}
Let $\tilde{\textbf{s}}$$=$$[\tilde{s}_1$,$\tilde{s}_2]^\top\!$$=$$[\frac{1}{L_1^2}$,$\frac{(v_1^\top v_2)^2}{3\|v_1\|^2}]^\top$. Note that by Assumption \ref{asm5},\vspace{-0.5em}
\begin{align*}
	&0<\beta<\min\{\min_{i\in\mathcal{V}}\{\frac{1}{\sum_{j\in\mathcal{N}_{\mathcal{C},i}^{+}}\hspace{-0.5em}\mathcal{C}_{ji}}\},\min_{l=2,\dots,n}\{\frac{\text{Re}(\varpi_l^{(2)})}{1+|\varpi_l^{(2)}|^2}\}\},\cr
	&0<\alpha<\min\{\min_{i\in\mathcal{V}}\{\frac{1}{\sum_{\hspace{-0.15em}j\in\mathcal{N}_{\mathcal{R},i}^{-}}\hspace{-0.65em}\mathcal{R}_{ij}}\}, \min_{l=2,\dots,n}\{\frac{\text{Re}(\varpi_l^{(1)})}{1+|\varpi_l^{(1)}|^2}\},\cr
	&~~~~~~~~~~~~~~~~\frac{\sqrt{2}(v_1^\top v_2)r_2\beta}{12\rho(\mathcal{L}_1)\|v_1\|L_1}\},\cr &0<\gamma<\min\{\frac{r_1\alpha}{2\|v_2\|L_1}\sqrt{\frac{\mu}{12L_1+2\mu}+\frac{\mathbb{I}_{\{\mu=0\}}}{2}}\},\cr
	&~~~~~~~~~~~~~~~~\frac{\sqrt{6}(v_1^\top v_2)r_2\beta}{12\|v_1\|\|v_2\|L_1}\sqrt{\frac{\mu}{36L_1+7\mu}+\frac{\mathbb{I}_{\{\mu=0\}}}{7}}\}.
\end{align*} Then, we have $\textbf{M}_K\tilde{\textbf{s}}<\tilde{\textbf{s}}$. By Lemma \ref{lemma a2}(i), $\rho(\textbf{M}_K)<1$. Thus, by Gelfand formula (\!\!\cite[Cor. 5.6.16]{horn2012matrix}), $I_2-\textbf{M}_K$ is invertible and its inverse matrix is $(I_2-\textbf{M}_K)^{-1}=\sum_{k=0}^{\infty}\textbf{M}_K^k$. Hence, \eqref{t1c2,4} can be rewritten as\vspace{-0.5em}
\begin{align}\label{t1c2,5}
	\hspace{-1em}\sum_{k=0}^{K+1}\!\!\textbf{V}_{k}\!\leq\!(I_2\!-\!\textbf{M}_K)^{-1}(\textbf{V}_0\!+\!\sum_{k=0}^K(\textbf{b}\E\|\nabla F(\bar{x}_k)\|^2\!+\!\textbf{u}_k)).
\end{align}
\indent{\bf Step 2:} In this step, we prove that $\lim_{K\to\infty}$$\|\nabla F(\bar{x}_{K+1})\|^2$ $=$$0$ a.s., $\lim_{K\to\infty}\E\|\nabla F(\bar{x}_{K+1})\|^2=0$. Let $\textbf{c}=[\textbf{c}_1,\textbf{c}_2]^\top=[A_K^{(31)},A_K^{(32)}]^\top$. Then, by \eqref{a8} we have\vspace{-0.5em}
\begin{align}\label{t1c2,5.1}
	&(\frac{(v_1^{\!\top} v_2)\gamma_K}{2n}-\frac{2(v_1^{\!\top} v_2)^2\gamma_K^2L_1}{n^2})\sum_{k=0}^K\!\E\|\nabla F(\bar{x}_k)\|^2\cr
	\leq&\E (F(\bar{x}_0)\!-\!F(x^*))+\sum_{k=0}^{K}u_k^{(3)}+\textbf{c}^\top\sum_{k=0}^{K}\textbf{V}_k.
\end{align}
Substituting \eqref{t1c2,5} into \eqref{t1c2,5.1} implies\vspace{-0.5em}
\begin{align}\label{t1c2,6}
	&(\frac{(v_1^{\!\top} v_2)\gamma_K}{2n}-\frac{2(v_1^{\!\top} v_2)^2\gamma_K^2L_1}{n^2})\sum_{k=0}^K\!\E\|\nabla F(\bar{x}_k)\|^2\notag\\
	\leq&\E(F(\bar{x}_0)\!\!-\!\!F(x^*))+\sum_{k=0}^{K}u_k^{(3)}\notag\\
	\noalign{\vskip -4pt}
	&+\textbf{c}^\top(I_2\!-\!\textbf{M}_K)^{-1}\textbf{V}_0\!+\!\sum_{k=0}^K\textbf{c}^\top(I_2\!-\!\textbf{M}_K)^{-1}\textbf{u}_k\notag\\
	&+\textbf{c}^\top(I_2\!-\!\textbf{M}_K)^{-1}\textbf{b}\!\sum_{k=0}^K\E\|\nabla F(\bar{x}_k)\|^2.
\end{align}Rearranging \eqref{t1c2,6} gives\vspace{-0.5em}
\begin{align}\label{t1c2,8}
	&(\frac{(v_1^{\!\top} v_2)\gamma}{2n}\!-\!\frac{2(v_1^\top v_2)^2\gamma^2L_1}{n^2}\!-\!\textbf{c}^\top\!\!(I_2\!-\!\textbf{M}_K)^{-1}\textbf{b})\!\!\sum_{k=0}^K\!\E\|\nabla F(\bar{x}_k)\|^2\notag\\
	\leq&\E(F(\bar{x}_0)\!\!-\!\!F(x^*))\!+\!\sum_{k=0}^{K}u_k^{(3)}+\textbf{c}^\top(I_2\!-\!\textbf{M}_K)^{-1}\textbf{V}_0\notag\\
	\noalign{\vskip -4pt}
	&+\sum_{k=0}^K\textbf{c}^\top(I_2\!-\!\textbf{M}_K)^{-1}\textbf{u}_k.
	\end{align}
{\vskip -3pt}\noindent Note that by Assumption \ref{asm5},
\begin{align*}
	&0<\gamma<\min\{\frac{r_1\alpha}{2\|v_2\|L_1}\sqrt{\frac{\mu}{12L_1+2\mu}+\frac{\mathbb{I}_{\{\mu=0\}}}{2}},\frac{n\sqrt{3n}r_1\alpha}{24\|v_2\|L_1},\cr
	&~~~~~~~~~~~~~~~~\frac{\sqrt{3}(v_1^\top v_2)r_2\beta}{36\|v_1\|\|v_2\|L_1},\frac{\sqrt{6}(v_1^\top v_2)r_1r_2\beta}{144\rho(\mathcal{L}_1)\|v_1\|\|v_2\|L_1}\}.
\end{align*}
Then, we have
\begin{align}\label{t1c2,9}
	\det(I_2\!-\!\textbf{M}_K)=&(1-A_K^{(11)})(1-A_K^{(22)})-A_K^{(12)}A_K^{(21)}\cr
	>&\frac{5}{6}r_1r_2\alpha\beta.
\end{align}

Moreover, note that by Assumption \ref{asm5},
\begin{align*}
	&0<\gamma<\min\{1,\frac{n\sqrt{3n}r_1\alpha}{24\|v_2\|L_1},\frac{\sqrt{3}(v_1^\top v_2)r_2\beta}{36\|v_1\|\|v_2\|L_1},\cr
	&~~~~~~~~~~~~~~~~\frac{\sqrt{6}(v_1^\top v_2)r_1r_2\beta}{144\rho(\mathcal{L}_1)\|v_1\|\|v_2\|L_1},\frac{\sqrt{3}r_2\beta}{6nL_1}\}.
\end{align*}
Then, by \eqref{t1c2,9}, we have
\begin{align}\label{t1c2,10}
	\textbf{c}^\top\!(I_2\!-\!\textbf{M}_K)^{-1}\textbf{b}=&\frac{1}{\det(I_2\!-\!\textbf{M}_K)}(\textbf{c}_1\textbf{b}_1(1\!-\!A_K^{(22)})\!+\!\textbf{c}_2\textbf{b}_1A_K^{(21)}\notag\\
	&+\textbf{c}_1\textbf{b}_2A_K^{(12)}\!+\!\textbf{c}_2\textbf{b}_2(1\!-\!A_K^{(11)}))\notag\\
	<&\frac{2(v_1^\top v_2)}{5n}\gamma.
\end{align}
By $\gamma$$<$$\frac{n}{20(v_1^\top v_2)L_1}$ in Assumption \ref{asm5}, we have $\frac{2(v_1^\top v_2)^2\gamma^2L_1}{n^2}$$<$ $\frac{(v_1^{\!\top} v_2)\gamma}{10n}$. Thus, combining this inequality and \eqref{t1c2,10} leads to
\begin{align*}
	&\frac{(v_1^{\!\top} v_2)\gamma}{2n}-\frac{2(v_1^\top v_2)^2\gamma^2L_1}{n^2}-\textbf{c}^\top(I_2-\textbf{M}_K)^{-1}\textbf{b}\cr
	>&\frac{(v_1^{\!\top}v_2)\gamma}{2n}-\frac{(v_1^{\!\top}v_2)\gamma}{10n}-\frac{2(v_1^{\!\top}v_2)\gamma}{5n}\cr
	=&0.
\end{align*} 

\noindent Since $m_K=\lfloor p_m^K\rfloor+1$ and the definition of $\textbf{u}_k$, there exists $G_5>0$ such that for any $K$$=$$0,1,\dots$, $\E(F(\bar{x}_0)\!\!-\!\!F(x^*))\!+\!\sum_{k=0}^{K}u_k^{(3)}$ $+$$\textbf{c}^\top\!(I_2\!-\!\textbf{M}_K)^{-1}\textbf{V}_0$$+$$\sum_{k=0}^K\textbf{c}^\top\!(I_2\!-\!\textbf{M}_K)^{-1}$ $\textbf{u}_k\!\leq\! G_5$. Then, for any $K=0,1,\dots$, by \eqref{t1c2,8} we have
\begin{align}\label{t1c2,11}
	&\sum_{k=0}^K\!\E\|\nabla F(\bar{x}_k)\|^2\cr
	\leq&\frac{G_5}{\frac{(v_1^{\!\top}v_2)\gamma}{2n}\!\!-\!\!\frac{2(v_1^\top v_2)^2\gamma^2L_1}{n^2}\!\!-\!\textbf{c}^\top\!(I_2\!\!-\!\!\textbf{M}_K)^{-1}\textbf{b}}.
\end{align}Since step-sizes $\alpha_K=\alpha,\beta_K=\beta,\gamma_K=\gamma$ are constants under \emph{Scheme (S2)}, the matrix $\textbf{M}_K$ is a constant matrix. Then, \eqref{t1c2,11} is uniformly bounded for any $K=0,1,\dots$, and thus, $\lim_{K\to\infty}\E\|\nabla F(\bar{x}_{K+1})\|^2$$=$ $\lim_{K\to\infty}\E\|\nabla F(\bar{x}_K)\|^2$$=$$0$.

By the monotone convergence theorem (\!\!\cite[Th. 4.2.2(i)]{chow2012probability}), we have
\begin{align}\label{t1c2,11.1}
	&\E\sum_{K=0}^{\infty}\|\nabla F(\bar{x}_K)\|^2\cr
	\noalign{\vskip -5pt}
	\leq&\frac{G_5}{\frac{(v_1^{\!\top}v_2)\gamma}{2n}\!\!-\!\!\frac{2(v_1^\top v_2)^2\gamma^2L_1}{n^2}\!-\!\textbf{c}^\top\!(I_2\!\!-\!\!\textbf{M}_K)^{-1}\textbf{b}}.
\end{align}Then, \eqref{t1c2,11.1} implies $\sum_{K=0}^{\infty}\|\nabla F(\bar{x}_K)\|^2<\infty$ a.s., and thus, $\lim_{K\to\infty}$$\|\nabla F(\bar{x}_{K\!+\!1})\|^2$$=$$0$, a.s..

\indent{\bf Step 3:} In this step, we prove that $\lim_{K\to\infty}$$\|\nabla F(x_{i,K\!+\!1})\|^2$ $=$$0$ a.s., $\lim_{K\to\infty}$$\E\|\nabla F(x_{i,K+1})\|^2$$=$$0$ for any $i\in\mathcal{V}$. By \eqref{t1c2,5} and \eqref{t1c2,11}, the following inequality holds for any $K=0,1,\dots$:
\begin{align}\label{t1c2,12}
	&\sum_{k=0}^{K+1}\E\|(W_1\!\!\otimes\!\! I_d) x_k\|^2\cr
	\leq&\frac{\textbf{c}^\top\!(I_2\!-\!\textbf{M}_K)^{-1}(\textbf{V}_0\!+\!\sum_{k=0}^K(\textbf{b}\E\|\nabla F(\bar{x}_k)\|^2\!+\!\textbf{u}_k))}{\textbf{c}_1}\cr
	\leq&\frac{(\frac{(v_1^{\!\top} v_2)\gamma}{2n}\!-\!\frac{2(v_1^\top v_2)^2\gamma^2L_1}{n^2})G_5}{\textbf{c}_1(\frac{(v_1^{\!\top} v_2)\gamma}{2n}\!\!-\!\!\frac{2(v_1^\top v_2)^2\gamma^2L_1}{n^2}\!\!-\!\textbf{c}^\top\!(I_2\!-\!\textbf{M}_K)^{-1}\textbf{b})}.
\end{align}Since the matrix $\textbf{M}_K$ is a constant matrix, \eqref{t1c2,12} is uniformly bounded for any $K=0,1,\dots$. Then, we have $\lim_{K\to\infty}\E\|(W_1\otimes\! I_d) x_{K+1}\|^2$$=$$0$. 

By \eqref{t1c2,12} and the monotone convergence theorem, we have
\begin{align}\label{t1c2,13}
	&\E\sum_{K=0}^{\infty}\|(W_1\!\otimes\! I_d) x_K\|^2\cr
	\noalign{\vskip -5pt}
	\hspace{-2em}\leq&\frac{(\frac{(v_1^{\!\top} v_2)\gamma}{2n}\!-\!\frac{2(v_1^\top v_2)^2\gamma^2L_1}{n^2})G_5}{\textbf{c}_1(\frac{(v_1^{\!\top} v_2)\gamma}{2n}\!\!-\!\!\frac{2(v_1^\top v_2)^2\gamma^2L_1}{n^2}\!\!-\!\textbf{c}^\top\!(I_2\!-\!\textbf{M}_K)^{-1}\textbf{b})}.
\end{align}Then, \eqref{t1c2,13} implies $\sum_{K=0}^{\infty}\!\|(W_1\!\otimes\! I_d) x_K\|^2<\infty$ a.s., and thus, $\lim_{K\to\infty}\|(W_1\!\otimes\! I_d) x_{K\!+\!1}\|^2$$=$$0$, a.s.. Therefore, by \eqref{s4,1}, the almost sure and mean square convergence of Algorithm \ref{algorithm1} with \emph{Scheme (S2)} is proved.$\hfill\blacksquare$
\section{Proof of Theorem \ref{thm2}}\label{appendix c}
Let $0$$<$$\Gamma$$<$$1$ and $\omega_K$$=$$\Gamma\min\{r_1\alpha_K,r_2\beta_K,\frac{(v_1^\top v_2)\mu\gamma_K}{n}\}$. Then, the following four steps are given to prove Theorem~\ref{thm2}.

\indent{\bf Step 1:} First, we prove that there exists a positive integer $K_0$ such that for any $K=K_0,K_0+1,\dots$,\vspace{-0.7em}
\begin{align}\label{42}
	\rho(A_K)\leq1-\omega_K.
\end{align}{\vskip -5pt}\noindent Since $2p_\alpha\!\!-\!p_\beta\!\!-\!2\max\{\max_{i\in\mathcal{V}}\{p_{\zeta_i}\},\!0\}\!\geq\!1$, $\frac{1}{2}<p_\beta< p_\alpha< p_\gamma<1$, and $2p_\gamma-p_\alpha\geq1$ in Assumption \ref{asm4}, there exists a positive vector $\tilde{u}\in\mathbb{R}^3$ and a positive integer $K_0$ such that for any $K=K_0,K_0+1,\dots$, the following inequality holds:\vspace{-0.5em}
\begin{align}\label{44b}
	A_K\tilde{u}\leq (1-\omega_K)\tilde{u}.
\end{align}
{\vskip -5pt}\noindent Then, by \eqref{44b} and Lemma \ref{lemma a2}(i), \eqref{42} holds for any $K=K_0,K_0+1,\dots$.

{\bf Step 2:} In this step, we prove that there exists a positive vector $\tilde{t}=[\tilde{t}_1,\tilde{t}_2,\tilde{t}_3]^\top$ such that for any $K=0,1,\dots$, $\E(\tilde{t}^\top V_{K+1})=O(\frac{1}{(K+1)^{\theta-p_\gamma}})$. Note that for any $K=K_0,K_0+1,\dots$, \eqref{42} holds. Then, by Lemma \ref{lemma a2}(ii), there exists a positive vector $\tilde{t}=[\tilde{t}_1,\tilde{t}_2,\tilde{t}_3]^\top$ such that $\tilde{t}^\top A_K=\rho(A_K)\tilde{t}^\top\leq(1-\omega_K)\tilde{t}^\top$. Moreover, by Assumptions \ref{asm1}-\ref{asm4}, \eqref{41b} in Lemma \ref{lemma a10} holds. Then, multiplying $\tilde{t}^\top$ on both sides of \eqref{41b} implies that for any $k=0,\dots,K$,\vspace{-0.3em}
\begin{align}\label{46b}
	\E(\tilde{t}^\top V_{k+1})\leq&\tilde{t}^\top A_K\E V_k + \tilde{t}^\top u_k\cr
	\leq&(1-\omega_K)\E(\tilde{t}^\top V_k)+\tilde{t}^\top u_k.
\end{align}
{\vskip -3pt}\noindent By Assumption \ref{asm4}, $\tilde{t}^\top u_k$$=$$O(\frac{(a_4+1)(\rho(\mathcal{R})^2+\rho(\mathcal{C})^2+1)}{a_4(K+1)^\theta})$ holds for any $k=0,\dots$, $K$. Thus, iteratively computing \eqref{46b} results~in\vspace{-0.6em}
\begin{align}\label{48b}
	&\E(\tilde{t}^\top V_{K\!+\!1})\notag\\
	\hspace{-2em}=&(1\!\!-\!\!\omega_K)^{K\!+\!1}\E(\tilde{t}^{\top}\!V_0)\cr
	&+O(\sum_{k=0}^{K}\!(1\!-\!\omega_K)^k\!\frac{(a_4\!+\!1)(\rho(\mathcal{R})^2\!+\!\rho(\mathcal{C})^2\!+\!1)}{a_4(K\!\!+\!\!1)^\theta})\notag\\
	\hspace{-2em}=&(1\!\!-\!\!\omega_K)^{K\!+\!1}\E(\tilde{t}^{\top}\!V_0)\!+\!O(\frac{(a_4\!+\!1)(\rho(\mathcal{R})^2\!+\!\rho(\mathcal{C})^2\!+\!1)}{a_4\omega_K (K\!\!+\!\!1)^\theta}).
\end{align}
{\vskip -5pt}\noindent By  the definition of $\omega_K$, it can be seen that\vspace{-0.4em}
\begin{align}\label{49b}
	&O(\frac{1}{\omega_K (K\!+\!1)^\theta})=O(\frac{1}{(K\!+\!1)^{\theta-p_\gamma}}),\cr
	&(1-\omega_K)^{K+1}=\exp\left((K+1)\ln(1-\omega_K)\right)\cr
	&\leq\exp(-(K\!+\!1)\omega_K)=\exp(-O\left((K\!+\!1)^{1-p_\gamma}\right))\cr
	&=o(\frac{1}{(K\!+\!1)^{\theta-p_\gamma}}).
\end{align}
{\vskip -4pt}\noindent By \eqref{49b}, we have $\E$$($$\tilde{t}^\top$$ V_{K\!+\!1}$$)$$=$$O(\frac{1}{(K+1)^{\theta-p_\gamma}})$ for any $K$$=$$K_0$, $K_0$$+$$1$, $\dots$. Thus, there exists $S_0>0$ such that $\E(\tilde{t}^{\top}\!V_{K+1})$$\leq$$\frac{S_0}{(K+1)^{\theta-p_\gamma}}$. Let $S=\max\{\E(\tilde{t}^{\top}V_1)$, $2^{\theta-p_\gamma}\E(\tilde{t}^{\top}V_2),\dots,(K_0$$-$$1)^{\theta-p_\gamma}$$\E(\tilde{t}^{\top}$$V_{K_0\!-\!1})$,$S_0\}$. Then, for any $K=0,1,\dots$, we have $\E(\tilde{t}^\top V_{K+1})$$\leq$$\frac{S}{(K+1)^{\theta-p_\gamma}}$, which leads to\vspace{-0.7em}
\begin{align}\label{50b}
	\E(\tilde{t}^\top V_{K\!+\!1})=&O(\frac{(a_4\!+\!1)(\rho(\mathcal{R})^2\!+\!\rho(\mathcal{C})^2\!+\!1)}{a_4(K\!+\!1)^{\theta-p_\gamma}})\cr
	\noalign{\vskip -2pt}
	=&O(\frac{1}{(K\!+\!1)^{\theta-p_\gamma}}).
\end{align}{\vskip -6pt}

{\bf Step 3:} In this step, we prove that for any $i\in\mathcal{V}$ and $K$$=$$0,1,\dots$, $\E\|\nabla F(x_{i,K\!+\!1})\|^2$$=$$O(\frac{1}{(K\!+\!1)^{\theta-p_\gamma}})$. By Lemma \ref{lemma a1}(i), we have\vspace{-0.4em}
\begin{align}\label{51b}
	&F(x_{i,K+1})\!\!-\!\!F(\bar{x}_{K\!+\!1})\cr
	\noalign{\vskip -4pt}
	\leq&\left\langle\nabla F(\bar{x}_{K\!+\!1}) ,x_{i,{K\!+\!1}}\!-\!\bar{x}_{K\!+\!1} \right\rangle\!+\!\frac{L_1}{2}\|\bar{x}_{K\!+\!1}-x_{i,K\!+\!1}\|^2.
\end{align}
{\vskip -5pt}\noindent Note that for any $\mathbf{a},\mathbf{b}\in\mathbb{R}^d$, $\langle \mathbf{a},\mathbf{b}\rangle\leq\frac{\|\mathbf{a}\|^2+\|\mathbf{b}\|^2}{2}$ holds. Then, \eqref{51b} can be rewritten as\vspace{-0.5em}
\begin{align}\label{52}
	&F(x_{i,K\!+\!1})-F(\bar{x}_{K\!+\!1})\cr
	\leq&\frac{\|\nabla F(\bar{x}_{K\!+\!1})\|^2+\|\bar{x}_{K\!+\!1}\!-\!x_{i,K\!+\!1}\|^2}{2}\!+\!\frac{L_1}{2}\|\bar{x}_{K\!+\!1}\!-\!x_{i,K\!+\!1}\|^2\cr
	=&\frac{L_1+1}{2}\|\bar{x}_{K\!+\!1}-x_{i,K\!+\!1}\|^2+\frac{\|\nabla F(\bar{x}_{K\!+\!1})\|^2}{2}.
\end{align}
{\vskip -3pt}\noindent By Lemma \ref{lemma a1}(ii), $\|\nabla F(\bar{x}_{K\!+\!1})\|^2\!\leq\!2L_1(F(\bar{x}_{K\!+\!1})-F(x^*))$. Substituting it into \eqref{52} gives $F(x_{i,K\!+\!1})\!-\!F(\bar{x}_{K\!+\!1})\!\leq\!\frac{L_1\!+\!1}{2}$ $\|\bar{x}_{K\!+\!1}-x_{i,K\!+\!1}\|^2\!+\!L_1(F(\bar{x}_{K\!+\!1})\!-\!F(x^*))$. Thus, we have\vspace{-0.5em}
\begin{align}\label{53b}
	&F(x_{i,K\!+\!1})-F(\bar{x}_{K\!+\!1})\cr
	\leq&\frac{L_1+1}{2}\sum_{i=1}^{n}\|\bar{x}_{K\!+\!1}-x_{i,K\!+\!1}\|^2+L_1(F(\bar{x}_{K\!+\!1})-F(x^*))\cr
	=&\frac{L_1+1}{2}\|(W_1\otimes I_d)x_{K\!+\!1}\|^2+L_1(F(\bar{x}_{K\!+\!1})-F(x^*)).
\end{align}{\vskip -3pt}\noindent Then, by \eqref{53b} it can be seen that\vspace{-0.5em}
\begin{align}\label{54b}
	&F(x_{i,K\!+\!1})-F(x^*)\cr
	=&\left(F(x_{i,K\!+\!1})-F(\bar{x}_{K\!+\!1})\right)+\left(F(\bar{x}_{K\!+\!1})-F(x^*)\right)\cr
	\leq&(L_1+1)\left(\mathbf{1}_3^\top \E V_{K+1}\right)=O\left(\E(\tilde{t}^\top V_{K\!+\!1})\right).
\end{align}
{\vskip -5pt}\noindent Thus, combining \eqref{50b} and \eqref{54b} gives $\E(F(x_{i,K+1})-F(x^*))=O\left(\frac{(a_4+1)(\rho(\mathcal{R})^2\!+\!\rho(\mathcal{C})^2\!+\!1)}{a_4(K+1)^{\theta-p_\gamma}}\right)$. By Lemma \ref{lemma a1}(ii), we have\vspace{-0.5em}
\begin{align}\label{54c}
	\E\|\nabla F(x_{i,K\!+\!1})\|^2\leq& 2L_1\E(F(x_{i,K+1})-F(x^*))\cr
	\noalign{\vskip -2pt}
	=&O(\frac{(a_4+1)(\rho(\mathcal{R})^2\!+\!\rho(\mathcal{C})^2\!+\!1)}{a_4(K+1)^{\theta-p_\gamma}})\cr
	\noalign{\vskip -3pt}
	=&O(\frac{1}{(K+1)^{\theta-p_\gamma}}).
\end{align}{\vskip -5pt}\noindent Hence, the polynomial mean square convergence rate \mbox{is achieved.}

{\bf Step 4:} In this step, we prove that the oracle complexity of Algorithm \ref{algorithm1} with \emph{Scheme (S1)} is $O(\varphi^{-\frac{165(1+\varphi)}{\max\{55-27\varphi,46\}}})$ for any $\varphi$$>$$0$. Let $p_\alpha=\max\{1$$-$$\frac{\varphi}{5}, \frac{9}{10}\}$, $p_\beta=\max\{\frac{2}{3}(1$$-$$\frac{5\varphi}{5}),\frac{3}{5}\}$, $p_\gamma=\max\{1$$-$$\frac{\varphi}{10},\frac{19}{20}\}$, $p_m=\max\{2-\frac{\varphi}{10}$,$\frac{39}{20}\}$, $p_{\zeta_i}=p_{\eta_i}=$ $\max\{\frac{\varphi}{10},\frac{1}{20}\}$. Then, by {\bf Step~3}, $\E\|\nabla F(x_{i,K\!+\!1})\|^2=O(\frac{1}{(K\!+\!1)^{\frac{\max\{20-22\varphi,9\}}{60}}})$ for any $i$$\in$$\mathcal{V}$ and $K$$=$$0,1,\dots$. Thus, there exists $\Phi>0$ such that the following inequality holds:\vspace{-0.3em}
\begin{align}\label{3t1}
	\E\|\nabla F(x_{i,K+1})\|^2\leq\frac{\Phi}{(K+1)^{\frac{\max\{20-22\varphi,9\}}{60}}}.
\end{align} 
{\vskip -5pt}\noindent Let $K=\lfloor(\frac{\Phi}{\varphi})^{\frac{60}{\max\{20-22\varphi,9\}}}\rfloor$. Then, by \eqref{3t1} we have\vspace{-0.7em}
\begin{align}\label{3t2}
	\hspace{-1em}\E\|\nabla F(x_{i,K\!+\!1})\|^2\!\leq\!\frac{\Phi}{(\frac{\Phi}{\varphi})^{\frac{60}{\max\{20-22\varphi,9\}}\frac{\max\{20-22\varphi,9\}}{60}}}\!=\!\varphi.
\end{align}
{\vskip -3pt}\noindent Thus, by \eqref{3t2} and Definition \ref{def1}, $x_{K+1}$ is a $\varphi$-suboptimal solution. Since $N(\varphi)$ is the smallest integer such that $x_{N(\varphi)}$ is a $\varphi$-suboptimal solution, we have\vspace{-0.6em}
\begin{align}\label{3t3}
	N(\varphi)\leq\lfloor(\frac{\Phi}{\varphi})^{\frac{60}{\max\{20-22\varphi,9\}}}\rfloor+1.
\end{align}
{\vskip -6pt}\noindent Since $m_K$$=$$\lfloor a_4 K^{p_m}\rfloor$$+$$1$$=$$\lfloor a_4 \lfloor(\frac{\Phi}{\varphi})^{\frac{60}{\max\{20-22\varphi,9\}}}\rfloor^{p_m}\rfloor$$+$$1$, by Definition \ref{def2} and \eqref{3t3}, the oracle complexity of Algorithm~\ref{algorithm1} with \emph{Scheme (S1)} is given as follows:\vspace{-0.6em}
\begin{align*}
	&\sum_{k=0}^{N(\varphi)}m_K\cr
	\noalign{\vskip -7pt}
	=&(N(\varphi)+1)(\lfloor a_4 \lfloor(\frac{\Phi}{\varphi})^{\frac{60}{\max\{20-22\varphi,9\}}}\rfloor^{p_m}\rfloor+1)\cr
	\noalign{\vskip -8pt}
	\leq&(\lfloor(\frac{\Phi}{\varphi})^{\frac{60}{\max\{20-22\varphi,9\}}}\rfloor+2)(a_4 \lfloor(\frac{\Phi}{\varphi})^{\frac{60}{\max\{20-22\varphi,9\}}}\rfloor^{p_m}+1)\cr
	=&O(\varphi^{-\frac{177+3\max\{1-2\varphi,0\}}{9-11\max\{1-2\varphi,0\}}}).
\end{align*}
{\vskip -5pt}\noindent Therefore, this theorem is proved. \hfill$\blacksquare$

\section{Proof of Theorem \ref{thm3}}\label{appendix e}
The following two steps are given to prove Theorem \ref{thm3}.

{\bf Step 1:} First, we prove that Algorithm \ref{algorithm1} with \emph{Scheme (S2)} achieves the exponential mean square convergence rate. Let $\tilde{s}=[\frac{1}{L_1^2},\frac{(v_1^\top v_2)^2}{3\|v_1\|^2},\frac{3}{\mu}]^\top$. Then, by Assumption \ref{asm5}, we have $A_K\tilde{s}<\tilde{s}$. Thus, by Lemma \ref{lemma a2}(i), $\rho(A_K)$$<$$1$. By Lemma \ref{lemma a2}(ii), there exists a positive vector $\tilde{\textbf{r}}$$=$$[\tilde{\textbf{r}}_1,\tilde{\textbf{r}}_2,\tilde{\textbf{r}}_3]^\top$ such that $\tilde{\textbf{r}}^\top A_K$$=$$\rho(A_K)\tilde{\textbf{r}}^\top$. By Assumptions \ref{asm1}-\ref{asm3}, \ref{asm5}, \eqref{41b} in Lemma \ref{lemma a10} holds. Then, multiplying $\tilde{\textbf{r}}^\top$ on both sides of \eqref{41b} implies that for any $k=0,\dots,K$,\vspace{-0.5em}
\begin{align}\label{55}
	\E(\tilde{\textbf{r}}^\top\! V_{k\!+\!1})\leq&\tilde{\textbf{r}}^\top\! A_K\E V_k \!+\! \tilde{\textbf{r}}^\top\! u_k\cr
	=&\rho(A_K)\E(\tilde{\textbf{r}}^\top V_k)\!+\!\tilde{\textbf{r}}^\top u_k.
\end{align}
{\vskip -7pt}\noindent Iteratively computing \eqref{55} gives\vspace{-0.4em}
\begin{align}\label{56}
	\E\tilde{\textbf{r}}^\top V_{K+1}\leq&\rho(A_K)^{K+1}\E\tilde{\textbf{r}}^\top V_{K}\cr
	&+\sum_{k=0}^{K}\rho(A_K)^{K-k}\tilde{\textbf{r}}^\top u_k.
\end{align}\noindent By Assumption \ref{asm5}, $\tilde{\textbf{r}}^\top u_k$$=$$O((p_m^{-K}$$+$$p_{\zeta_i}^K$$+$$p_{\eta_i}^K)(\rho(\mathcal{R})^2\!+\!\rho(\mathcal{C})^2\!+\!1))$ for any $k=0,\dots,K$. Then, \eqref{56} can be rewritten~as\vspace{-0.7em}
\begin{align}\label{57}
&\E\tilde{\textbf{r}}^\top V_{K+1}\notag\\
=&\rho(A_K)^{K+1}\E\tilde{\textbf{r}}^\top V_{0}\notag\\
&+O((\rho(\mathcal{R})^2\!+\!\rho(\mathcal{C})^2\!+\!1)\max\{\rho(A_K),\frac{1}{p_m},p_{\zeta_i},p_{\eta_i}\}^K)\notag\\
\noalign{\vskip -2pt}
=&O(\max\{\rho(A_K),\frac{1}{p_m},p_{\zeta_i},p_{\eta_i}\}^K).
\end{align}
{\vskip -4pt}\noindent By {\bf Step 3} of Appendix \ref{appendix c}, $F(x_{i,K+1})-F(x^*)=O\left(\E(\tilde{\textbf{r}}^\top V_{K\!+\!1})\right)$. Then by \eqref{57} and Lemma \ref{lemma a1}(ii), the exponential mean convergence rate of Algorithm \ref{algorithm1} is achieved.

{\bf Step 2:} Next, we prove that the oracle complexity of Algorithm \ref{algorithm1} with \emph{Scheme (S2)} is $O(\frac{1}{\varphi}\ln\frac{1}{\varphi})$. For any $\varphi$$>$$0$, let $\beta$$=$$\min\{\frac{1}{2},\frac{n}{40(v_1^\top v_2)L},\min_{i\in\mathcal{V}}$$\{\!\frac{1}{2\sum_{j\in\mathcal{N}_{\mathcal{C},i}^{+}}\!\!\!\!\mathcal{C}_{ji}}\!\},\min_{l=2,\dots,n}\{$ $\frac{\text{Re}(\varpi_l^{(2)})}{2+2|\varpi_l^{(2)}|^2}\}\}$, $\alpha=\min\{\beta,\min_{i\in\mathcal{V}}$$\{\!\frac{1}{2\sum_{j\in\mathcal{N}_{\mathcal{R},i}^{-}}\!\!\!\!\mathcal{R}_{i j}}\!\},\min_{l=2,\dots,n}\{$ $\frac{\text{Re}(\varpi_l^{(1)})}{2+2|\varpi_l^{(1)}|^2}\},\frac{\sqrt{2}(v_1^\top v_2)r_2\beta}{12\rho(\mathcal{L}_1)\|v_1\|L_1}$$\}$, $\gamma=$$\min\{\frac{1}{2},\frac{n}{40(v_1^{\!\top}v_2)L},\frac{Q_1\alpha}{2},\frac{Q_2\beta}{2}\}$, $p_m$$=$$\min\{\frac{1}{\varphi},\frac{1}{\rho(A_K)}\},p_{\zeta_i}$$=$$p_{\eta_i}$$=$$\min\{\varphi,\rho(A_K)\}$. Then, by Theorem \ref{thm2}, there exists $\Phi>0$ such that for any $i\in\mathcal{V}$, $K=0,1,\dots$,\vspace{-0.4em}
\begin{align}\label{4t1}
	\E\|\nabla F(x_{i,K+1})\|^2\!\leq\!\Phi \max\{\rho(A_K),\varphi\}^K.
\end{align}
{\vskip -3pt}\noindent Let $K$$=$$\lfloor\max\{\frac{\ln \varphi-\ln \Phi}{\ln \rho(A_K)},\frac{\ln \varphi-\ln \Phi}{\ln \varphi}\}\rfloor$$+$$1$. Then, by \eqref{4t1} we have $\E\|\nabla F(x_{i,K\!+\!1})\|^2<\varphi$. By Definition~\ref{def1}, $x_{K+1}$ is a $\varphi$-suboptimal solution. Thus, by the definition of $N(\varphi)$, we have\vspace{-0.3em}
\begin{align}\label{4t2}
	N(\varphi)\!\leq\!\lfloor\max\{\frac{\ln \varphi-\ln \Phi}{\ln \rho(A_K)},\frac{\ln \varphi-\ln \Phi}{\ln \varphi}\}\rfloor\!+\!2.
\end{align}
{\vskip -2pt}\noindent Since $m_K=\lfloor\min\{\frac{1}{\varphi},\frac{1}{\rho(A_K)}\}^K\rfloor+1$. Thus, by Definition \ref{def2} and \eqref{4t2}, the oracle complexity of Algorithm \ref{algorithm1} with \emph{Scheme (S2)} is given as follows:\vspace{-0.7em}
\begin{align*}
	&\sum_{k=0}^{N(\varphi)}m_K\cr
	\noalign{\vskip -5pt}
	=&(N(\varphi)\!+\!1)(\lfloor\min\{\frac{1}{\varphi},\!\frac{1}{\rho(A_K)}\}^{\lfloor\max\{\frac{\ln \varphi-\ln \Phi}{\ln \rho(A_K)},\frac{\ln \varphi-\ln \Phi}{\ln \varphi}\}\rfloor\!+\!1}\rfloor\!+\!1)\cr
	\leq&(\lfloor\max\{\frac{\ln \varphi-\ln \Phi}{\ln \rho(A_K)},\frac{\ln \varphi-\ln \Phi}{\ln \varphi}\}\rfloor\!+\!3)\cdot\cr
	&~~~~~~~(\min\{\frac{1}{\varphi},\!\frac{1}{\rho(A_K)}\}^{\max\{\frac{\ln \varphi-\ln \Phi}{\ln \rho(A_K)},\frac{\ln \varphi-\ln \Phi}{\ln \varphi}\}\!+\!1}\!+\!1)\cr
	=&O(\frac{|\ln\varphi|}{\varphi}).
\end{align*}{\vskip -5pt}\noindent Therefore, this theorem is proved. \hfill$\blacksquare$
\section{Proof of Lemma \ref{lemma4}}\label{appendix f}
Let
\begin{align*}
	\|\Delta x_{i,k}\|_1=&\begin{cases}
		\sup\limits_{\substack{\mathcal{O}\subseteq\mathbb{R}^{2nd},\\(\breve{x}_0,\breve{y}_0)\in\mathcal{O},\\\text{Adj}(\mathcal{D}_i,\mathcal{D}_i^\prime)}}\hspace{-0.3em}\|x_{i,0}-x_{i,0}^\prime\|_1,\hspace{-0.6em}&\text{if }k=0;\\
		\sup\limits_{\substack{\mathcal{O}\subseteq\mathbb{R}^{2nd},\\(\breve{x}_{k\!-\!1},\breve{y}_{k\!-\!1})\in\mathcal{O},\\\text{Adj}(\mathcal{D}_i,\mathcal{D}_i^\prime)}}\hspace{-1.5em}\|x_{i,k}-x_{i,k}^\prime\|_1,\hspace{-0.6em}&\text{if }k=1,\dots,K,
	\end{cases}\\
	\|\Delta y_{i,k}\|_1=&\begin{cases}
		\sup\limits_{\substack{\mathcal{O}\subseteq\mathbb{R}^{2nd},\\(\breve{x}_0,\breve{y}_0)\in\mathcal{O},\\\text{Adj}(\mathcal{D}_i,\mathcal{D}_i^\prime)}}\hspace{-0.3em}\|y_{i,0}-y_{i,0}^\prime\|_1,\hspace{-0.6em}&\text{if }k=0;\\
		\sup\limits_{\substack{\mathcal{O}\subseteq\mathbb{R}^{2nd},\\(\breve{x}_{k\!-\!1},\breve{y}_{k\!-\!1})\in\mathcal{O},\\\text{Adj}(\mathcal{D}_i,\mathcal{D}_i^\prime)}}\hspace{-1.5em}\|y_{i,k}-y_{i,k}^\prime\|_1,\hspace{-0.6em}&\text{if }k=1,\dots,K.
	\end{cases}
\end{align*}
Then, by Definition \ref{def5}, we have
\begin{align*}
	\Delta_{i,k}^{q}=\|\Delta x_{i,k}\|_1+\|\Delta y_{i,k}\|_1.
\end{align*}
Thus, the following two steps are given to prove Lemma \ref{lemma4}.

{\bf Step 1:} We compute $\|\Delta y_{i,k}\|_1$ for any $k=0,\dots,K$ and $i\in\mathcal{V}$. When $k=0$, we have\vspace{-0.4em}
\begin{align}\label{58}
	\|\Delta y_{i,0}\|_1=&\hspace{-0.2em}\sup_{\substack{\mathcal{O}\subseteq\mathbb{R}^{2nd},\\(\breve{x}_0,\breve{y}_0)\in\mathcal{O},\\\text{Adj}(\mathcal{D}_i,\mathcal{D}_i^\prime)}}\hspace{-0.3em}\|y_{i,0}\!-\!y_{i,0}^\prime\|_1=\sup_{\substack{\mathcal{O}\subseteq\mathbb{R}^{2nd},\\(\breve{x}_0,\breve{y}_0)\in\mathcal{O},\\\text{Adj}(\mathcal{D}_i,\mathcal{D}_i^\prime)}}\hspace{-0.3em}\|g_{i,0}\!-\!g_{i,0}^{\prime}\|_1\notag\\
	=&\sup_{\text{Adj}(\mathcal{D}_i,\mathcal{D}_i^\prime)}\|g_{i,0}-g_{i,0}^\prime\|_1.
\end{align}
Note that by Step 5 of Algorithm~\ref{algorithm1}, $m_K$ different data samples are taken uniformly from $\mathcal{D}_i$, $\mathcal{D}_i^\prime$, respectively. Then, there exists at most one pair of data samples $\lambda_{i,0,l_1},\lambda_{i,0,l_1}^\prime$ such that $\lambda_{i,0,l_1}=\xi_{i,l_0}$, $\lambda_{i,0,l_1}^\prime=\xi_{i,l_0}^\prime$. Thus, by \eqref{avg_sampled_grad}, \eqref{58} can be rewritten as\vspace{-0.3em}
\begin{align}\label{60}
	&\|\Delta y_{i,0}\|_1\cr
	\noalign{\vskip -2pt}
	=&\sup_{\text{Adj}(\mathcal{D}_i,\mathcal{D}_i^\prime)}\!\left\|\frac{1}{m_K}\!\sum_{l=1}^{m_K}(g_{i}(x_{i,0},\lambda_{i,0,l})\!-\!g_{i}(x_{i,0},\lambda_{i,0,l}^{\prime}))\right\|_1\cr
	=&\sup_{\text{Adj}(\mathcal{D}_i,\mathcal{D}_i^\prime)}\!\left\|\frac{1}{m_K}(g_{i}(x_{i,0},\lambda_{i,0,l_1})- g_{i}(x_{i,0},\lambda_{i,0,l_1}^{\prime}))\right\|_1\cr
	\noalign{\vskip -3pt}
	\leq&\frac{1}{m_K}\left\|g_{i}(x_{i,0},\xi_{i,l_0})-g_{i}(x_{i,0},\xi_{i,l_0}^{\prime})\right\|_1.
\end{align}As mentioned in Remark \ref{rmk6}, Assumption \ref{asm2}(i) ensures there exists a constant $C$ such that $\text{Adj}(\mathcal{D}_i,\mathcal{D}_i^\prime)$, and then, there exists exactly one pair of data samples $\xi_{i,l_0},\xi_{i,l_0}^{\prime}$ such that \eqref{eq3} holds. Thus, \eqref{60} can be rewritten as
\begin{align}\label{60.5}
	\|\Delta y_{i,0}\|_1\leq\frac{C}{m_K}.
\end{align}

{\vskip -3pt}\indent When $k=1$, we have\vspace{-0.4em}
\begin{align}\label{61}
	&\|\Delta y_{i,1}\|_1\cr
	\hspace{-1.5em}=&\hspace{-0.2em}\sup_{\substack{\mathcal{O}\subseteq\mathbb{R}^{2nd},\\(\breve{x}_0,\breve{y}_0)\in\mathcal{O},\\\text{Adj}(\mathcal{D}_i,\mathcal{D}_i^\prime)}}\hspace{-0.3em}\|y_{i,1}-y_{i,1}^\prime\|_1\cr
	\noalign{\vskip -5pt}
	\hspace{-1.5em}=&\hspace{-0.2em}\sup_{\substack{\mathcal{O}\subseteq\mathbb{R}^{2nd},\\(\breve{x}_0,\breve{y}_0)\in\mathcal{O},\\\text{Adj}(\mathcal{D}_i,\mathcal{D}_i^\prime)}}\hspace{-0.3em}\|(1\!-\!\beta_K\hspace{-0.5em}\sum_{j\in\mathcal{N}_{\mathcal{C},i}^{+}}\hspace{-0.5em}\mathcal{C}_{ji})(y_{i,0}-y_{i,0}^{\prime})\cr
	\noalign{\vskip -2pt}
	&-\beta_K\hspace{-0.8em}\sum_{j\in \mathcal{N}_{\mathcal{C},i}^{-}}\hspace{-0.8em}\mathcal{C}_{ji}(\breve{y}_{j,0}\!-\!\breve{y}_{j,0}^{\prime})\!\!+\!\!(g_{i,1}\!\!-g_{i,1}^{\prime})\!\!+\!\!(g_{i,0}\!-\!g_{i,0}^{\prime})\|_{\!1}.
\end{align}
{\vskip -3pt}\noindent By Definition \ref{def5}, $\breve{y}_{j,0}=z_2=\breve{y}_{j,0}^\prime$ holds for any $j\in\mathcal{N}_{\mathcal{C},i}^{+}$. Then, \eqref{61} can be rewritten as\vspace{-0.3em}
\begin{align}\label{61.1}
	\|\Delta y_{i,1}\|_1=&\hspace{-0.2em}\sup_{\substack{\mathcal{O}\subseteq\mathbb{R}^{2nd},\\(\breve{x}_0,\breve{y}_0)\in\mathcal{O},\\\text{Adj}(\mathcal{D}_i,\mathcal{D}_i^\prime)}}\hspace{-0.3em}\|(1\!-\!\beta_K\hspace{-0.5em}\sum_{j\in\mathcal{N}_{\mathcal{C},i}^{+}}\hspace{-0.5em}\mathcal{C}_{ji})(y_{i,0}-y_{i,0}^{\prime})\cr
	\noalign{\vskip -5pt}
	&+(g_{i,1}-g_{i,1}^{\prime})+(g_{i,0}-g_{i,0}^{\prime})\|_1.
\end{align}
{\vskip -5pt}\noindent Since $y_{i,0}=g_{i,0}$ and $y_{i,0}^\prime=g_{i,0}^\prime$ hold for any $i\in\mathcal{V}$, by \eqref{60.5}, \eqref{61.1} can be rewritten as\vspace{-0.2em}
\begin{align}\label{61.2}
	&\|\Delta y_{i,1}\|_1\cr
	\hspace{-1em}=&\hspace{-0.2em}\sup_{\text{Adj}(\mathcal{D}_i,\mathcal{D}_i^\prime)}\hspace{-0.3em}\|(1\!-\!\beta_K\hspace{-0.5em}\sum_{j\in\mathcal{N}_{\mathcal{C},i}^{+}}\hspace{-0.5em}\mathcal{C}_{ji})(y_{i,0}-y_{i,0}^{\prime})+(g_{i,1}-g_{i,1}^{\prime})\cr
	\noalign{\vskip -5pt}
	&+(g_{i,0}-g_{i,0}^{\prime})\|_1\cr
	\hspace{-1em}\leq&\hspace{-0.2em}\sup_{\text{Adj}(\mathcal{D}_i,\mathcal{D}_i^\prime)}\hspace{-0.3em}\|(1\!-\!\beta_K\hspace{-0.5em}\sum_{j\in\mathcal{N}_{\mathcal{C},i}^{+}}\hspace{-0.5em}\mathcal{C}_{ji})(y_{i,0}-y_{i,0}^{\prime})\|_1\cr
	&+\hspace{-0.2em}\sup_{\text{Adj}(\mathcal{D}_i,\mathcal{D}_i^\prime)}\hspace{-0.3em}\|g_{i,1}-g_{i,1}^{\prime}\|_1+\hspace{-0.2em}\sup_{\text{Adj}(\mathcal{D}_i,\mathcal{D}_i^\prime)}\hspace{-0.3em}\|g_{i,0}-g_{i,0}^{\prime}\|_1\cr
	\hspace{-1em}\leq&\hspace{-0.2em}\sup_{\text{Adj}(\mathcal{D}_i,\mathcal{D}_i^\prime)}\hspace{-0.3em}\|(1\!-\!\beta_K\hspace{-0.5em}\sum_{j\in\mathcal{N}_{\mathcal{C},i}^{+}}\hspace{-0.5em}\mathcal{C}_{ji})(y_{i,0}-y_{i,0}^{\prime})\|_1\cr
	&+\hspace{-0.2em}\sup_{\text{Adj}(\mathcal{D}_i,\mathcal{D}_i^\prime)}\hspace{-0.3em}\|g_{i,1}-g_{i,1}^{\prime}\|_1+\frac{C}{m_K}.
\end{align}
By $\text{Adj}(\mathcal{D}_i,\mathcal{D}_i^\prime)$, there exists at most one pair of data samples $\lambda_{i,1,l_2},\lambda_{i,1,l_2}^{\prime}$ such that $\lambda_{i,1,l_2}=\xi_{i,l_0}$, $\lambda_{i,1,l_2}^{\prime}=\xi_{i,l_0}^{\prime}$. Then by \eqref{60.5}, \eqref{61.2} can be rewritten as\vspace{-0.6em}
\begin{align}\label{62}
	\hspace{-0.56em}\|\Delta y_{i,1}\|_1\leq&\hspace{-0.2em}\sup_{\text{Adj}(\mathcal{D}_i,\mathcal{D}_i^\prime)}\hspace{-0.3em}\|(1\!-\!\beta_K\hspace{-0.9em}\sum_{j\in\mathcal{N}_{\mathcal{C},i}^{+}}\hspace{-0.75em}\mathcal{C}_{ji})(y_{i,0}\!-\!y_{i,0}^{\prime})\|_1\!+\!\frac{2C}{m_K}\notag\\
	\noalign{\vskip -8pt}
	=&|1-\beta_K\hspace{-0.5em}\sum_{j\in\mathcal{N}_{\mathcal{C},i}^{+}}\hspace{-0.5em}\mathcal{C}_{ji}|\|\Delta y_{i,0}\|_1+\frac{2C}{m_K}\cr
	\noalign{\vskip -5pt}
	\leq&|1-\beta_K\hspace{-0.5em}\sum_{j\in\mathcal{N}_{\mathcal{C},i}^{+}}\hspace{-0.5em}\mathcal{C}_{ji}|\frac{C}{m_K}+\frac{2C}{m_K}.
\end{align}

{\vskip -5pt}\indent When $k=2,\dots,K$, we have\vspace{-0.6em}
\begin{align}\label{63}
	\|\Delta y_{i,k}\|_1=&\hspace{-0.2em}\sup_{\substack{\mathcal{O}\subseteq\mathbb{R}^{2nd},\\(\breve{x}_{k\!-\!1},\breve{y}_{k\!-\!1})\in\mathcal{O},\\\text{Adj}(\mathcal{D}_i,\mathcal{D}_i^\prime)}}\hspace{-0.3em}\|y_{i,k}-y_{i,k}^\prime\|_1\cr
	\noalign{\vskip -5pt}
	=&\hspace{-0.2em}\sup_{\substack{\mathcal{O}\subseteq\mathbb{R}^{2nd},\\(\breve{x}_{k\!-\!1},\breve{y}_{k\!-\!1})\in\mathcal{O},\\\text{Adj}(\mathcal{D}_i,\mathcal{D}_i^\prime)}}\hspace{-0.3em}\|(1\!-\!\beta_K\hspace{-0.75em}\sum_{j\in\mathcal{N}_{\mathcal{C},i}^{+}}\hspace{-0.5em}\mathcal{C}_{ji})(y_{i,k\!-\!1}\!-\!y_{i,k\!-\!1}^{\prime})\cr
	\noalign{\vskip -2pt}
	&-\beta_K\hspace{-0.75em}\sum_{j\in \mathcal{N}_{\mathcal{C},i}^{-}}\hspace{-0.75em}\mathcal{C}_{ji}(\breve{y}_{j,k\!-\!1}\!-\!\breve{y}_{j,k\!-\!1}^{\prime})+(g_{i,k}-g_{i,k}^{\prime})\cr
	\noalign{\vskip -5pt}
	&+(g_{i,k-1}-g_{i,k-1}^{\prime})\|_1.
\end{align}
{\vskip -3pt}\noindent By Definition \ref{def5}, $\breve{y}_{j,k}=z_2=\breve{y}_{j,k}^\prime$ holds for any $j\in\mathcal{N}_{\mathcal{C},i}^{+}$. Then, \eqref{63} can be rewritten as
\begin{align}\label{63.1}
	&\|\Delta y_{i,k}\|_1\notag\\
	\hspace{-1em}=&\hspace{-0.2em}\sup_{\substack{\mathcal{O}\subseteq\mathbb{R}^{2nd},\\(\breve{x}_{k\!-\!1},\breve{y}_{k\!-\!1})\in\mathcal{O},\\\text{Adj}(\mathcal{D}_i,\mathcal{D}_i^\prime)}}\hspace{-0.3em}\|(1\!-\!\beta_K\hspace{-1em}\sum_{j\in\mathcal{N}_{\mathcal{C},i}^{+}}\hspace{-0.75em}\mathcal{C}_{ji})(y_{i,k\!-\!1}\!-\!y_{i,k\!-\!1}^{\prime})\notag\\
	&+(g_{i,k}-g_{i,k}^{\prime})+(g_{i,k\!-\!1}-g_{i,k\!-\!1}^{\prime})\|_1\cr
	\hspace{-2em}\leq&\hspace{-0.2em}\sup_{\text{Adj}(\mathcal{D}_i,\mathcal{D}_i^\prime)}\hspace{-0.5em}\|(1\!-\!\beta_K\hspace{-1em}\sum_{j\in\mathcal{N}_{\mathcal{C},i}^{+}}\hspace{-0.75em}\mathcal{C}_{ji})(y_{i,k\!-\!1}\!-\!y_{i,k\!-\!1}^{\prime})\|_1\notag\\
	&+\hspace{-0.5em}\sup_{\text{Adj}(\mathcal{D}_i,\mathcal{D}_i^\prime)}\hspace{-0.5em}\|g_{i,k}-g_{i,k}^{\prime}\|_1+\hspace{-0.5em}\sup_{\text{Adj}(\mathcal{D}_i,\mathcal{D}_i^\prime)}\hspace{-0.5em}\|g_{i,k\!-\!1}-g_{i,k\!-\!1}^{\prime}\|_1.
\end{align}
By $\text{Adj}(\mathcal{D}_i,\mathcal{D}_i^\prime)$, there exists at most one pair of data samples $\lambda_{i,k\!-\!1,l_k},\lambda_{i,k\!-\!1,l_k}^{\prime}$ in $\mathcal{D}_{i,k\!-\!1},\mathcal{D}_{i,k\!-\!1}^\prime$ such that $\lambda_{i,k\!-\!1,l_k}=\xi_{i,l_0}$, $\lambda_{i,k\!-\!1,l_k}^{\prime}=\xi_{i,l_0}^{\prime}$, and there exists at most one pair of data samples $\lambda_{i,k,l_{k\!+\!1}},\lambda_{i,k,l_{k\!+\!1}}^{\prime}$ such that $\lambda_{i,k,l_{k\!+\!1}}=\xi_{i,l_0}$, $\lambda_{i,k,l_{k\!+\!1}}^{\prime}=\xi_{i,l_0}^{\prime}$. Thus, \eqref{63.1} can be rewritten as\vspace{-0.6em}
\begin{align}\label{63.5}
	\|\Delta y_{i,k}\|_1\leq|1\!-\!\beta_K\hspace{-0.75em}\sum_{j\in\mathcal{N}_{\mathcal{C},i}^{+}}\hspace{-0.75em}\mathcal{C}_{ji}|\|\Delta y_{k-1}\|_1+\frac{2C}{m_K}.
\end{align}
{\vskip -5pt}\noindent Iteratively computing \eqref{63.5} implies\vspace{-0.6em}
\begin{align}\label{64}
	\hspace{-1.2em}\|\Delta y_{i,k}\|_1\!\!\leq\!\!\sum_{l=0}^{k-1}\!|1\!\!-\!\!\beta_K\hspace{-0.75em}\sum_{j\in\mathcal{N}_{\mathcal{C},i}^{+}}\hspace{-0.75em}\mathcal{C}_{ji}|^l\frac{2C}{m_K}\!+\!|1\!\!-\!\!\beta_K\hspace{-0.75em}\sum_{j\in\mathcal{N}_{\mathcal{C},i}^{+}}\hspace{-0.75em}\mathcal{C}_{ji}|^k\!\frac{C}{m_K}.
\end{align}

{\vskip -7pt}\noindent{\bf Step 2:} Next, we compute $\|\Delta x_{i,k}\|_1$ for any $k=0,\dots,K$ and $i\in\mathcal{V}$. When $k=0$, since the initial value $x_{i,0}=x_{i,0}^\prime$ for any $i\in\mathcal{V}$, we have $\|\Delta x_{i,0}\|_1=0$. When $k=1$, we have\vspace{-0.6em}
\begin{align}\label{65}
	\hspace{-1.5em}\|\Delta x_{i,1}\|_1=&\hspace{-0.2em}\sup_{\substack{\mathcal{O}\subseteq\mathbb{R}^{2nd},\\(\breve{x}_0,\breve{y}_0)\in\mathcal{O},\\\text{Adj}(\mathcal{D}_i,\mathcal{D}_i^\prime)}}\hspace{-0.3em}\left\|x_1-x_1^\prime\right\|_1\cr
	\noalign{\vskip -4pt}
	\hspace{-1.5em}=&\hspace{-0.2em}\sup_{\substack{\mathcal{O}\subseteq\mathbb{R}^{2nd},\\(\breve{x}_0,\breve{y}_0)\in\mathcal{O},\\\text{Adj}(\mathcal{D}_i,\mathcal{D}_i^\prime)}}\hspace{-0.3em}\|(1\!-\!\alpha_K\hspace{-0.75em}\sum_{j\in\mathcal{N}_{\mathcal{R},i}^{-}}\hspace{-0.75em}\mathcal{R}_{ij})(\!x_{i,0}\!-\!x_{i,0}^{\prime}\!)\cr
	\hspace{-1.5em}&-\alpha_K\hspace{-0.9em}\sum_{j\in \mathcal{N}_{\mathcal{R},i}^{-}}\hspace{-0.75em}\mathcal{R}_{ij}\!(\breve{x}_{j,0}\!-\!\breve{x}_{j,0}^{\prime})\!-\!\gamma_K(y_{i,0}\!-\!y_{i,0}^{\prime}\!)\|_{\!1}\!.
\end{align}
{\vskip -8pt}\noindent By Definition \ref{def5}, $\breve{x}_{j,0}=z_1=\breve{x}_{j,0}^{\prime}$ holds for any $i\in\mathcal{V}$, $j\in \mathcal{N}_{\mathcal{R},i}^{-}$. Moreover, since the initial value $x_{i,0}=x_{i,0}^\prime$,  \eqref{65} can be \mbox{rewritten as}\vspace{-0.5em}
\begin{align}\label{65.5}
	\|\Delta x_{i,1}\|_1=&\hspace{-0.2em}\sup_{\substack{\mathcal{O}\subseteq\mathbb{R}^{2nd},\\(\breve{x}_0,\breve{y}_0)\in\mathcal{O},\\\text{Adj}(\mathcal{D}_i,\mathcal{D}_i^\prime)}}\hspace{-0.3em}\|\gamma_K(y_{i,0}-y_{i,0}^{\prime})\|_1\cr
	=&\gamma_K\sup_{\text{Adj}(\mathcal{D}_i,\mathcal{D}_i^\prime)}\|\Delta y_{i,0}\|_1.
\end{align}
{\vskip -8pt}\noindent Then, substituting \eqref{60.5} into \eqref{65.6} implies\vspace{-0.2em}
\begin{align}\label{65.6}
	\|\Delta x_{i,1}\|_1\leq\frac{\gamma_K C}{m_K}.
\end{align}

{\vskip -7pt}\indent When $k=2,\dots,K$, we have\vspace{-0.4em}
\begin{align}\label{66}
	&\|\Delta x_{i,k}\|_1\cr
	=&\hspace{-0.2em}\sup_{\substack{\mathcal{O}\subseteq\mathbb{R}^{2nd},\\(\breve{x}_{k\!-\!1},\breve{y}_{k\!-\!1})\in\mathcal{O},\\\text{Adj}(\mathcal{D}_i,\mathcal{D}_i^\prime)}}\hspace{-0.3em}\left\|x_k-x_k^\prime\right\|_1\cr
	\noalign{\vskip -3pt}
	=&\hspace{-0.2em}\sup_{\substack{\mathcal{O}\subseteq\mathbb{R}^{2nd},\\(\breve{x}_{k\!-\!1},\breve{y}_{k\!-\!1})\in\mathcal{O},\\\text{Adj}(\mathcal{D}_i,\mathcal{D}_i^\prime)}}\hspace{-0.3em}\|(1\!-\!\alpha_K\hspace{-0.75em}\sum_{j\in\mathcal{N}_{\mathcal{R},i}^{-}}\hspace{-0.75em}\mathcal{R}_{ij})(x_{i,k\!-\!1}-x_{i,k\!-\!1}^{\prime})\cr
	\noalign{\vskip -3pt}
	&\!-\!\alpha_K\hspace{-0.75em}\sum_{j\in \mathcal{N}_{\mathcal{R},i}^{-}}\hspace{-0.8em}\mathcal{R}_{ij}(\breve{x}_{j,k\!-\!1}\!\!-\!\!\breve{x}_{j,k\!-\!1}^{\prime})\!\!-\!\gamma_K(y_{i,k\!-\!1}\!\!-\!\!y_{i,k\!-\!1}^{\prime})\|_1\!.
\end{align}
{\vskip -8pt}\noindent By Definition \ref{def5}, $\breve{x}_{j,k-1}=z_1=\breve{x}_{j,k-1}^\prime$ holds for any $j\in\mathcal{N}_{\mathcal{R},i}^{-}$, $i\in\mathcal{V}$. Then, \eqref{66} can be rewritten as\vspace{-0.5em}
\begin{align}\label{66.1}
	\|\Delta x_{i,k}\|_1=&\sup_{\text{Adj}(\mathcal{D}_i,\mathcal{D}_i^\prime)}\!\|(1\!-\!\alpha_K\hspace{-0.5em}\sum_{j\in\mathcal{N}_{\mathcal{R},i}^{-}}\hspace{-0.5em}\mathcal{R}_{ij})(\!x_{i,k\!-\!1}\!-\!x_{i,k\!-\!1}^{\prime})\notag\\
	&-\gamma_K(y_{i,k\!-\!1}\!-\!y_{i,k\!-\!1}^{\prime})\|_1\cr
	\leq&\sup_{\text{Adj}(\mathcal{D}_i,\mathcal{D}_i^\prime)}\!\|(1\!-\!\alpha_K\hspace{-0.5em}\sum_{j\in\mathcal{N}_{\mathcal{R},i}^{-}}\hspace{-0.5em}\mathcal{R}_{ij})(\!x_{i,k\!-\!1}\!-\!x_{i,k\!-\!1}^{\prime})\|_1\cr
	&+\sup_{\text{Adj}(\mathcal{D}_i,\mathcal{D}_i^\prime)}\|\gamma_K(y_{i,k\!-\!1}\!-\!y_{i,k\!-\!1}^{\prime})\|_1\cr
	=&|(1\!-\!\alpha_K\hspace{-0.75em}\sum_{j\in\mathcal{N}_{\mathcal{R},i}^{-}}\hspace{-0.75em}\mathcal{R}_{ij})|\sup_{\text{Adj}(\mathcal{D}_i,\mathcal{D}_i^\prime)}\|x_{i,k\!-\!1}-x_{i,k\!-\!1}^{\prime}\|_1\cr
	&+\gamma_K\sup_{\text{Adj}(\mathcal{D}_i,\mathcal{D}_i^\prime)}\|y_{i,k\!-\!1}\!-\!y_{i,k\!-\!1}^{\prime}\|_1.
\end{align}
{\vskip -5pt}\noindent Note that $\sup_{\text{Adj}(\mathcal{D}_i,\mathcal{D}_i^\prime)}\|x_{i,k-1}-x_{i,k-1}^{\prime}\|_1=\|\Delta x_{i,k-1}\|_1$ and $\sup_{\text{Adj}(\mathcal{D}_i,\mathcal{D}_i^\prime)}\|y_{i,k-1}-y_{i,k-1}^{\prime}\|_1=\|\Delta y_{i,k-1}\|_1$. Then, \eqref{66.1} can be rewritten as\vspace{-0.5em}
\begin{align}\label{67}
	\hspace{-1em}\|\Delta x_{i,k}\|_1\leq&|1\!-\!\alpha_K\hspace{-0.75em}\sum_{j\in\mathcal{N}_{\mathcal{R},i}^{-}}\hspace{-0.75em}\mathcal{R}_{ij}|\|\Delta x_{i,k\!-\!1}\|_1\!+\!\gamma_K\|\Delta y_{i,k\!-\!1}\|_1.
\end{align}
{\vskip -5pt}\noindent Iteratively computing \eqref{67} implies\vspace{-0.5em}
\begin{align}\label{68}
	\|\Delta x_{i,k}\|_1\!\leq\gamma_K\hspace{-0.3em}\sum_{l=0}^{k-1}\!|1\!-\!\alpha_K\hspace{-0.75em}\sum_{j\in\mathcal{N}_{\mathcal{R},i}^{-}}\hspace{-0.75em}\mathcal{R}_{ij}|^{k-l-1}\|\Delta y_{i,l}\|_1.
\end{align}
{\vskip -5pt}\noindent Therefore, by \eqref{60.5}, \eqref{62}, \eqref{64}, \eqref{65.6} and \eqref{68}, this lemma \mbox{is proved.} \hfill$\blacksquare$\vspace{-1em}
\section{Proof of Lemma \ref{lemma5}}\label{appendix g}
For any $i$$\in$$\mathcal{V}$ and observation set $\mathcal{O}$$\subseteq$$\mathbb{R}^{2(K+1)d}$, let $\mathcal{K}_{\mathcal{D}_i,\mathcal{O}}=\{(\zeta_{i,0},\eta_{i,0},\dots,\zeta_{i,K},\eta_{i,K})\text{: }\mathcal{M}(\mathcal{D}_i)$$\in$$\mathcal{O}\}$, $\mathcal{K}_{\mathcal{D}_i^\prime,\mathcal{O}}=\{(\zeta_{i,0}^\prime,\eta_{i,0}^\prime,\dots,\zeta_{i,K}^\prime$, $\eta_{i,K}^\prime)\text{: }\mathcal{M}(\mathcal{D}_i^\prime)$$\in$$\mathcal{O}\}$ be sets of all possible state and tracking variables under the observation set $\mathcal{O}$ for $\text{Adj}(\mathcal{D}_i,\mathcal{D}_i^\prime)$, respectively. Then, for any $(\zeta_{i,0},\eta_{i,0},\dots,\zeta_{i,K},\eta_{i,K})$$\in$$\mathcal{K}_{\mathcal{D}_i,\mathcal{O}}$ there exists a unique $(\zeta_{i,0}^\prime,\eta_{i,0}^\prime,\dots,\zeta_{i,K}^\prime,\eta_{i,K}^\prime)$$\in$$\mathcal{K}_{\mathcal{D}_i^\prime,\mathcal{O}}$ such that $(\breve{x}_{i,0},\breve{y}_{i,0},\dots,\breve{x}_{i,K},\breve{y}_{i,K})$$=$$(\breve{x}_{i,0}^\prime,\breve{y}_{i,0}^\prime,\dots,\breve{x}_{i,K}^\prime,\breve{y}_{i,K}^\prime)$. Thus, we can define a bijection $\mathcal{B}:\mathcal{K}_{\mathcal{D}_i,\mathcal{O}}\rightarrow\mathcal{K}_{\mathcal{D}_i^\prime,\mathcal{O}}$ such that $\mathcal{B}((\zeta_{i,0},\eta_{i,0},\dots,\zeta_{i,K},\eta_{i,K}))$$=$$(\zeta_{i,0}^\prime,\eta_{i,0}^\prime,\dots,\zeta_{i,K}^\prime,\eta_{i,K}^\prime)$ satisfies
\begin{align}\label{68.5}
		&(x_{i,0}+\zeta_{i,0},y_{i,0}+\eta_{i,0},\dots,x_{i,K}+\zeta_{i,K},y_{i,K}+\eta_{i,K})\cr
		\hspace{-1em}=&(\breve{x}_{i,0},\breve{y}_{i,0},\dots,\breve{x}_{i,K},\breve{y}_{i,K})=(\breve{x}_{i,0}^\prime,\breve{y}_{i,0}^\prime,\dots,\breve{x}_{i,K}^\prime,\breve{y}_{i,K}^\prime)\cr
		\hspace{-1em}=&(x_{i,0}^\prime+\zeta_{i,0}^\prime, y_{i,0}^\prime+\eta_{i,0}^\prime,\dots,x_{i,K}^\prime+\zeta_{i,K}^\prime, y_{i,K}^\prime+\eta_{i,K}^\prime).
\end{align}
\noindent Let $x_{i,k}^{(l)}$, $y_{i,k}^{(l)}$, $\zeta_{i,k}^{(l)}$, $\eta_{i,k}^{(l)}$, $x_{i,k}^{(l)\prime}$, $y_{i,k}^{(l)\prime}$, $\zeta_{i,k}^{(l)\prime}$, $\eta_{i,k}^{(l)\prime}$ be the $l$-th coordinate of $x_{i,k}$, $y_{i,k}$, $\zeta_{i,k}$, $\eta_{i,k}$, $x_{i,k}^\prime$, $y_{i,k}^\prime$, $\zeta_{i,k}^\prime$, $\eta_{i,k}^\prime$, respectively. Then, by \eqref{68.5}, the following holds for any $k=0,\dots,K$ and $l=1,\dots,d$:
\begin{align}\label{69}
	x_{i,k}^{(l)}-x_{i,k}^{(l)\prime}=&\zeta_{i,k}^{(l)\prime}-\zeta_{i,k}^{(l)},\cr
	y_{i,k}^{(l)}-y_{i,k}^{(l)\prime}=&\eta_{i,k}^{(l)\prime}-\eta_{i,k}^{(l)}.
\end{align}
\noindent Note that probability density functions of $(\zeta_{i,0},\eta_{i,0},\dots$, $\zeta_{i,K},\eta_{i,K})$ and $(\zeta_{i,0}^\prime,\eta_{i,0}^\prime,\dots,\zeta_{i,K}^\prime,\eta_{i,K}^\prime)$ are given as follows, respectively:
\begin{align}\label{70}
		p(\zeta_i,\eta_i)=&\prod_{k=0}^{K}\prod_{l=1}^{d}p(\zeta_{i,k}^{(l)};\sigma_k^{(\zeta_i)})p(\eta_{i,k}^{(l)};\sigma_k^{(\eta_i)}),\cr
		\noalign{\vskip -3pt}
		p(\zeta_i^\prime,\eta_i^\prime)=&\prod_{k=0}^{K}\prod_{l=1}^{d}p(\zeta_{i,k}^{(l)\prime};\sigma_k^{(\zeta_i)})p(\eta_{i,k}^{(l)\prime};\sigma_k^{(\eta_i)}).
\end{align}
\noindent Then, by \eqref{70}, $\frac{p(\zeta_i,\eta_i)}{p(\mathcal{B}(\zeta_i,\eta_i))}$ can be rewritten as
\begin{align}\label{71}
		&\frac{p(\zeta_i,\eta_i)}{p(\mathcal{B}(\zeta_i,\eta_i))}\cr
		=&\prod_{k=0}^{K}\prod_{l=1}^{d}\frac{p(\zeta_{i,k}^{(l)};\sigma_k^{(\zeta_i)})p(\eta_{i,k}^{(l)};\sigma_k^{(\eta_i)})}{p(\zeta_{i,k}^{(l)\prime};\sigma_k^{(\zeta_i)})p(\eta_{i,k}^{(l)\prime};\sigma_k^{(\eta_i)})}\cr
		=&\prod_{k=0}^{K}\prod_{l=1}^{d}\exp\!\left(\!\frac{|\zeta_{i,k}^{(l)\prime}|\!-\!|\zeta_{i,k}^{(l)}|}{\sigma_k^{(\zeta_i)}}\!\right)\exp\!\left(\!\frac{|\eta_{i,k}^{(l)\prime}|\!-\!|\eta_{i,k}^{(l)}|}{\sigma_k^{(\eta_i)}}\!\right)\cr
		\leq&\prod_{k=0}^{K}\prod_{l=1}^{d}\exp\!\left(\!\frac{|\zeta_{i,k}^{(l)\prime}\!-\!\zeta_{i,k}^{(l)}|}{\sigma_k^{(\zeta_i)}}\!\right)\exp\!\left(\!\frac{|\eta_{i,k}^{(l)\prime}\!-\!\eta_{i,k}^{(l)}|}{\sigma_k^{(\eta_i)}}\!\right)\!.
\end{align}
\noindent Substituting \eqref{69} into \eqref{71} implies\vspace{-0.5em}
\begin{align}\label{72}
	\hspace{-1em}\frac{p(\zeta_i,\eta_i)}{p(\mathcal{B}(\zeta_i,\eta_i))}\leq&\prod_{k=0}^{K}\prod_{l=1}^{d}\!\exp\!\!\left(\!\frac{|x_{i,k}^{(l)}\!-\!x_{i,k}^{(l)\prime}|}{\sigma_k^{(\zeta_i)}}\!\right)\!\exp\!\!\left(\!\frac{|y_{i,k}^{(l)}\!-\!y_{i,k}^{(l)\prime}|}{\sigma_k^{(\eta_i)}}\!\right)\notag\\
	=&\prod_{k=0}^{K}\exp\!\left(\!\frac{\|x_{i,k}\!-\!x_{i,k}^\prime\|_1}{\sigma_k^{(\zeta_i)}}\!\right)\exp\!\left(\!\frac{\|y_{i,k}\!-\!y_{i,k}^\prime\|_1}{\sigma_k^{(\eta_i)}}\!\right)\notag\\
	\leq&\exp\left(\sum_{k=0}^{K}\left(\frac{\|\Delta x_{i,k}\|_1}{\sigma_k^{(\zeta_i)}}+\frac{\|\Delta y_{i,k}\|_1}{\sigma_k^{(\eta_i)}}\right)\right).
\end{align}
\noindent For any $i\in\mathcal{V}$, let $\varepsilon_i=\sum_{k=0}^{K}(\frac{\|\Delta x_{i,k}\|_1}{\sigma_k^{(\zeta_i)}}+\frac{\|\Delta y_{i,k}\|_1}{\sigma_k^{(\eta_i)}})$. Then, by \eqref{72} we have
\begin{align*}
	&\frac{\p(\mathcal{M}(\mathcal{D}_i)\in \mathcal{O})}{\p(\mathcal{M}(\mathcal{D}_i^\prime)\in \mathcal{O})}=\frac{\int_{\mathcal{K}_{\mathcal{D}_i,\mathcal{O}}}p(\zeta_i,\eta_i)\text{d}\zeta_i\text{d}\eta_i}{\int_{\mathcal{K}_{\mathcal{D}_i^\prime,\mathcal{O}}}p(\zeta_i^\prime,\eta_i^\prime)\text{d}\zeta_i^\prime\text{d}\eta_i^\prime}\cr
	=&\frac{\int_{\mathcal{K}_{\!\mathcal{D}_i\!,\mathcal{O}}}\!p(\zeta_i,\eta_i)\text{d}\zeta_i\text{d}\eta_i}{\int_{\mathcal{K}_{\!\mathcal{D}_i^\prime\!,\mathcal{O}}}\!p(\mathcal{B}(\zeta_i,\eta_i))\text{d}\zeta_i^\prime\text{d}\eta_i^\prime}	=\frac{\int_{\mathcal{K}_{\!\mathcal{D}_i\!,\mathcal{O}}}\!p(\zeta_i,\eta_i)\text{d}\zeta_i\text{d}\eta_i}{\int_{\mathcal{B}^{\!-\!1}\!(\mathcal{K}_{\!\mathcal{D}_i^\prime\!,\mathcal{O}})}\!p(\mathcal{B}(\zeta_i,\eta_i))\text{d}\zeta_i\text{d}\eta_i}\cr
	=&\frac{\int_{\mathcal{K}_{\!\mathcal{D}_i\!,\mathcal{O}}}\!p(\zeta_i,\eta_i)\text{d}\zeta_i\text{d}\eta_i}{\int_{\mathcal{K}_{\!\mathcal{D}_i\!,\mathcal{O}}}\!p(\mathcal{B}(\zeta_i,\eta_i))\text{d}\zeta_i\text{d}\eta_i}	\leq\frac{e^{\varepsilon_i}\!\!\int_{\mathcal{K}_{\!\mathcal{D}_i\!,\mathcal{O}}}\!p(\mathcal{B}(\zeta_i,\eta_i))\text{d}\zeta_i\text{d}\eta_i}{\int_{\mathcal{K}_{\!\mathcal{D}_i\!,\mathcal{O}}}\!p(\mathcal{B}(\zeta_i,\eta_i))\text{d}\zeta_i\text{d}\eta_i}\cr
	=&e^{\varepsilon_i}.
\end{align*}
\noindent Therefore, this lemma is proved. \hfill$\blacksquare$

\end{document}